%% file: tolstoyhilltosi_resub.tex
\documentclass{ar2e}
\usepackage{natbib}
\usepackage{graphicx}  
\usepackage{url}
\usepackage{color}
\usepackage{amsmath,amssymb}
\usepackage{lscape}
\usepackage[hyperindex,breaklinks]{hyperref}

\begin{document}

\input psfig.sty

\def\aapr{{Astron. \& Astrophys. Rev.}}
\def\nat{{ Nature }}
\def\aap{{ Astron. \& Astrophys. }}
\def\aaps{{ Astron. \& Astrophys.~Supp.}}
\def\aj{{ Astron.~J. }}
\def\apj{{ Astrophys.~J. }}
\def\araa{{ Ann. Rev. Astron. Astrophys. }}
\def\apjl{{ Astrophys.~J.~Letters }}
\def\apjs{{ Astrophys.~J.~Suppl. }}
\def\apss{{ Astrophys.~Space~Sci. }}
\def\icarus{{ Icarus }}
\def\mnras{{ MNRAS }}
\def\pasp{{ Pub. Astron. Soc. Pacific }}
\def\planss{{ Plan. Space Sci. }}
\def\physrep{{ Phys. Rep.}}
\def\bain{{ Bull.~Astron.~Inst.~Netherlands }}
\def\lesssim{\mathrel{\hbox{\rlap{\hbox{\lower4pt\hbox{$\sim$}}}\hbox{$<$
}}}}

\def\gsim{\;\lower.6ex\hbox{$\sim$}\kern-7.75pt\raise.65ex\hbox{$>$}\;}
\def\lsim{\;\lower.6ex\hbox{$\sim$}\kern-7.75pt\raise.65ex\hbox{$<$}\;}
                                     
\def\cc{\mbox{cm$^{-3}$}}
\def\tauv{\mbox{$\tau_V$}}
\def\av{\mbox{$A_V$}}
\def\ra{\mbox{$\rightarrow$}}
\def\d{^\circ}
\def\h{^{\rm h}}
\def\mi{^{\rm m}}
\def\s{^{\rm s}}
\def\mum{\ts \mu{\rm m}}
\def\mm{\ts {\rm mm}}
\def\cm{\ts {\rm cm}}
\def\percm{\ts {\rm cm}^{-1}}
\def\m{\ts {\rm m}}
\def\kms{\rm{\, km \, s^{-1}}}
\def\K{\ts {\rm K}}
\def\Kkms{\ts {\rm K\ts km\ts s^{-1}}}
\def\kHz{\ts {\rm kHz}}
\def\MHz{\ts {\rm MHz}}
\def\GHz{\ts {\rm GHz}}
\def\pc{\ts {\rm pc}}
\def\kpc{\ts {\rm kpc}}
\def\Mpc{\ts {\rm Mpc}}
\def\cmsq{\ts {\rm cm^2}}
\def\pcsq{\ts {\rm pc^2}}
\def\dsq{\ts {\rm deg^2}}
\def\debye{\ts10^{-18}\ts {\rm esu}\ts {\rm cm}}
\def\swash2o{$1_{10} - 1_{01}$}             

\let\ap=\approx
\let\ts=\thinspace

\def\an{{ Astronomische Nachrichten }}
\def\sci{{ Science }}
\def\prl{{ Phys. Rev. Lett. }}
\def\zfa{{ Zeitschrift fur Astrophysik }}
\def\ba{{ Baltic Astronomy }}
\def\rmp{{ Rev. Mod. Phys. }}
\def\rpp{{ Rep. Prog. Phys. }} 
\def\pasj{{ Pub. Astron. Soc. Japan }}
\def\pr{{ Phys. Rev. }}
\def\grg{{ Gen. Rel. Grav. }}
\def\sitz{{ Sitzungsber. K. Akad. }}

\def\gsim{\;\lower.6ex\hbox{$\sim$}\kern-7.75pt\raise.65ex\hbox{$>$}\;}
\def\lsim{\;\lower.6ex\hbox{$\sim$}\kern-7.75pt\raise.65ex\hbox{$<$}\;}

\jname{Annu. Rev. Astron. Astrophys.}
\jyear{2009}
\jvol{47}
\ARinfo{1056-8700/97/0610-00}

\date \today

\title{Star Formation Histories, Abundances and Kinematics of Dwarf Galaxies 
in the Local Group}

\markboth{Tolstoy, Hill \& Tosi}{Dwarf Galaxies}

\author{Eline Tolstoy, 
\affiliation{Kapteyn Astronomical Institute, University of Groningen, 
Postbus 800, 9700AV Groningen, the Netherlands; email:etolstoy@astro.rug.nl
}
Vanessa Hill 
\affiliation{Departement Cassiopee, Universit\'{e} de Nice Sophia-Antipolis, 
Observatoire de la C\^{o}te d'Azur, 
CNRS, bd. de l'Observatoire, B.P. 4229, F-06304 Nice Cedex 4, France; 
email: Vanessa.Hill@oca.eu}
Monica Tosi
\affiliation{INAF - Osservatorio Astronomico di Bologna, Via Ranzani 1, I-40127 Bologna, Italy; email: monica.tosi@oabo.inaf.it}
}


\begin{abstract}

Within the Local Universe galaxies can be studied in great detail star
by star, and here we review the results of quantitative studies 
in nearby dwarf galaxies.  The
Color-Magnitude Diagram synthesis method is well established
as the most accurate way to determine star formation
history of galaxies back to the earliest times. This approach received
a large boost from the exceptional data sets that wide field CCD
imagers on the ground and the Hubble Space Telescope could provide.
Spectroscopic studies using large ground based telescopes such as VLT,
Magellan, Keck and HET have allowed the determination of abundances
and kinematics for significant samples of stars in nearby dwarf
galaxies. These studies have shown how the properties of stellar
populations can vary spatially and temporally. This leads to important
constraints to theories of galaxy formation and evolution.  The
combination of spectroscopy and imaging and what they have taught us about
dwarf galaxy formation and evolution is the aim of this review.


\newpage

\end{abstract}

\maketitle
\newpage
{\it ... Les gens ont des \'etoiles qui ne sont pas les m\^emes. Pour les uns,
qui voyagent, les \'etoiles sont des guides. Pour d'autres elle ne sont rien
que de petites lumi\`eres. Pour d'autres qui sont des savants elles sont des
probl\`emes.} 

\noindent\small{(Antoine de Saint-Exup\'ery, Le Petit Prince, XXVI)}

\section{Introduction}\label{intro}

What is a dwarf galaxy? Past definitions always focus on size
\citep[e.g.,][]{Hodge71, Tammann94}, and the presence of a dark matter
halo \citep[e.g.,][]{Mateo98}.  Is there any other physical property
that distinguishes a dwarf galaxy from bigger galaxies?  Are the
differences merely due to the amount of baryonic matter that is
retained by a system during its evolution? In general, large late-type
galaxies sit on the constant central surface brightness ridge defined
by \citet{Freeman70}, and appear to have managed to retain most of the
baryons they started with. Conversely for galaxies which lie below
this limit it seems that the fainter they are, the higher the fraction
of baryons they have lost. This could be due to Supernova winds and/or
tidal interactions, which are effective when a galaxy lacks a suitably
deep potential well to be able to hold onto to its gas and/or metals.
The galaxies above this central surface brightness limit are either
currently forming stars very actively, such as Blue Compact Dwarfs
(BCDs), or they have had very active star formation activity in the
past (e.g., Elliptical galaxies).

\begin{figure}[ht]
\centerline{\psfig{file=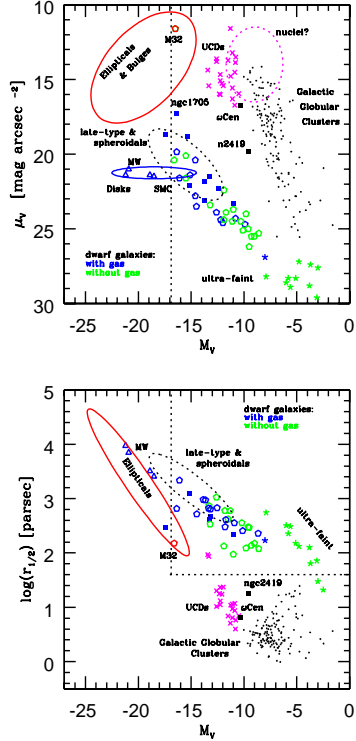,height=10.7cm}}
\begin{center}
\caption{
Here are plotted the relationships between structural properties for
different types of galaxies \citep[after][]{Kormendy85b, Binggeli94,
Kormendy08}, including as dotted lines the classical limits of the
dwarf galaxy class as defined by \citet{Tammann94}.  In the upper plot
we show the absolute magnitude, M$_V$, vs. central surface brightness,
$\mu_V$, plane, and in the lower plot the M$_V$ vs.  half light radius,
r$_{1/2}$, plane.  Marked with coloured ellipses are the typical
locations of Elliptical galaxies \& bulges (red), spiral galaxy
disks (blue), galactic nuclei (dashed magenta) and large early
(spheroidals) and late-type systems (dashed black).  Galactic globular
clusters are plotted individually as small black points.  M~31, the
Milky Way (MW), M~33 and LMC are shown as blue open triangles.  Some
of the BCDs with well studied CMDs are marked as blue solid
squares. The peculiar globular clusters $\omega$~Cen and NGC~2419 are
marked close to the globular cluster ellipse; M~32 in the region of
Elliptical galaxies; the SMC near the border of the dwarf class.  The
Ultra-compact dwarfs (UCDs) studied in the Virgo and Fornax clusters
are marked with magenta crosses. Local Group dwarf galaxies are
plotted as open pentagons, blue for systems with gas and green for
systems without gas.  The recently discovered uFds are given star
symbols, and the same colour code. For references and discussion see text.
}
\label{fig-bing}
\end{center}
\end{figure}

The definition of a galaxy as a dark matter halo naturally excludes
globular clusters, which are believed not to contain any dark matter,
and also do not contain complex stellar populations or any evidence of
enrichment. The structural properties of globular clusters (see
Fig.~\ref{fig-bing}) tend to support the idea that they are
distinct from galaxies.  This definition also excludes tidal dwarfs,
and indeed we do not consider them here as they are more a probe of
the disruption of large systems. They are a different category
of objects that formed much later than the epoch of galaxy
formation. There are also no obvious nearby examples of tidal dwarfs
where the resolved stellar population can be accurately studied.

Here we aim to build upon the outstanding review of \citet{Mateo98}
and leave behind the idea that dwarf galaxies are in any way special
systems.  Many galactic properties (e.g., potential well, metallicity,
size) correlate with mass and luminosity, and all types of galaxies
show continuous relations in structural, kinematic and population
features between the biggest and the smallest of their kind (e.g., see
Fig.~\ref{fig-bing}).  Part of our aim in this review is to
investigate these trends and learn from them.  The only justification
to segregate dwarf galaxies from other types is to study specific
aspects of galaxy formation and evolution on a small scale.

The taxonomy of dwarf galaxies typically opens a {\it Pandora's box}.  At a
very influential conference held at the Observatoire de Haute-Provence
in 1993 G. Tammann gave a working definition: all galaxies that are
fainter than M$_B \le -16$ (M$_V \le -17$) and more spatially extended
than globular clusters (see dotted lines in Fig.~\ref{fig-bing}) are
dwarf galaxies \citep{Tammann94}. This is broadly consistent with the
limit of mass at which outflows tend to significantly affect the
baryonic mass of a galaxy.  This includes a number of different types:
early-type dwarf spheroidals (dSphs); late-type star-forming dwarf
irregulars (dIs); the recently discovered very low surface brightness,
ultra-faint, dwarfs (uFd); centrally concentrated actively
star-forming BCDs. The new
class of even more extreme ultra-compact dwarfs (UCDs) are
identified as dwarf galaxies form spectra but are of a similar
compactness to globular clusters (see Fig.~\ref{fig-bing}).

As has been stated throughout the years \citep[][and references
therein]{Kormendy08}, a morphological classification is only useful if
it incorporates a physical understanding of the processes
involved. However, at present this understanding is not complete and
hence structural parameters and their relations may give us clues to
the underlying physics. But we also have to be careful not to
over-interpret these global measures, especially when we cross
over from structurally simple to more complex systems (e.g., from
early type spheroidals to late type disk-halo star-forming systems). 
This requires care to establish a
meaningful comparison of the same properties of such different
systems.  This has most commonly been done using basic parameters such
as surface brightness and absolute magnitude and physical size of the
systems.  In Fig.~\ref{fig-bing} we show these familiar
relations. These kinds of plots were first made by
\citet{Kormendy85b}, and have been used to great effect by
\citet{Binggeli94} and more recently by \citet{Belokurov07}. 

Fig.~\ref{fig-bing} illustrates how dwarfs compare with all other
galaxies with no real evidence of a discontinuity, as already noted by
\cite{Kormendy85b}.  From a comparison of the absolute magnitude
(M$_V$) and central surface brightness ($\mu_V$) of galaxies (upper
plot in Fig.~\ref{fig-bing}), the early and late-type dwarfs
\citep[from][]{Irwin95, Mateo98, Whiting99, Hunter06} appear to fall
along similar relations, overlapping with BCDs and other larger
late-type systems \citep[from][]{Hunter06} as well as faint spiral
galaxy disks and those galaxies defined as spheroidals by
\citet{Kormendy08}.  
This means systems which resemble late-type
galaxies in their structural properties but are no longer forming stars.
The uFds are clearly separated but arguably
follow the same relation as the other dwarfs \citep[from][]{Simon07,
Martin08struct}.  There are clear distinctions in Fig.~\ref{fig-bing}
between elliptical galaxies \citep[from][]{Faber97, Kormendy08} and
other types, with the exception of spiral galaxy bulges. Similarly
there are also clear distinctions between Globular clusters
\citep[from][]{Harris96} and any other type of galaxy, with the
exception of galactic nuclei. The position of M~32 in the Elliptical
galaxy region is consistent with it being a low-luminosity Elliptical
galaxy \citep[e.g.,][]{Wirth84} and not a dwarf galaxy, or even a
tidally stripped larger system.  There is evidence that $\omega$~Cen,
with its clear spread in Main Sequence Turn-offs (MSTOs), Red Giant
Branch (RGB) sequences and chemical abundances, maybe be the stripped
central remnant of an early-type system \citep[e.g.,][]{Lee99,
Pancino00, Bekki03}, and its position in Fig.~\ref{fig-bing} is 
consistent with that of galactic nuclei. It is interesting to note
that $\omega$~Cen and nuclei also lie in the same region as the UCDs
\citep[from][]{Evstig08}.

The Magellanic Clouds move in and out of the dwarf galaxy class, which
is not surprising as at least the Small Magellanic Cloud 
(SMC) lies near the boundary of the
luminosity definition of dwarf class (see Fig.~\ref{fig-bing}).  The
fact that the Magellanic Clouds are interacting with each other and
our Galaxy makes it more difficult to determine their intrinsic
properties.  The Large Magellanic Cloud (LMC) appears to be similar to
low luminosity spiral galaxies, such as M~33, in terms of mass,
luminosity and size. The SMC on the other hand more resembles the
larger dIs in the Local Group (e.g., NGC~6822, IC~1613), with similar mass,
luminosity and metallicity of star-forming regions.

In the lower plot of Fig.\ref{fig-bing} the varying physical size
scales of different galaxy types and globular clusters are shown by
plotting M$_V$ against the half-light radius r$_{1/2}$, after
\citet{Belokurov07}.  In this plot there is a clear (and unsurprising)
trend for more luminous galaxies to be larger.  The Ellipticals
clearly fall on a distinct narrow sequence (which is a projection of
the fundamental plane). Dwarf galaxies, i.e., BCDs,  
late-type and spheroidal galaxies fall along a similar, although
offset, tilted and more scattered relation to the Elliptical galaxies.
``Classical'' Local Group dSphs clearly overlap with Irregular and BCD
types. The uFds appear in a somewhat offset position. This is perhaps
due to difficulties in accurately measuring the size of such diffuse
objects, or it may be a real difference with other dwarf galaxies.

From Fig.~\ref{fig-bing} it can be seen that there is no clear
separation between dwarf galaxies and the larger late-type and
spheroidal systems.  The dIs, BCDs, dSphs, late-type and spheroidal
galaxies tend to overlap with each other in this parameter space.  The
overlapping properties of early and late-type dwarfs has long been
shown as convincing evidence that early-type dwarfs are the same as
late-type systems that have been stripped of their gas
\citep{Kormendy85b}. This is quite different from the distinction
between Ellipticals and Spirals (and Spheroidals), which show a more
fundamental difference \citep{Kormendy08}.  There is no clear break
which distinguishes a dwarf from a larger galaxy, and hence the most
simple definition does not have an obvious physical meaning, as
recognised by \cite{Tammann94}.


Whatever the precise definition of sub-classes, dwarf galaxies cover a
large range of size, surface brightness and distance, and so they are
usually studied with different techniques with varying sensitivity to
detail.  Some galaxies are just easier to study than others
(due to distance, size, concentration, location in the sky,
heliocentric velocity etc.).  This also leads to biases in 
understanding the full distribution of properties of a complete 
sample \citep[e.g.][]{Koposov08}.
Because of this the properties and
inter-relations of the various types of dwarf galaxies are not always
easy to understand.

The classic dichotomy is between early and late-type dwarf galaxies.
It is not easy to compare the properties of dwarf galaxies which have
on-going star formation (e.g., dIs, BCDs), with those that do not
(e.g., dSphs). Indeed, the properties which can be measured, and then
compared, are often different from one type of galaxy to another.  The
dSphs do not contain gas and so their internal kinematics can only be
determined from stellar velocity dispersions. In gas rich dIs, on the
other hand, the internal kinematics can be easily determined from the
gas and their distance makes them challenging targets to determine
stellar velocities from RGB stars. Likewise abundances in dIs
are typically [O/H] measurements in young HII regions whereas in
dSphs they are usually [Fe/H] coming from individual red giant branch
(RGB) stars over a range of ages.

Galaxies in which the individual stars can be resolved are those which
can be studied in the greatest detail. These are primarily to be found
in the Local Group, where individual stars can be resolved and photometered
down to the oldest main sequence turnoffs (MSTOs). This provides the
most accurate star formation histories (SFHs) going back
to the earliest times.  In the Local Group spectra can be taken of individual
RGB stars at high and intermediate resolution, providing 
information about the chemical content as well as the kinematics of a
stellar population.  The most accurate studies of
resolved stellar populations have been made in Local Group dwarf galaxies which
are the numerically dominant constituent \citep[e.g.,][]{Mateo98}.

Historically the first dwarfs to be noticed in the Local Group, leaving out the
Magellanic Clouds which are clearly visible to the naked
eye, were the early-type dwarf satellites of M~31. M~32 and NGC~205
(M~110) were first catalogued by C. Messier in the 1770, NGC~185 by
W. Herschel in 1787 and NGC~147 by J. Herschel in 1829. The spatially
extended but low surface brightness dIs were first noticed somewhat
later, e.g., NGC~6822 (1881, by E.E. Barnard), IC~1613 and WLM (early
1900s, by M. Wolf). In all cases these galaxies were catalogued as
``faint nebulae''.  It was not until the discovery (by H. Leavitt in
1912) and application to NGC~6822 (by E. Hubble 1926) of the Cepheid
distance scale that they were realised to be (dwarf) extra-galactic
systems. In 1938 H. Shapley discovered the first low surface
brightness dwarf spheroidal galaxies, Sculptor (Scl) and Fornax (Fnx),
around the MW.  From the 1930s onwards extensive observing
campaigns led to the compilation of large catalogs of dwarf galaxies
extending beyond the Local Group, such as Zwicky's catalogs and the Uppsala
General Catalog (UGC) initiated by E. Holmberg.

Over the last 50 years there has been a steady stream of new
discoveries of dwarf galaxies in the Local Group, and also in other nearby
groups and clusters.  The Local Group discovery rate has dramatically increased
recently thanks to the Sloan Digital Sky Survey \citep[SDSS;
e.g.,][]{Adelman07} and a new class of uFds
\citep[e.g.,][]{Willman05a, Zucker06b, Belokurov07} has been found
around the MW.  However there remains some uncertainty about the true
nature of these systems.  

Thus dwarf galaxies provide an overview of galaxy evolution in
miniature which will also be relevant to understand the early years of
their larger cousins and important physical processes which govern
star formation and its impact on the surrounding interstellar medium.
There remain issues over the inter-relations between different types
of dwarf galaxies, and what (if any) is the connection with globular
clusters. When these relations are better understood we will be a
significant step closer to understanding the formation and evolution
of all galaxies.


\section{Detailed Star Formation Histories}\label{sfhs}

The field of resolved stellar population studies was initiated by
W. Baade in the 1940s when he first resolved dwarf 
satellites of M~31 into individual stars and from their color
distribution he realised that they were a different ``population''
from what is typically seen in our Galaxy \citep{Baade44a, Baade44b}.
These were simplified using the terms Population~I for young stars and
Population~II for old stars.  Thus the importance of determining
accurate star formation histories (SFHs) of dwarf galaxies was
recognized long ago and over the years many different approaches have
been followed. The earliest quantitative results came from the
determination by \citet{Searle73} of how the color of the integrated
light of different galaxy types reflected their SFH.  This work was
extended and improved upon in an extensive series of papers by
Gallagher, Hunter and collaborators starting in the early 1980s,
\citep[e.g.,][and references therein]{Gallagher84a, Hunter86}.  They
used various different indicators (e.g., colors and spectrophotometry,
H$\alpha$ luminosity, and emission line ratios) to estimate the star formation
rates (SFRs) at different epochs for large samples of late-type
irregular galaxies, including dwarfs.

The transformation in this field occurred around 15 years ago, when
the power and resolution of a new generation of telescopes
(particularly the Hubble Space Telescope, HST) and detectors (large
format CCDs) allowed accurate photometry and thus detailed
Color-Magnitude Diagrams (CMDs) of individual stars in crowded fields
of external galaxies to be made.  The CMD of a stellar system retains
information about the past SFH, as it preserves the imprint of
fundamental evolutionary parameters such as age, metallicity and
Initial Mass Function (IMF) in such a way that it is possible to
disentangle them. 

\begin{figure}[t]
\centerline{\psfig{file=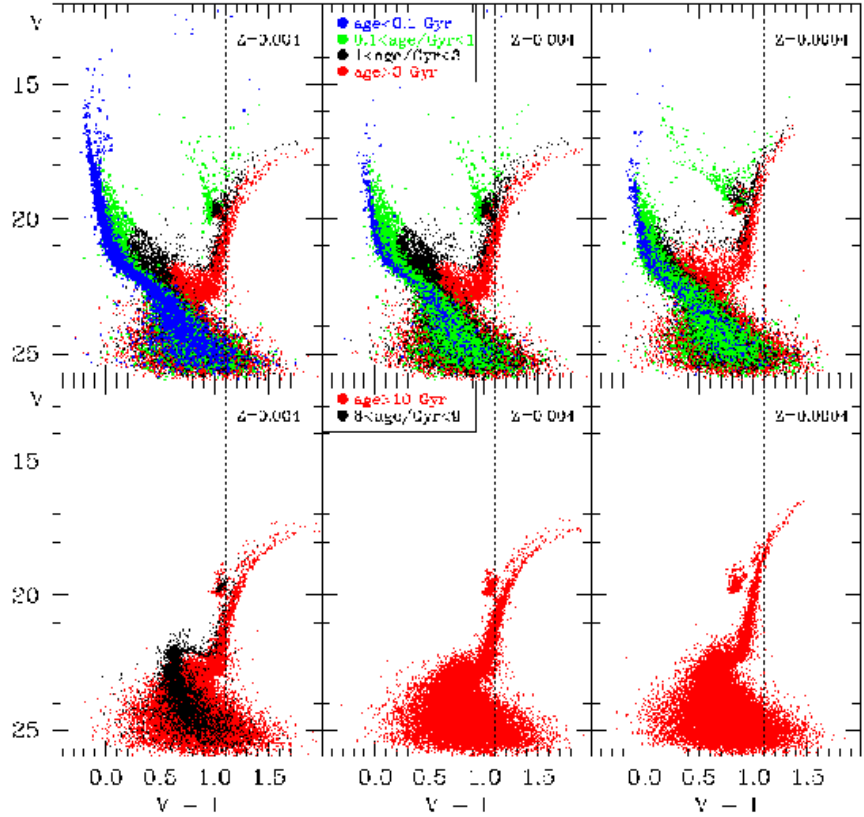,height=4.745in}}
\caption{
The effect on the CMD of different SFHs for a hypothetical galaxy (see
text for more details). All CMDs contain 50000 stars, assume
Salpeter's IMF and are based on the Padova stellar evolution models
\citep{Fagotto94a, Fagotto94b}. A constant metallicity is assumed, and the
value is indicated in the top-right corner of each panel.  In all panels
the colors correspond to different stellar ages. The color codes
for the top and bottom CMDs are shown in the central panel of each
row.  The dotted lines are drawn to help visualise the differences
between the various cases.  Top-central panel: the SFR is constant
from 13 Gyr ago to the present.  Top-left panel: the effect of
concentrating recent SFR into the last 20 Myr.  Top-right panel: the
same SFH as in the top-central panel, but with a ten times lower
metallicity.  Bottom-central panel: an old burst of star
formation overlying a constant SFR from 13 to 10 Gyr ago.  Bottom-left
panel: two old bursts, one with a constant SFR from 13 to 10 Gyr ago
and the other from 9 to 8 Gyr ago, where only 10\% of the stars were
born in the younger burst.  Bottom-right panel: the same old SFH as in
the bottom-central panel, but with a ten times lower metallicity.
}
\label{syn_b}       
\end{figure}

\subsection{Techniques: Synthetic CMD analysis}

At the beginning of last century, stars were found to group themselves
in temperature-luminosity ranges (observed as color and
magnitude), in the Hertzsprung-Russell Diagram; and it was later
understood that the positions of stars in a CMD represent the
evolutionary sequences of stellar populations. Since the 1950s large
numbers of detailed CMDs have been derived for star clusters and
nearby dwarf galaxies \citep[e.g.,][and references therein]{Hodge71}.
However it was not until the advent of modern CCDs and analysis
techniques of the early 1980s that the field really took off for
complex galactic systems, like dwarf galaxies.

Until about twenty years ago all stellar age dating used isochrone
fitting, which is appropriate for simple stellar populations such as
star clusters, but a serious over-simplification for the
interpretation of the composite stellar populations of galaxies. In
galaxies numerous generations of stars, with different metallicities
and ages contribute to the appearance of the observed CMD.  Thus a new
approach was needed to make the most of the new and accurate CMDs, a
method to determine a quantitative SFH: the synthetic CMD method.

The synthetic CMD method determines the variation of the SFR within
the look-back time reached by the available photometry, namely the
SFH. It is based on comparing observed with theoretical CMDs created
via Monte-Carlo based extractions from stellar evolution tracks, or
isochrones, for a variety of star formation laws, IMFs, binary
fractions, age-metallicity relations, etc. Photometric errors,
incompleteness and stellar crowding factors also have to be estimated
and included in the procedure to fully reproduce an observed CMD
\citep[e.g.,][]{Tosi91,Aparicio96,Tolstoy96a,Dolphin97}.  A
combination of assumed parameters is acceptable only if the resulting
synthetic CMD satisfactorily reproduces all the main features of the
observational one. This means morphology, luminosity, color
distribution, and number of stars in specific evolutionary phases.
Different authors use different approaches to assess the quality of
the fit, typically using a form of likelihood analysis comparing
the model and the data within the uncertainties of the measurement
errors. The method is intrinsically statistical in nature and cannot
provide a unique solution for the SFH for a number of reasons, but it
usefully limits the range of possible scenarios
\citep[e.g.,][]{Tolstoy96b, Hernandez00, Dolphin02, Aparicio04}.  The
theoretical uncertainties in the stellar evolution models also
influence the numerical results and have to be treated carefully
\citep[see][for a review]{Gallart05}.

\subsubsection{An Example:}
Fig.~\ref{syn_b} shows examples of how CMDs reflect different SFHs in
a hypothetical galaxy.  Here we have assumed a distance modulus of
(m-M)$_0$=19, reddening E(B-V)=0.08 and photometric errors and
incompleteness typical of HST photometry with the Wide Field Planetary
Camera 2 (WFPC2).  Thus these CMDs could apply to a typical SMC field
observed with the WFPC2.  In all panels the number of stars and the
IMF are the same, and what changes from panel to panel is the
metallicity and the SFH. 
In the top panels
all stellar evolution phases are visible: the blue plume typical of
late-type galaxies, populated by massive and intermediate-mass stars
on the main-sequence, and in the most metal poor case also by brighter
blue loop stars; the red clump and blue loops of stars in the core
helium burning phase; the Asymptotic Giant Branch (AGB) and RGB; the
sub-giant branch; the oldest MSTOs and the main-sequence of the lower
mass stars.
In the lower panels of Fig.~\ref{syn_b} we see a much simpler old
SFH. On the left is the effect of a burst on top of this old
population, and on the right a different metallicity.
Fig.~\ref{syn_b} (the top panels in particular) emphasizes the
challenge in interpreting real CMDs: observed data points don't have
convenient labels indicating their age and metallicity, and unraveling
different sub-populations overlying each other is challenging

One important issue in the derivation of the SFH from a CMD is the
metallicity variation of the stellar population.  If no spectroscopic
abundance information is available (which is unfortunately frequently
the case), the metallicity is assumed to be that of the stellar
evolution models with colors and CMD morphology in best agreement with
the empirical CMD. This is often a particularly uncertain assumption
because some of the key evolutionary sequences in the CMD can be
heavily affected by age-metallicity degeneracy. For instance,
metal-rich RGB stars from a relatively young (a few Gyrs old)
population can occupy the same region in a CMD as a more metal-poor,
but older population.  Without a spectroscopic estimate of the
metallicity, it is impossible to break this degeneracy, unless the
MSTOs are also observed. This is often not the case and one has to
deal with the inevitable uncertainty.  A further aspect of the effect
of metallicity is related to the rather coarse grids of different
initial chemical compositions that stellar evolution models are
actually computed for.  Despite the commendable efforts by stellar
evolution modellers, complete sets of homogeneous models covering the
entire stellar mass range are limited to a few key metallicities
(e.g., Z=0.02, 0.008, 0.004, 0.0004, 0.00004). Thus synthetic CMDs
assuming a smoothly varying age-metallicity relation have to be
created by interpolating between the available metallicities, and this
adds to the uncertainty in the derived SFH.

\begin{figure}[ht]
\centerline{\psfig{file=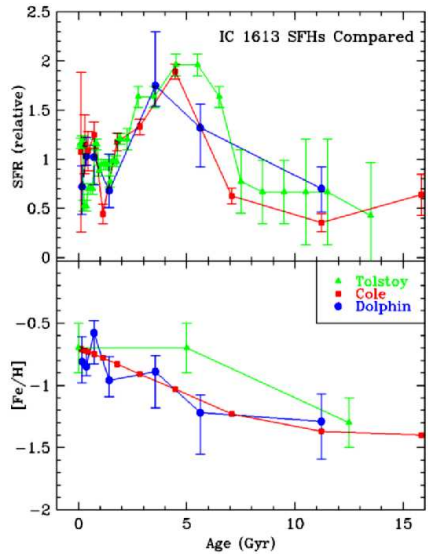,height=3in}}
\caption{
Comparison of SFHs for dI IC~1613 derived via three different methods 
(Cole, Dolphin \& Tolstoy), from
\citet{Skillman03}. Note the enhanced levels of SFR between 3 and 6 Gyr ago 
which appear in all three models. Also shown are the age-metallicity
relations which were derived in each case.
}
\label{sfh_ic1613}
\end{figure}

\subsubsection{Testing the Reliability:}
In 2001 the different procedures to statistically determine SFHs from
about 10 groups, using a variety of different assumptions and stellar
evolution models, were compared in the {\it Coimbra Experiment}
\citep[see,][and references therein]{Skillman02}.  This experiment
showed that, despite all the different assumptions, modelling
procedures and even stellar evolution models, most synthesis methods
provided consistent results within their uncertainties. This was again
shown in the HST/WFPC2 study of the dI IC~1613 \citep{Skillman03}
where a synthetic CMD analysis was carried out independently by 3
different people (representing 3 independent modelling approaches) and
again the results were reassuringly similar (see
Fig.~\ref{sfh_ic1613}). These WFPC2 data, however, did not reach the
oldest MSTOs in IC~1613 and the SFH at the earliest epochs therefore
remained uncertain with these data.  More recent deep observations
with the HST/ACS, as part of the LCID project \citep{Gallart07}, do
reach the faint oldest MSTOs in IC~1613, but in a different field.
Interestingly its CMD is quite different from that of
\citet{Skillman03}, as it lacks the dominant younger component and it
is reproduced by an almost constant SFR with time.  This new LCID ACS
field is situated at a galacto-centric radius similar to that of the
WFPC2 CMD, but, by their nature, irregular galaxies are often
asymmetric and these differences are not surprising. This reminds us
of the dangers of looking at a
small fraction of these complex systems.

With a sufficient investment in telescope time we can resolve
individual stars down to the oldest MSTOs in all the galaxies of the
Local Group, and use the resulting CMDs to infer their SFHs over the
entire Hubble time. This kind of analysis has been obtained for only a
handful of galaxies to-date.  If the oldest MSTOs are not reached, then
the look-back time depends on which features of the CMD can be
resolved: Horizontal Branch (HB) stars are $>$10 Gyr old but hard to
interpret in terms of a SFH, except to say that there are ancient
stars. RGB stars are at least 1$-$2 Gyr old, but without further
information it is impossible to be quantitative about the SFH because
of the age-metallicity degeneracy.  Of course younger populations are
much brighter and obtaining the SFH over the last Gyr, especially in
actively star-forming galaxies, is possible even well beyond the Local Group.

\subsection{Observations: Dwarf Galaxies in the Local Group}

In spite of all the uncertainties, and perhaps because most of them
are well treated in a Monte-Carlo approach, the first applications of
the synthetic CMD method immediately showed the powerful capability to
provide detailed new perspectives \citep[e.g.,][]{Ferraro89, Tosi91,
Greggio93, Marconi95, Gallart96a, Tolstoy96a}.  These early studies
found that the SFH not only differs significantly from one galaxy to
another, but also according to where one looks within the same galaxy.
It was shown that star formation in late-type dwarfs usually occurs in
long episodes of moderate intensity separated by short quiescent
phases \citep[{\it gasping} regime;][]{Marconi95}, rather than in
short episodes of strong intensity separated by long intervals ({\it
bursting} regime).

When WFPC2 became available after the first HST refurbishment at the
end of 1993, it created a tremendous amount of interest and enthusiasm
in the field of SFH research \citep[e.g., see reviews,][and references
therein]{Tolstoy03hst, Dolphin05} because WFPC2 provided such
accurate, well defined and deep CMDs.  In 2002, the Advanced Camera
for Surveys (ACS) on HST yet again improved the possibilities,
reaching a photometric depth and resolution that is likely to remain
unequalled for quite a long time \citep[e.g., see review,][and
references therein]{Tosi07hst}.

To date, a significant fraction of Local Group galaxies have been studied using
the synthetic CMD method to infer their SFH with varying degrees of
depth and accuracy, see Tables~\ref{sfh_late} \& \ref{sfh_early}.
Many of the galaxies in these tables, and several not included, for
which HST/WFPC2 data exist, have been compiled by \cite{Holtzman06},
and their SFH homogeneously derived by \cite{Dolphin05}. Homogeneous
data sets and analyses are valuable to obtain a uniform overview of
dwarf galaxy properties in the Local Group.

Tables~\ref{sfh_late} \& \ref{sfh_early} are presented in a uniform
way to allow an easy comparison between the synthetic CMD analyses for
different galaxies over a range of distances.  In these Tables we do
note if there is supporting evidence for an ancient population (i.e.,
RR~Lyr variable stars) that is not clearly seen in the CMD (e.g., the
HB and/or the oldest MSTOs are not visible). Leo~A is a good example
of this, where the presence of RR~Lyr variable stars show that there
is an ancient stellar population that is not apparent from the CMD
(see Fig.~\ref{sfh_lcid}).  This is also true for NGC~6822, and for
some of the early-type galaxies.

In Tables~\ref{sfh_late} \& \ref{sfh_early} we have included the most
recent distance measurements, with references, in column~2. We have
then updated the absolute magnitude, M$_V$, typically from
\citet{Mateo98}, in column~3. We include the
physical size (the Holmberg radius, r$_h$) of each galaxy in
arc-minutes, in column~4. This is to highlight the fraction of the
area of the galaxy covered by the instrument used to image the galaxy,
which is given in column~5. In columns 6$-$9 we have given an overview
of the depth and detail of the CMD analysis allowed by the different
data sets. Sometimes there is more than one data set per galaxy:
column~6 lists the faintest feature detectable in the CMD; column~7
indicates if populations of $\le 10$~Myr were detected or not; column
8 indicates if populations in the range 2$-$8~Gyr are detected; and
column~9 indicates if populations older than 10~Gyr were detected. In
the case where a column~contains a ``?''  this means that the CMD was
not deep enough to determine if stars in this age range exist. A
column~which contains an ``x'' means that stars of this age were
explicitly not detected. In the last three columns we give an overview of
the spectroscopic measurements that exist for individual stars
(columns 10, 11) and HII regions in column~12.  Column~10 indicates,
with a reference, if individual stars in the galaxy have been observed
at low resolution (R$<$10~000), typically to determine metallicities,
from a single indicator, or kinematics. Column~11 indicates, with a
reference, if individual stars in the galaxy have been observed at
high resolution (R$>$18~000), to determine abundances of 
different elements. In late-type galaxies typically these analyses are
carried out on young massive stars (e.g., B super-giants), and for the
closer by early-type galaxies, which do not contain young stars, this
typically means RGB stars. In column~12 we indicate if
HII region spectroscopy has been carried out (obviously this is only
possible in galaxies with recent star formation).  It should be noted
that we have not included {\it all} synthetic analyses.  In some cases
there are multiple studies of one system, and in this case usually the
most recent is quoted. Sometimes however the older study is not
superseded (usually because it covers a more significant fraction of
the system), and in this case more than one study is listed. In the 
particular case of studies of several large dSphs (Table~\ref{sfh_early}),
we have included more than one study based on the same HST data where the
results were not the same (e.g., Carina dSph).

\begin{figure}[ht]
\centerline{\psfig{file=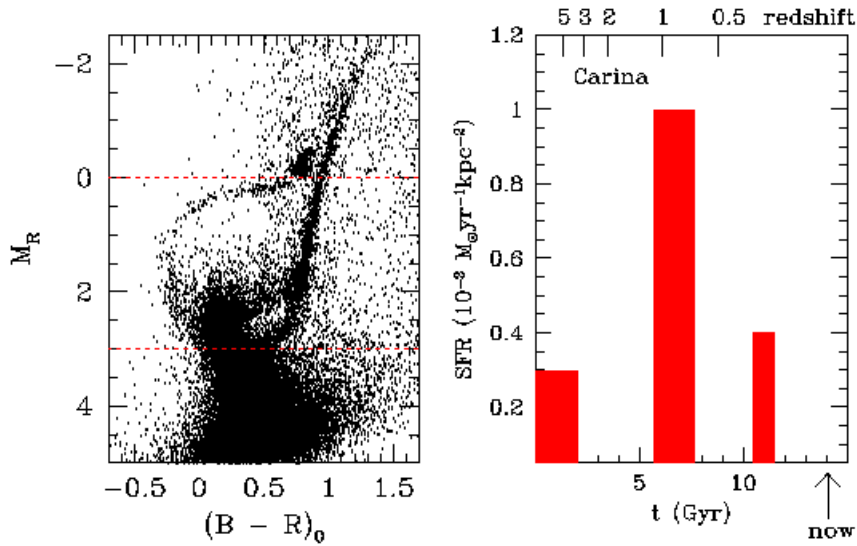,height=8cm}}
\caption{
On the left-hand side is a CMD of the Carina dSph (taken by M. Mateo
with the CTIO4m and MOSAIC camera, private communication) in the
central 30$'$ of the galaxy.  This clearly shows the presence of at
least 3 distinct MSTOs.  On the right-hand side is shown the SFH of
the central region of Carina determined by \citet{Hurley98}, showing
the relative strength of the different bursts.
}
\label{carina}
\end{figure}

The 3D physical spatial distribution of the different
types of dwarf galaxy in the Local Group has been displayed in 
increasing detail over the last years \citep[e.g.,][]{Grebel99} and
most recently by \citet{Mateo08rev}, including the newly discovered
uFds as well as globular clusters and also the most recent version of the
morphology-density relation. This shows that galaxies which are
currently forming stars are preferentially to be found more than
$\sim$300~kpc away from the MW, and thus the difference
between the distribution of dSph and dI around the MW gives
a clear indication of the possibility of morphological transformation. 

\subsubsection{Early-type Dwarf Galaxies:}

Early-type galaxies, such as dSphs, are typically associated with large
galaxies like our own. They are among the systems closest to us, with
the majority at distances $<$~130~kpc, although there are also several
more distant examples. Arguably the new uFds are an
extension of the dSph class down to much lower
luminosities.  The dSph systems typically look very much like the old
extended stellar populations which appear to underlie most late-type
systems. This suggests that the major difference is that they lack gas
and recent star formation, an hypothesis supported by their
overlapping structural properties (see Fig.~\ref{fig-bing}). They have
typically not formed stars for at least several 100~Myr (e.g.,
Fnx), and in several cases for much longer (e.g., the Scl dSph
apparently formed the majority of its stars more than 10~Gyrs ago).

The proximity of dSphs makes it easier to carry out studies of their
resolved stellar populations, although this requires wide field
instrumentation to efficiently gain an overview as they are typically
$>1$~degree across on the sky.  The most famous example is Carina,
which has been much studied over the years. It was one of the first
galaxies shown, from deep wide field imaging on the CTIO~4m telescope,
to have completely distinct episodes of star formation
\citep{Smecker96, Hurley98}, identified by three distinct MSTOs in the
CMD (see Fig.~\ref{carina}, left-hand side).  These distinct MSTOs
translate into a SFH (see Fig.~\ref{carina}, right-hand side) which
consists of three separate episodes of star formation, with the SFR
apparently going to zero in between. The existence of a complex SFH
was already inferred from the properties of its variable stars
\citep{Saha92}, and the red clump and HB morphology
\citep{Smecker94} but it took synthesis analysis of a CMD going down
to the oldest MSTOs to quantify it \citep{Hurley98}. The resulting
SFH is displayed in Fig.~\ref{carina}.

There have been a number of consistent studies of the Galactic dSphs
using HST \citep{Hernandez00,Dolphin02}. These analyses are typically
hampered by the small field of view of HST, compared to the size of
the galaxies. Especially as we now know that even these small systems
have population gradients, the small field-of-view HST studies are
very dependent upon where the telescope is pointing.

\begin{figure}[ht]
\centerline{\psfig{file=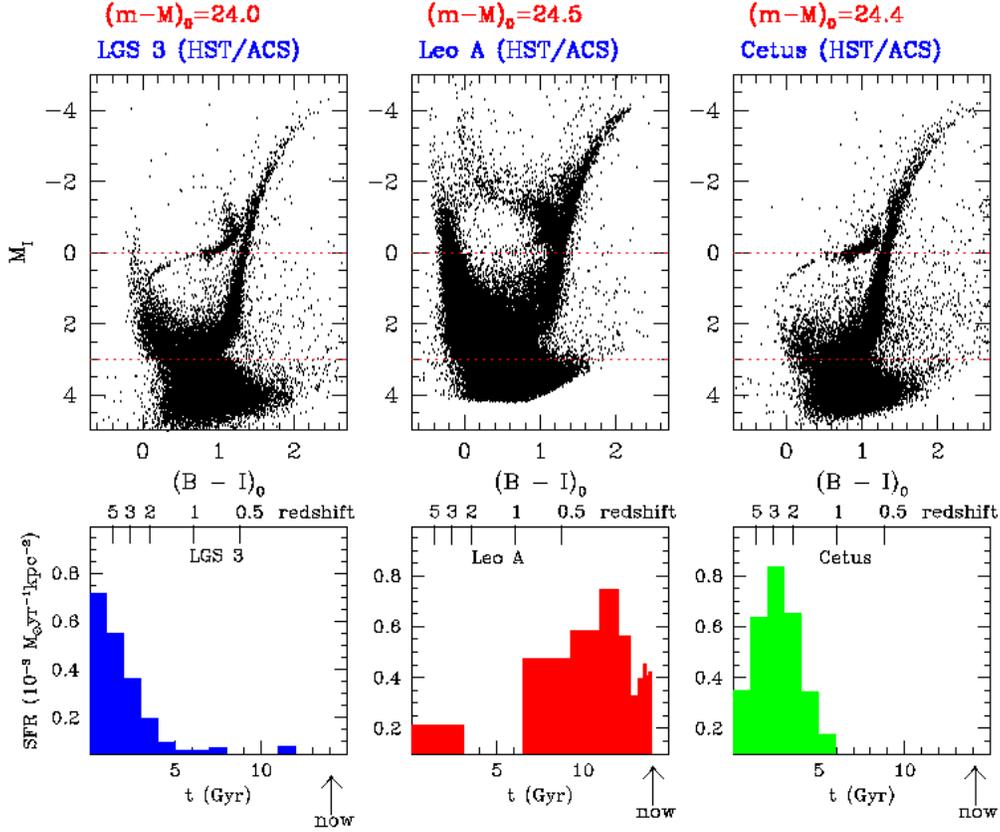,height=12cm}}
\caption{ HST/ACS CMDs and SFHs for three Local Group dwarf galaxies:
LGS~3 a transition type dwarf galaxy (Hildago et al. in prep); Leo~A a
dwarf irregular \citep{Cole07}; and Cetus a distant dwarf spheroidal
galaxy (Monelli et al. in prep.).  These results come from the LCID
project \citep{Gallart07, Cole07}, which is a large programme designed
to exploit the exquisite image quality of the HST/ACS to obtain
uniquely detailed CMDs going back to the oldest MSTOs for a sample of
dwarf galaxies. The SFHs in the lower panels come from synthetic CMD
analysis.  
}
\label{sfh_lcid}
\end{figure}

There are also more distant dSph galaxies, such as Cetus and Tucana,
which display all the characteristics found in the closer-by dSphs,
but they are at distances much beyond the halo of the MW and
M~31. Tucana is at a distance of 880~kpc and Cetus is at 775~kpc (see
Table~\ref{sfh_late}). Both these galaxies have been looked at in
great depth by the LCID HST/ACS programme \citep{Gallart07}.
Preliminary results for Cetus can be seen in the right hand panel of
Fig.~\ref{sfh_lcid} (Monelli et al., in prep.).  It looks very similar
to a predominantly old dSph, like Scl, and it is likely not to
have formed any stars over the last 8~Gyr. The small hint of blue
plume in the CMD (in Fig.~\ref{sfh_lcid}) is most likely due to blue
stragglers. These are old stars that are known to exist in Galactic
dSphs \citep[e.g.,][]{Mapelli07, Momany07}, and which have undergone
mass transfer and appear rejuvenated, but should not be confused with
more recent star formation activity.

\citet{Wirth84}, first made the distinction between
diffuse and compact dEs, namely between NGC~205-like and M~32-like
galaxies, immediately confirmed by \citet{Kormendy85b} for a larger
sample. The issue has been  
comprehensively reviewed by \citet{Kormendy08}, who show that the  physical
properties of M~32 place it as a low-luminosity
Elliptical galaxy (see Fig.~\ref{fig-bing}).
This also suggests that M~32 is not compact
because of any kind of tidal pruning, but because of their intrinsic
evolutionary history and/or formation scenario.

Compact objects like M~32 are rare (there is only one in the Local Group),
whilst more diffuse dwarfs, like dSphs and NGC~205 are much more
common.   Thus the three compact systems around M~31 NGC~205, NGC~185
\& NGC~147 are all big spheroidals, not small ellipticals, as is clear
from their position in Fig.~\ref{fig-bing}. These systems have
typically not had much attention from CMD synthesis modelling, see
Table~\ref{sfh_early}. This is probably due to the fact that they are
quite distant, and compact, which makes accurate photometry very
challenging even with the help of HST.

There is in addition the class of UCDs
which appear to be found only in nearby galaxy clusters, such as
Fnx \citep[e.g.,][]{Evstig08}. They may be objects like
$\omega$~Cen, which is now often considered to be the tidally stripped nucleus
of a compact system. The structural properties of $\omega$~Cen and UCDs 
clearly overlap (see Fig.~\ref{fig-bing}).
They have also been proposed to be
low-luminosity Ellipticals like M~32, but Fig.~\ref{fig-bing} would tend
to argue against this.

\subsubsection{Late-type Dwarf Galaxies:} 

These galaxies have long been well studied in the Local Group, and they have
proved themselves valuable tools for understanding the wider Universe,
starting from the monitoring of Cepheid variable stars in NGC~6822 by
E. Hubble in 1925, and the subsequent realisation that a larger
Universe existed beyond our MW.  The dIs have also long been used as
probes of metal-poor star formation, both young and old. They 
still retain HI gas, and are thus, with a few curious exceptions,
typically forming stars at the present time as they have probably done
over their entire history, with a variety of rates, from extremely low
(e.g., Pegasus) to zero (e.g., transition systems DDO~210, LGS~3) to
relatively high (e.g., NGC~6822, SMC).  The dIs were the first systems
to which synthetic CMD analysis was applied (e.g., WLM, Sextans~B).
They are a numerous and often fairly luminous class within the
Local Group. They are typically at a distance $>$400~kpc (the SMC being a
notable exception), see Table~\ref{sfh_late}.  Studies down
to the oldest MSTOs of dIs typically require HST-like sensitivity and
image stability.

HST has had a large impact on studies of these systems.  The
exceptionally detailed CMDs from WFPC2 allowed for the first time the
clear distinction between the main sequence and the blue loop sequence
in young metal-poor systems \citep[e.g., Sextans~A,][]{Dohm97}. 
Photometric errors previously blended these sequences 
in  ``the blue plume'' and there
were debates about the reliability of the theoretical predictions of 
 blue loop stars. These stars have been
subsequently shown to be powerful tools for mapping the spatial
variations in the SFR over the last 800~Myr  \citep{Dohm98,
Dohm02}. 
 The resulting space/time variations are intriguingly
reminiscent of the predictions of the stochastic self-propagating star
formation proposed by \citet{Seiden79} 30 years ago, with star
formation coming and going in different regions over periods of
several hundred Myr.

The HST/ACS CMD of Leo~A \citep[from][]{Cole07}, see
Fig.~\ref{sfh_lcid} (central panels), 
is one of the deepest and most accurate ever made
for a dI.  The SFR as a function of time over the entire history of
the galaxy was determined using synthetic CMD analysis (see
Fig.\,\ref{sfh_lcid}), and it was found that 90\% of the star
formation in Leo~A happened during the last 8~Gyr. There is a peak in
the SFR 1.5$-$3~Gyr ago, when stars were forming at a level 5$-$10
times the current rate.  The CMD analysis of Leo~A only required a
very slight metallicity evolution with time. The mean inferred
metallicity in the past is consistent with measurements of the
present-day gas-phase oxygen abundance.  There appears to have been
only a small and uncertain amount of star formation in Leo~A at the
earliest times, as the HB is very weak in the CMD in
Fig.~\ref{sfh_lcid}. The error bars on the SFH \citep[see][]{Cole07}
show that from CMD analysis alone this ancient population is not well
defined.  The only definite proof of truly ancient stars in Leo~A
comes from the detection of RR~Lyrae variable stars
\citep{Dolphin02leoa}.

Fig.\ref{sfh_lcid} also shows a preliminary HST/ACS CMD and SFH
derived for LGS~3 (Hildago et al., in prep), which is a transition
type galaxy. This means that it contains HI~gas, but no very young
stars (no HII regions, and no super-giants).  From the CMD it looks
like it has been forming stars at a low rate for a very long time with
a gradually declining rate, and the present day hiatus is just a
normal event in its very low average SFR.

\subsubsection{Ultra-faint dwarf galaxies}

An ever increasing number of extremely faint systems (uFds) are
being found by SDSS around the MW. As displayed in Fig.~\ref{fig-bing}
they appear somewhat offset in the M$_V$ - r$_{1/2}$ plane, although
this may be due to the observational difficulties in accurately
determining their physical extent.  In the M$_V$ - $\mu_V$ plane they
appear to be the extension of the dSph sequence to lower luminosity
rather than a new class of object. However it is clear from both plots
in Fig.~\ref{fig-bing} that especially the fainter of these new
systems exist in a region where the extension of classical dwarf
galaxies {\it and} globular cluster sequences may lie. In several
cases the properties of the uFds appear to resemble more 
diffuse (perhaps tidally disrupted) 
metal poor globular clusters rather than dwarf galaxies. From a
careful study of the structural properties of uFds 
\citep{Martin08struct} it
can be seen that these new systems range in absolute magnitude from
M$_V = -1.5$ (Segue~I) to M$_V = -8.0$ (Leo~T). Leo~T and CVn~I (M$_V
=-7.9$) are the two brightest of these new systems
although they
are measurably fainter and with lower surface brightness than any of
the ``classical'' dwarfs they are consistent with 
a lower luminosity extension of the dI and dSph type galaxies. 
There are $\sim$8 systems at $-7 <$ M$_V <
-4.0$ (Boo~I, UMa~I, UMa~II, Leo~IV, Leo~V, CVn~II, Coma and Her),
and most of the rest are at M$_V > -3.0$ (e.g., Wil~I, Segue~I,
Segue~II and Boo~II \& III).  
The more luminous and populous 
CVn~I contains a mix of Oosterhoff type I \& II RR~Lyr variable
stars \citep{Kuehn08}, as is typical for dSph, whereas the fainter
systems do not. 
So far most of these new uFds have been found
in the immediate vicinity of the MW. The bright systems Leo~T, at
410~kpc, and CVn~I, at 218~kpc are the most distant, and the typical
distances of the fainter systems range between 23~kpc (Seg~I) \&
160~kpc (Leo~IV, CVn~II). These faint and diffuse systems are
challenging to study and it is virtually impossible to detect them
beyond these distances.

Some of these new systems have had their stellar populations analysed
using the synthetic CMD method \citep[e.g.,][]{deJong08sdss}.  However
the SFHs, and even the basic physical properties of the faintest of
these systems can be particularly sensitive to the effect that large
and uncertain Galactic contamination brings to small number statistics
\citep[e.g.,][]{Martin08struct}.  In several cases it is impossible to
distinguish the stars which are in uFds from those in the MW
without spectroscopic follow-up, and even then they are often found to
only contain few RGB stars, or to have kinematics almost
indistinguishable from either the Sagittarius (Sgr) tidal streams or the
Galaxy \citep[e.g.,][]{Geha08}.  This makes separating these systems
out from the surrounding stars and determining their
properties quite challenging.

\begin{figure}[ht]
\centerline{\psfig{file=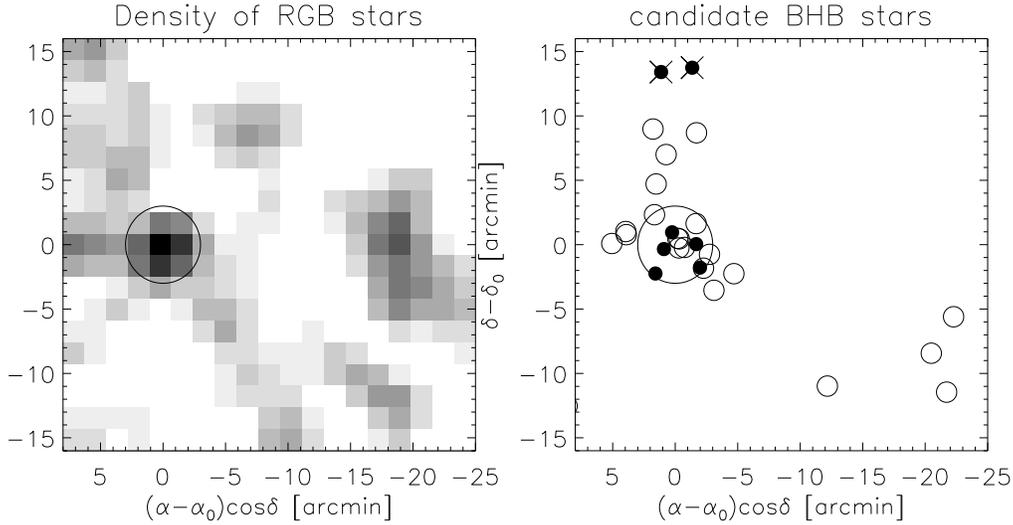,height=2.8in}}
\begin{center}
\caption{ From \citet{Belokurov08}.
Left: The density of RGB candidate members selected from 
photometry. The extent of Leo~V as judged from two half-light radii is 
marked. Right: The locations of BHB candidate members. 
Note that the BHB 
distribution is elongated and more extended than that of the RGB stars. Black 
dots are RGB stars with spectroscopy, $v_\odot \approx 173$ kms$^{-1}$ and low 
equivalent width of the MgT feature.
}
\label{fig-leov}
\end{center}
\end{figure}

Another approach is to look for distinctive stellar populations, such
as blue HB (BHB) stars, or RR~Lyr variable stars which
clearly stand out from the Galactic stellar population. These are
usually still small number tracer populations but at least they are
clear markers of the spatial extent and age of these small and faint
systems \citep[e.g., for Boo~I, ][]{DallOra06}.  In the case
of Leo~V it can be seen from the BHB stars that the galaxy has a much
more extended stellar component than the half-light radius would
suggest \citep[][see Fig.~\ref{fig-leov},]{Belokurov08}. These smallest
systems are clearly being disrupted and understanding what they
were before this process began is challenging.

It is possible that some of these systems are no more than over-density 
enhancements along a stream, and possibly along streams
related to Sgr.  For example, Segue~I has the same space and velocity
distribution of a supposed 
ancient leading arm of Sgr, wrapped 520$^o$ around the 
MW \citep[e.g.,][]{Geha08}. Similarly Boo~II and Coma are believed
to lie within streams originating from Sgr.
In the case of Leo~IV and Leo~V they lie
on top of (although clearly behind) the Orphan stream, and thus 
kinematic information is needed to hope to 
disentangle their stars from the complex fore/background
stellar populations lying in that direction. The newly discovered
Segue~II system \citep{Belokurov09} is also found to lie along the
edge of a Sgr stream and to perhaps be embedded in a
stream of its own. In this
case it is postulated to be evidence for groups of galaxies falling
onto the MW simultaneously. 

\subsubsection{The Small Magellanic Cloud:}

The closest galaxies (excepting some of the new uFds)
are the Magellanic Clouds, and their SFHs can be studied in quite some
detail.  The SMC is an irregular galaxy at the boundary of the dwarf 
class. As can be seen from Fig.~\ref{fig-bing}, this
does not have a clear physical distinction, and the cut-off is
arbitrary.  Hence, the SMC can be considered the closest late-type
dwarf.  It shares key properties with this type of galaxy: high gas
content, low current metallicity (Z$\simeq$0.004 in mass fraction) and 
low mass \citep[between 1 and 5 $\times 10^9M_{\odot}$][]{Kallivayalil06}, 
near the upper limit of the range of masses typical of late-type dwarfs.

The SMC hosts several hundreds star clusters 
and several populous
clusters, covering all ages from 11~Gyr \citep[NGC~121, e.g.,][]{Glatt08} to a
few Myr \citep[e.g., NGC~346 and NGC~602,][]{Sabbi07,Cignoni09}.

\begin{figure}[ht]
\centerline{\psfig{file=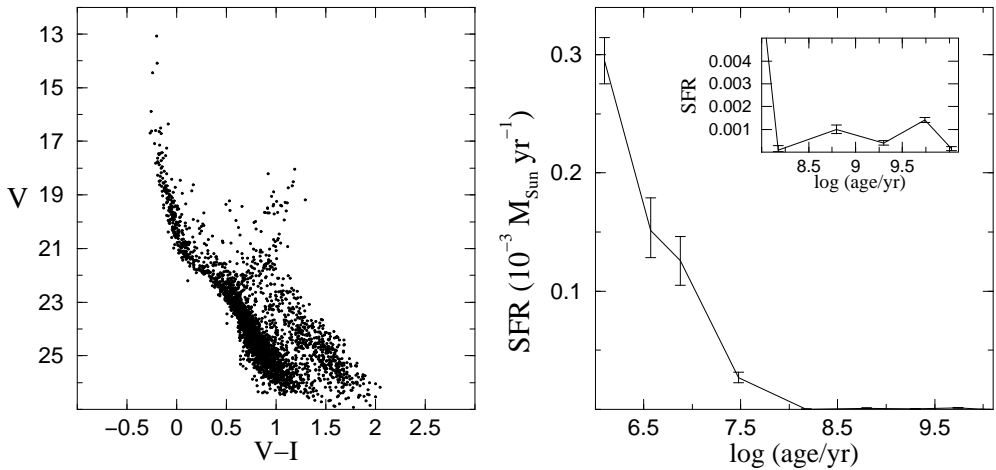,height=2.5in}}
\caption{ Left-hand panel: CMD of the HST/ACS field around the young 
cluster NGC~602 in the SMC. The bright blue plume contains the young cluster   
stars. The red sequence of pre-MS stars 
of lower-mass which have not yet made it on to the main sequence are
also easily recognisable.
Also visible are the old Main Sequence and evolved stars of the SMC field
population. Notice that the lower Main Sequence 
is only populated by field stars, since the
cluster stars with mass below $\sim$1 $M_{\odot}$ 
haven't yet had time to reach it.
Right-hand panel: corresponding SFH as derived with the
synthetic CMD method \citep{Cignoni09}. The oldest part of the SFH
is shown as an inset panel in the upper right.}
\label{sfh_in}
\end{figure}

Accurate photometry down to the oldest MSTOs is feasible from the
ground, although time consuming, and HST allows measurements
of both the oldest and the youngest objects, including pre-MS 
stars,
although with fields of view covering only a tiny fraction of the
galaxy. While stars at the oldest MSTOs and sub-giant branch are the
unique means to firmly establish the SFH at the earliest epochs,
pre-MS stars are precious tools to study the details of the most
recent SFH \citep{Cignoni09}, in terms of time and space
behaviour. The SMC regions of intense recent star formation can provide key
information on the star formation 
mechanisms in environments with metallicity much
lower than in any Galactic star forming region. As an example,
Fig.~\ref{sfh_in} shows the CMD of the young cluster NGC~602 in the
Wing of the SMC, observed with HST/ACS.  Both very young stars (either
on the upper MS or still on the pre-MS) and old stars are found. The
SFH of the cluster and the surrounding field is also shown, 
revealing that the cluster has formed most of its
stars around 2.5~Myr ago, while the surrounding field has formed stars
continuously since the earliest epochs.
The SFR in this SMC region appears to be quite
similar to that of Galactic star forming regions 
\citep[][and references therein]{Cignoni09}. 

Despite being the nearest dI system, the SMC has been
less studied than might be expected. For instance the SFHs derived from
synthetic CMDs, have so far only been based on a few ground-based studies 
\citep{Harris04,Chiosi06,Noel07}, and a few HST-based ones on 
small individual regions \citep{Dolphin01b,Mccumber05,Cignoni09}. New
extensive surveys to infer the SFH of the whole SMC back to the earliest epochs
are planned \citep{Cioni08,Tosi08}, both at visible and near
infrared wavelengths.

\cite{Harris04} were the first to apply the synthetic CMD method to the
derivation of the SMC SFH. They mapped the whole SMC from the ground
and concluded that 50\% of its
stars are older than 8.4~Gyr and diffused over the whole body of the galaxy. 
They also found an indication of a long period of moderate (possibly zero) 
activity between 3 and 8.4~Gyr ago. 
Their photometry however didn't reach the oldest MSTO and all the studies
\citep{Dolphin01b,Mccumber05,Noel07,Cignoni09,Tosi08} which do reach it
indicate that, although present, stars older than 8~Gyr
do not dominate the SMC population. From the latter studies, 
the population bulk seems to peak at ages
somewhat younger than 6--9~Gyr essentially everywhere in the SMC main body.

\subsection{Beyond the Local Group}\label{bcds}

The dwarf galaxies which have been studied using the CMD synthesis
method beyond the Local Group are predominantly actively star forming
BCDs (e.g., I~Zw~18; NGC~1705). These
galaxies are typically quite distant, but as there are no obvious BCDs
in the Local Group (with the possible exception of IC~10 hidden behind
a lot of foreground obscuration from the MW) there is no other
possibility to study this class of actively star forming, yet low
metallicity, systems.

In galaxies beyond the Local Group, distance makes crowding more
severe, and even HST cannot resolve stars as faint as the MSTO of old
populations. The further the distance, the worse the crowding
conditions, and the shorter the look-back time reachable even with the
deepest, highest resolution photometry.  Depending on distance and
intrinsic crowding, the reachable look-back time in galaxies more than
1~Mpc away ranges from several Gyrs (in the best cases, when the RGB
or even the HB  are clearly identified) to several
hundreds Myr (when AGB stars are recognized), to a few tens Myr (when
only the brightest super-giants are resolved). To date, the unique
performances of the HST/ACS have allowed us to resolve individual
stars on the RGB in some of the most metal-poor 
BCDs, e.g., SBS~1415+437 at 13.6~Mpc \citep{Aloisi05} and I~Zw~18 at
18~Mpc \citep{Aloisi07}. The discovery of stars several Gyrs old in
these extremely metal-poor galaxies is key information for
understanding these systems and placing them in the proper context of
galaxy formation and evolution studies.

\begin{figure}[ht]
\centerline{\psfig{file=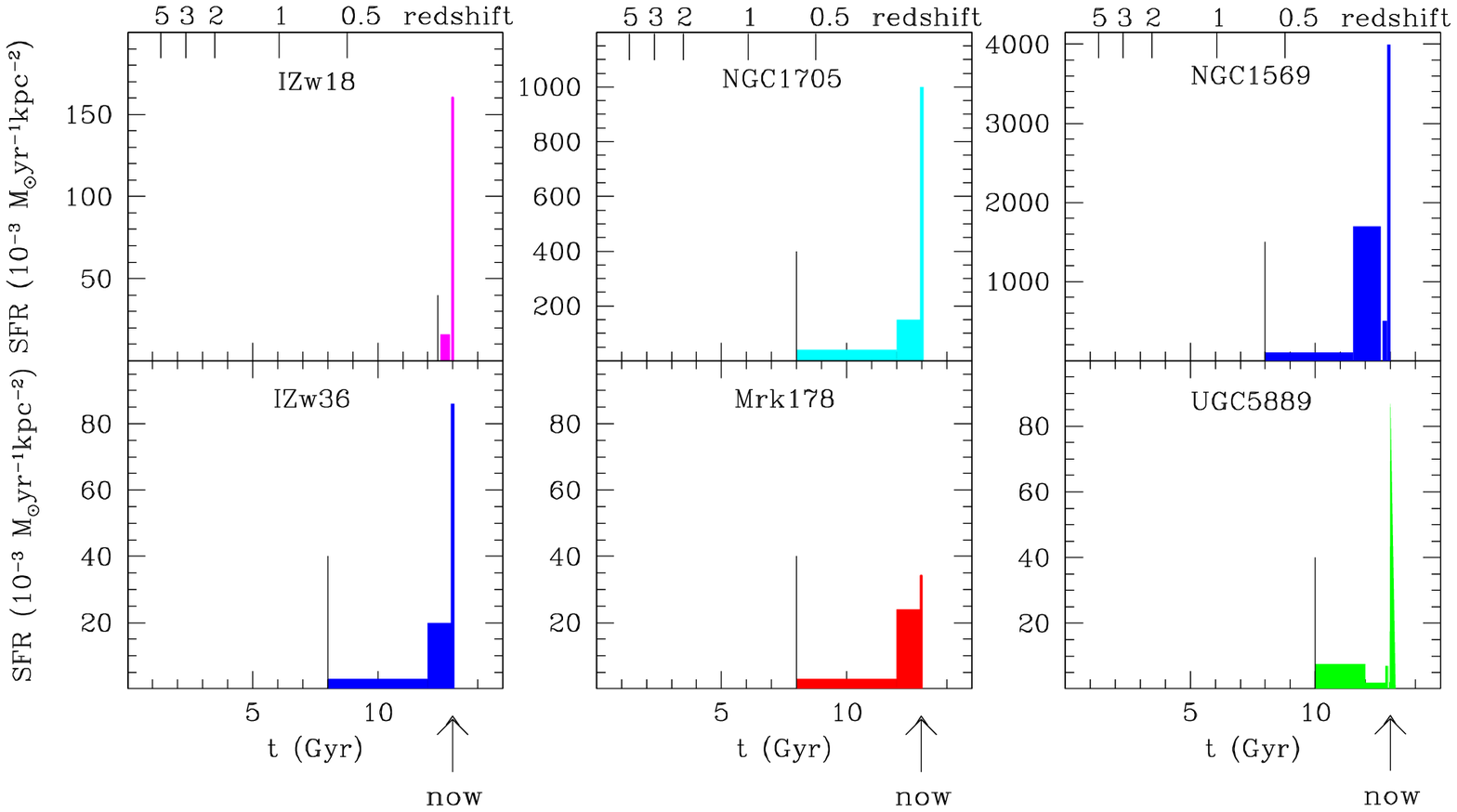,height=3.0in}}
\caption{SFHs of late-type dwarfs outside the Local Group. In all
panels the SFR per unit area as a function of time is plotted. The thin
vertical line indicates the look-back time reached by the adopted
photometry. References: NGC~1569, \citet{Greggio98,Angeretti05};  
NGC~1705, \citet{Annibali03}; 
I~Zw~18, \citet{Aloisi99}; I~Zw~36, \citet{Schulte01};
Mrk~178, \citet{Schulte00}; UGC~5889, \citet{Vallenari05}.
}
\label{sfh_bcd}
\end{figure}

Not many groups have embarked on the challenging application of the
synthetic CMD method beyond the Local Group \citep[for a summary see][]{Tosi07}
and most of them have concentrated their efforts on starbursting
late-type dwarfs.  In Fig.\ref{sfh_bcd}, some examples of the SFH of
external late-type dwarfs are shown.  All these SFHs have been derived
with the synthetic CMD method applied to HST/WFPC2 or NICMOS
photometry.  The look-back time reached by the photometry is indicated
by the thin vertical line in each panel, and in all cases stars of
that age were detected. For those galaxies that have subsequently been
observed with the HST/ACS, the look-back time is significantly longer,
and the further back we look we always find indisputable evidence of
star formation activity at that increasingly old epoch.  This means
that there is no evidence that any of these systems is younger than
the look-back time.  The sample of galaxies shown in Fig.\ref{sfh_bcd}
is not homogeneous: UGC~5889 is a low surface brightness galaxy (LSB),
whilst NGC~1705, I~Zw~18, I~Zw~36 and Mrk~178 are BCDs, and NGC~1569
is classified as dI. Nonetheless, all these dwarfs show qualitatively
similar behaviour, with a strong current burst superimposed on a
moderate and rather continuous underlying star formation activity.
Quantitatively, the actual SFRs differ between galaxies by orders of
magnitude. Notice that the least active system is one of the BCDs
(Mrk~178) and the most active is the dI (NGC~1569).  This is not what
one would have expected on the basis of their morphological
classification. This highlights the difficulties in making accurate
classification of the structural properties of these active, compact
systems. If NGC~1569 were at a distance of 20~Mpc it would most likely
have been classified as a BCD.  The SFR in NGC~1569 is actually about
a factor two higher than shown in Fig.~\ref{sfh_bcd}, as more recent
HST/ACS imaging has detected the RGB, and made a significant revision
of the distance to NGC~1569, to make it almost a magnitude farther
away than previously thought \citep{Grocholski08}.

An interesting result of the SFH studies both in the Local Group and beyond is
that the vast majority of dwarfs have, and have always had, fairly
moderate star formation activity. From an extensive H$\alpha$ study of 94
late-type galaxies \cite{Hunter04} found that the typical SFR of
irregular galaxies is $10^{-3}M_{\odot} \rm yr^{-1}kpc^{-2}$ and that
of BCDs is generally higher but not by much. They also found that
NGC~1569 and NGC~1705 are among the few systems with unusually high
SFRs (see Fig.\ref{sfh_bcd}).  Hunter \& Elmegreen
conclude that the star formation regions are not intrinsically
different in the various galaxy types, but they crowd more closely
together in the centers of BCDs.


\section{Stellar Kinematics and Metallicities}\label{lrspec}

Stellar abundances and kinematics have been shown to be excellent
tools for disentangling the properties of complex stellar systems like
our own Galaxy \citep[e.g.,][]{Eggen62}. This approach is the only
means we have to separate the diverse stellar populations in the solar
neighbourhood. These stars can be split up into disk and halo
components on the basis of their 3D velocities, and these subsets can
then be studied independently.  This concept has subsequently been
expanded and renamed ``Chemical Tagging'' \citep{Freeman02}.  As large
samples of stellar velocities and metallicities have become available
for other galaxies this approach remains the only way to obtain a
detailed understanding of a multi-component stellar system.

The kinematics and metallicities of early (dSph/dE) and late (dI) type
dwarf galaxies in the Local Group have almost always been measured using
different tracers. This is due to the different distances and stellar
densities which are typical for the two types of systems. It is also
because dIs contain an easily observable ISM in the form of HI gas and
dSphs do not.  Because early-type galaxies usually do not contain any
(observable) gas nor any young star forming regions, most of what we
know of their internal properties comes from studies of their evolved
stellar populations (e.g., RGB stars). Late-type galaxies are
typically further away (the SMC being a clear exception), which can
make the accurate study of individual RGB stars more challenging, and
they contain HI gas and several HII regions.  Thus, most of what we
know about the kinematics and metallicity of dIs comes from gas and
massive (young) star abundances.  RGB stars have the advantage that
they are all old ($> 1$~Gyr) and their properties are most likely to
trace the gravitational potential and chemical evolution throughout
the entire galaxy up to the epoch when they formed, and not the most
recent star formation processes and the final metallicity. It is only
with detailed studies of the same tracers that kinematics and
metallicities in these different dwarf galaxies types can be
accurately compared to make confident global statements about the
differences and similarities between early and late-type galaxies.

\subsection{Early-Type Dwarfs}

Dwarf Spheroidal galaxies are the closest early-type galaxies that
contain sufficient numbers of well distributed RGB stars to provide
useful kinematic and metallicity probes. Moreover, dSphs are
considered to be interesting places to search for dark matter, since
there is so little luminous matter to contribute to the gravitational
potential.  The velocity dispersion of individual stars can be used to
determine the mass of the galaxy.  It is also possible to determine
metallicities for the same stars. This allows a more careful
distinction of the global properties based on structural, kinematic
and metallicity information \citep[e.g.,][]{Battaglia07phd}.

\subsubsection{Galactic dSphs:}
It was originally thought that the luminosity profiles of
Galactic dSphs resembled globular clusters and showed systems
truncated by the gravitational field of the MW
\citep[e.g.,][]{Hodge71}.  As measurements improved and the discussion
focused on the possible presence of dark matter, \citet{Faber83}
showed that the profiles are exponential more like those of galaxy
disks and thus predicted that Galactic dSphs had a much higher
mass-to-light ratio (M/L), $\ge$30, than had previously been
thought. In parallel \citet{Aaronson83} found observational evidence
for this in radial velocity studies of individual stars in the Draco
dSph.  Thus it became clear that dSph are small galaxies, related to
late-type disk and irregular systems, and not globular clusters.

For a given M/L the central velocity dispersion of a self-gravitating
spheroidal system in equilibrium may scale with the characteristic
radial scale length and the central surface brightness
\citep{Richstone86}. Given that globular clusters typically have a
central velocity dispersion $\sim 2 - 8$ kms$^{-1}$ it was expected that
dwarf galaxies, which have scale lengths at least 10 times bigger and
surface brightnesses about 1000 times smaller (see
Fig.~\ref{fig-bing}), should have central velocity dispersions $<
2$kms$^{-1}$. This has been consistently shown not to be the case; all
galaxies have stellar velocity dispersions which are typically larger
than those of globular clusters ($\sim 8 - 15$kms$^{-1}$). This was first
shown by \citet{Aaronson83} for a sample of 3 stars in the Draco dSph.
This early tentative (and brave!)  conclusion has been verified and
strengthened significantly over the past decades, with modern samples
containing measurements for many hundreds of individual stars in Draco
\citep[e.g.,][]{Munoz05, Wilkinson04} and in all other Galactic
dSphs. If it can be assumed that this velocity dispersion is not
caused by tidal processes, then this is evidence that dSph galaxies
contain a significant amount of unseen (dark) matter
\citep[e.g.,][]{Mateo94,EdO98, Gilmore07}, or that we do not
understand gravity in these regimes (e.g., MOND applies).  There has
been some uncertainty coming from the possible presence of binary
stars but a number of studies have carried out observations over
multiple epochs and this effect has been found to be minimal
\citep[e.g.,][and references therein]{Battaglia08cat}.

\begin{figure}[ht]
\centerline{\psfig{file=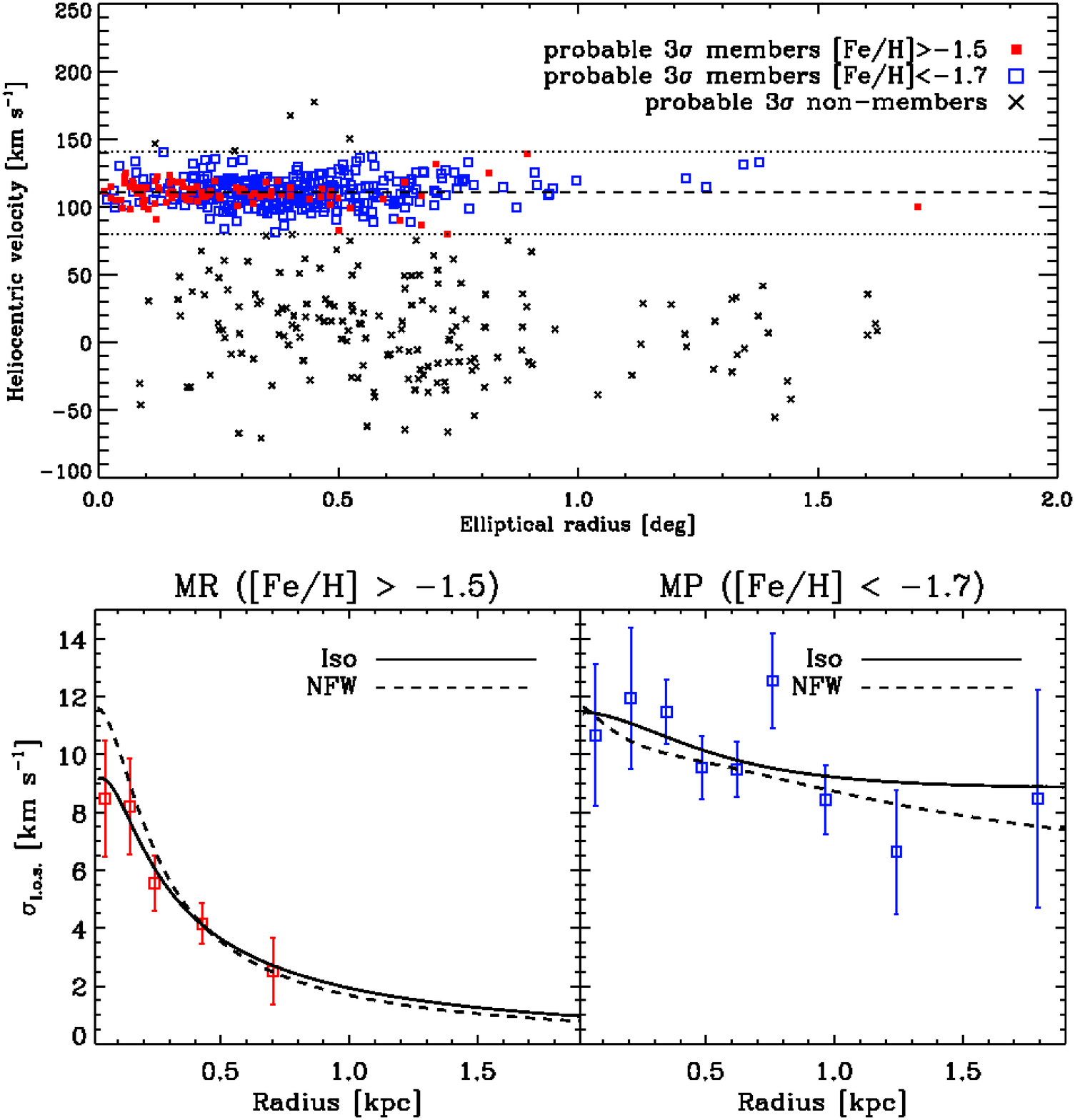,height=4.5in}}
\begin{center}
 \caption{
From the DART survey, these are VLT/FLAMES line-of-sight velocity
measurements for individual RGB stars in the Scl dSph
\citep{Battaglia07phd, Battaglia08mass}.  In the top panel elliptical
radii are plotted against velocity for each star.  The limits for
membership are given by dotted lines about v$_{\rm hel} = +110.6$
kms$^{-1}$, the heliocentric velocity, which is shown as a dashed line.
It is apparent that the velocity dispersion and central
concentration of the metal-rich (MR) stars, in red, can clearly
distinguish them from the metal-poor (MP) stars, in blue, which have
a larger velocity dispersion and are more uniformly distributed over
the galaxy. In the bottom 2 panels the line of sight velocity
dispersion profiles, with rotation removed, for the MR (red) and MP
(blue) stars are shown along with the best fitting pseudo-isothermal
sphere (solid line) and NFW model (dashed line), see
\citet{Battaglia08mass} for details.
}
\label{fig-batt07}
\end{center}
\end{figure}

As instrumentation and telescopes improved, a significant amount of
work on the kinematic properties of dSphs has become possible.  This
field has benefited particularly from wide-field multi-fibre
spectrographs on 6--8m-class telescopes (e.g., VLT/FLAMES and
Magellan/MIKE), but also WYFOS on the WHT and AAOmega on the AAT.
These facilities have allowed samples of hundreds of stars out to
the tidal radii 
\citep[e.g.,][]{Wilkinson04, Tolstoy04, Munoz05, Kleyna05,
Walker06sext, Walker06fnx, Battaglia06, Battaglia07phd,
Battaglia08mass}.  These velocity measurements 
often have sufficient
signal-to-noise to also obtain metallicities from the
Ca~II triplet, the Mg~B index or a combination of weak lines
\citep[e.g.,][]{Suntzeff93,Tolstoy04,Battaglia06,Munoz06,Koch06,
Kirby08meth, Battaglia08cat, Shetrone09}. This approach resulted in the discovery of
surprising complexity in the ``simple'' stellar populations in dSphs. It
was found that RGB stars of different metallicity range (and hence
presumably age range) in dSphs can have noticeably different kinematic
properties \citep[e.g.][]{Tolstoy04, Battaglia06}.  This has implications for
understanding the formation and evolution of the different components
in these small galaxies. It is also important for correctly
determining the overall potential of the system.  The presence of
multiple components allows more accurate modelling of the overall
potential of the system and so better constraints on the the
underlying dark matter profile and the mass content
\citep{Battaglia07phd, Battaglia08mass}, see Fig.~\ref{fig-batt07}.

The VLT/FLAMES DART survey \citep{Tolstoy06} determined 
kinematics and metallicities for large samples of individual stars in
nearby dSphs. There have also been similar surveys by other teams on
VLT and Magellan telescopes \citep[e.g.,][]{Gilmore07, Walker09}.
Traditionally the mass distribution of stellar systems has been
obtained from a Jeans analysis of the line-of-sight velocity
dispersion \citep[e.g.,][]{Mateo94}, assuming a single stellar
component embedded in a dark matter halo. This analysis suffers from a
degeneracy between the mass distribution and the orbital motions
presumed for the individual stars, the mass-anisotropy degeneracy.
From DART the mass of the Scl dSph was determined taking advantage of
the presence of the two separate components distinguished by
metallicity, spatial extent and kinematics \citep{Battaglia07phd,
Battaglia08mass}, see Fig.~\ref{fig-batt07}.  Here it was shown that
it is possible to partially break the mass-anisotropy degeneracy when
there are two components embedded in the same dark matter halo.  The
new dynamical mass of the Scl dSph is $M_{\rm dyn} = 3 \times 10^8
M_{\odot}$, within 1.8~kpc, which results in an M/L$\sim$160.  This is
a factor $\sim$10 higher than the previous value obtained from a much
smaller and more centrally concentrated sample of stars
\citep{Queloz95}.  This corresponds to a dark matter density
within 600~pc, for the best fitting model of
$0.22$M$_\odot$pc$^{-3}$.  
This result is largely
independent of the exact distribution of dark matter in the central
region of the Scl dSph, see Fig.~\ref{fig-batt07}.  This same study
also found evidence for a velocity gradient, of 7.6$^{+3.0}_{-2.2}$ km
s$^{-1}$ deg$^{-1}$, in Scl, which has been interpreted as a signature
of intrinsic rotation. This is the first time that rotation has been
detected in a nearby dSph, and it was a faint signal that required a
large data set going out to the tidal radius.

Another aspect of these surveys has been the determination of
metallicity distribution functions (MDFs), typically using the Ca~II
triplet metallicity indicator \citep{Battaglia06, Helmi06}.  This uses
the empirical relation between the equivalent width of the Ca~II
triplet lines and [Fe/H], and its accuracy was also tested
\citep{Battaglia08cat}.  The Ca~II triplet method will start to fail
at low metallicities, [Fe/H]$< -2.5$, but this is starting to be
better understood on physical \citep[e.g.,][]{Starkenburg08} as well
as empirical grounds from following up stars with Ca~II triplet
metallicities, [Fe/H]$< -2.5$.  This means that the effect can be
corrected for, and so far there is no significant change to the MDFs
in \citet{Helmi06}. This is because the fraction of the stellar
samples that may be affected by this uncertainty is very small
($\sim1-2$\%), and then only a fraction of these actually need to be
corrected. Thus it seems likely that 
there are very few, if any, extremely metal
poor stars ([Fe/H]$< -4$) in most classical dSph. But this result
has to be verified by careful follow up of Ca~II triplet measurements
with [Fe/H]$< -2.5$.

\begin{figure}[ht]
\centerline{\psfig{file=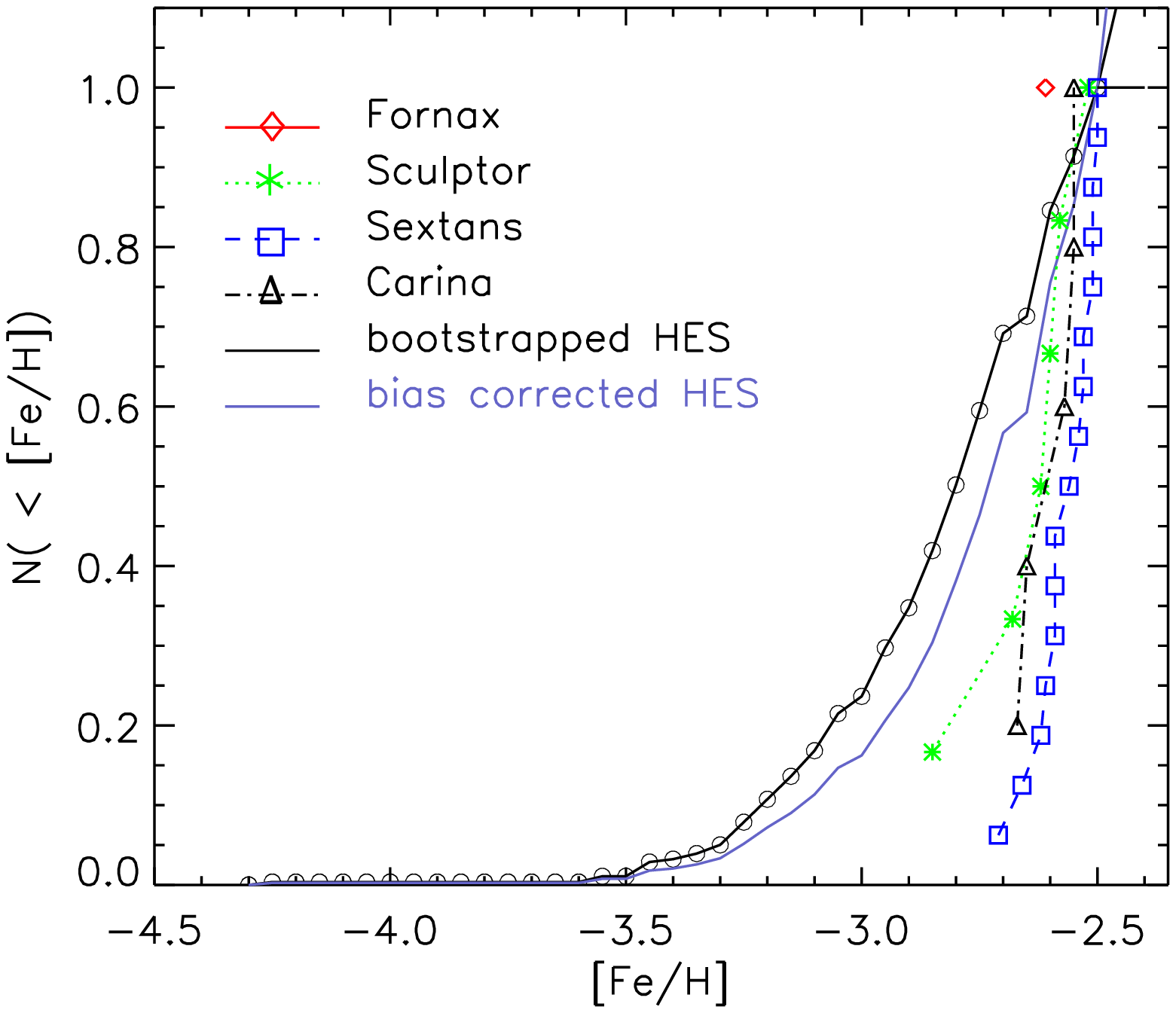,height=3in}}
\begin{center}
\caption{
Comparison of the cumulative metallicity distribution functions (MDFs)
of the stars in the mean bootstrapped Hamburg-ESO survey sample as a
solid black line, and the new bias corrected Galactic halo MDF from
\citet{Schoerck08} as a solid blue line. These are compared to the
MDFs for 4 dSphs from the DART survey \citep{Helmi06}.  The halo and
the dSph MDFs have been normalised at [Fe/H]=-2.5, which assumes that
the completeness of the halo MDF is well understood at this
metallicity and below.  Note that at present the Fornax dSph lacks
sufficient stars at [Fe/H] $< -2.5$ to be properly present on this
plot.
}
\label{fig-helmi06}
\end{center}
\end{figure}

These dSph MDFs were compared to the Galactic halo MDF from the
Hamburg-ESO survey \citep[HES,][]{Beers05}, and found to be
significantly different \citep{Helmi06}, see Fig.~\ref{fig-helmi06}
for an update.  The Galactic halo MDF has recently been revised
\citep[][submitted]{Schoerck08}, and this revision is also shown in
Fig.~\ref{fig-helmi06}. It can be seen that the difference between the
dSphs and the Galactic halo MDFs remains.  However, it is obviously of
critical importance that the different degrees of incompleteness are
well understood, and this is particularly complicated in the halo.
What is shown as the Galactic halo MDF in Fig.~\ref{fig-helmi06} may
still change, but it is most likely that the two halo MDFs shown
represent a reasonable range of possibilities.  This difference
provides a challenge to models where all of the Galactic halo builds
up from the early merging of dwarf galaxies, because it begs the
question: where have all the most metal-poor stars in the Galactic
halo come from? Was there a pre-enrichment of dwarf galaxies, perhaps
by the most metal-poor stars which we appear to find only in the halo
\citep{Salvadori08}?  This mismatch applies equally to any merging
scenario that extends over a significant fraction of a Hubble time, as
it means that dSph and dI galaxies could not have merged to form the halo
except at very select moments in the past
(see Section~\ref{hrspec}). Fig.~\ref{fig-helmi06}
does not include uFds, and there is evidence that they may
include more metal poor stars than are to be found in 
dSph \citep[e.g.,][]{Kirby08, Frebel09}.

\subsubsection{More Distant dSphs:}
There are also more isolated dSphs, within the Local Group but not obviously
associated to the MW or M~31, for example Antlia, Phoenix,
Cetus and Tucana.  They have also benefited from spectroscopic studies
\citep[e.g.,][]{Tolstoy00, Gallart01,
Irwin02, Lewis07, Fraternali09}. Antlia, Tucana and Phoenix have HI gas in their
vicinity, but after careful study, only for Phoenix and Antlia has the
association been confirmed. These spectroscopic
studies have shown evidence for rotation
in Cetus and Tucana at a similar magnitude to 
Scl \citep{Lewis07, Fraternali09}. This
rotation is consistent with the flattening of the galaxy.  In Phoenix
the kinematics and morphology of HI gas compared to the stellar
component suggests that the HI is being blown out by a recent star
formation episode \citep[e.g.,][]{Young07}. This supports the
theoretical predictions of this effect \citep[e.g.,][]{Larson74,
MacLow99}.  All these more distant dSphs are far enough away from the
MW that any strong tidal influence is likely to have been several Gyr in the
past.  

\subsubsection{Dwarf galaxies around M~31}
There are also diffuse dEs and dSphs around M~31 where spectra have
been taken of RGB stars. From a sample of 725 radial velocity
measurements in NGC~205, it was found to be rotating at 11 $\pm
5$~kms$^{-1}$ \citep{Geha06}.  A careful study of the structural
properties of the dSphs around M~31 shows that on average the scale
radii of the dSphs around M31 are about a factor 2 larger than those of dSphs
around the MW at all luminosities \citep{McC06}.  This could
either be due to small number statistics or it might suggest that the tidal
field of M31 is weaker than that of the MW or the environment
in the halo of M31 is different from that of the MW halo.

\subsection{Late-Type Dwarfs}

Most of what we know about the kinematics of dI galaxies comes from
observations of their HI gas \citep[e.g.,][]{Lo93, Young03}, which is
strongly influenced by recent events in the systems.  For example, the
velocity dispersion measured in the HI gas is predominantly
influenced by on-going star formation processes.  Thus the HI velocity
dispersion is almost always $\sim 10$kms$^{-1}$ in any system, from the
smallest dIs to the largest spiral galaxies, regardless of the mass or
rotation velocity of the HI.  This makes it difficult to compare the
kinematic properties of dI and dSph galaxies.

Likewise, most of the metallicity information comes from HII region
spectroscopy 
\citep[e.g.,][]{Pagel81, Hunter86, Skillman89a, Izotov99, Kunth00,
Hunter04} or spectroscopy of 
(young) massive stars \citep[e.g.,][]{Venn01,
Venn04conf}. For a few
BCDs the FUSE satellite has also provided
abundances for the neutral gas 
\citep[e.g.,][]{Thuan02, Aloisi03,Lebouteiller04}.  Thus the
metallicity measures come from sources which are only a few million
years old and the product of the entire history of star formation in a
galaxy.  By contrast, in dSphs the abundances are typically measured
for stars older than $\sim 1$~Gyr, and the value quoted is some
form of a mean of the values measured over the entire SFH.  This makes
it difficult to accurately compare the properties of early and late-type 
dwarf galaxies, as there are few common measurements that 
can be directly compared. Such comparisons have  
been attempted by \citet[e.g.,][]{Skillman89a, Grebel03}. 

From studies of the HI gas in these systems it has been
found that for the smallest and faintest dIs if rotation is detected
at all, it is at or below the velocity dispersion
\citep[e.g.,][]{Lo93, Young03}. Despite this, 
in all dIs 
the HII regions in a single galaxy, even those widely spaced, appear
to have, within the margins of error of the observations, identical
[O/H] abundances.  So it appears that either the enrichment process
progresses uniformly galaxy wide, or the oxygen abundance within an
HII region is affected by some internal, self-pollution process which
results in uniform [O/H] values \citep[e.g.,][]{Olive95}.  However,
the clear gradients in HII region abundances seen in spiral galaxies
argue against this explanation.

One dwarf galaxy which has both HI and stellar kinematic information
is the faint transition type dwarf LGS~3 (M$_{V}=-9.9$). The
stellar component looks like a dSph, dominated by (old) RGB stars (see
Fig.~\ref{sfh_lcid}). There are no HII regions \citep{Hodge95}, and the
youngest stars are around 100~Myr old \citep{Miller01}.  However, the
galaxy also contains 2$\times 10^5$ M$_\odot$ of HI \citep{Lo93,
Young97}, which is more extended than the optical galaxy, and with no
convincing evidence of rotation.  \citet{Cook99} measured the radial
velocities of 4 RGB stars in LGS~3 at the same systemic velocity as
for the HI, confirming the association. They found the stellar
velocity dispersion of these 4 stars to be 7.9$^{+5.3}_{-2.9}$ kms$^{-1}$.
This leads to a high M/L ($> 11$, perhaps as high as 95), similar to
other dSphs.  However this sample of radial velocities is hardly
sufficient.

LGS~3 used to be the lowest luminosity galaxy with HI, but that was
before the recent discovery of Leo~T \citep{Irwin07}.  Leo~T contains
$2.8 \times 10^5$M$_\odot$ of HI gas \citep{RyanWeber08}, and it does
not contain HII regions or young stars. 
The total dynamical mass was determined to be $8.2 \pm 3.6 \times
10^6 \rm M_\odot$, with M/L$\sim 140$.
Leo~T seems to be a
particularly faint dwarf (M$_V = -8$), at a distance of $\sim$420~kpc
(see Table~\ref{sfh_late}). It is about 2~magnitudes fainter than the
other transition type systems like LGS~3 and Phoenix, and
0.6~magnitudes fainter than the faintest dwarf spheroidal system,
Draco.

Leo~T is another of the very few systems for which kinematics have
been derived from HI and velocities of individual stars.
\citet{Simon07} measured the radial velocities and metallicities of 19
RGB stars, and found the average metallicity to be
[Fe/H]$\sim -2.3$ with a range of $\pm 0.35$. 
They found a central optical velocity of
$\sim +38 \pm 2$kms$^{-1}$, a velocity dispersion $\sigma = 7.5 \pm 1.6$
kms$^{-1}$ and no obvious sign of rotation.  This is comparable to the HI
value, $\sigma_{HI} = 6.9$kms$^{-1}$, also with 
no sign of rotation \citep{RyanWeber08}.
This is the smallest and lowest luminosity galaxy with fairly recent star
formation known.  The inferred past SFR of $1.5-2\times
10^{-5} \rm M_\odot$/year might be sufficiently low that gas is neither
heated nor blown out in this system, thus allowing it to survive
\citep{DeJong08}.

\subsection{Ultra-Faint Dwarfs}\label{LR-ufaint}

The stellar kinematics and metallicities of individual stars 
play an important role in determining what kind of systems uFds
are. These measurements can attempt to quantify the degree of
disruption uFds may have undergone and if they should be considered
faint galaxies or some kind of
diffuse globular clusters, such as are seen around
M~31 \citep[e.g.,][]{Mackey06}.  These systems are so
embedded in the foreground of our Galaxy, both in position and in
velocity, and the total number of their stars is so often so small
(many have M$_V \gsim -4$) that studies can easily get different
results for even the most fundamental properties, like their size and
their dark matter content depending upon membership selection
\citep[e.g.,][]{Ibata06, Simon07, Siegel08, Geha08}.

The basic kinematic properties of these systems
\citep[e.g.,][]{Simon07} show evidence that they are much more dark
matter dominated than previously known systems, with M/L$\sim 140 -
1700$, although this study did not try to correct for tidal effects
which are almost certainly present. Many of these galaxies also have
very small numbers of stars, and thus test particles, and so the
properties of the dark matter content are often extrapolated from
a small central region.

\citet{Simon07} also determined the average stellar metallicities
([Fe/H] $\le -2$) of uFds and found them to be lower than in most
globular clusters, and with a scatter that is not expected in globular
clusters. The average metallicities are also lower than in other more
luminous dwarf galaxies \citep[see also,][]{Kirby08}, and the lowest
metallicity stars appear to be more metal-poor than the most metal
poor stars found in the brighter ``classical'' dSphs \citep{Norris08,
Frebel09}.  \citet{Norris08} also found evidence for carbon-rich metal-poor
stars in Boo~I, which suggests that the metal-poor stars in uFds maybe more
similar to those found in the MW halo, where a 
large fraction of stars more metal poor than [Fe/H] $< -4$ are carbon
rich.

CVn~I, at a distance of 220~kpc, at M$_V \sim -7.9$ \citep{Zucker06b},
is one of the brighter examples of uFds, and bears much similarity to
classical dSphs, both in structural properties, kinematics and SFH.
From a CMD analysis \citet{Martin08CanVen} found that the galaxy is
dominated by an ancient population ($> 10$~Gyr old), with about 5\% of
its stars in a young blue plume $\sim 1.4-2$~Gyr old.  It has well
populated, broad, RGB and HB which have been studied spectroscopically
\citep{Ibata06, Simon07}.  With a sample of 44~stars \citet{Ibata06} 
detected the presence of two components with different
metallicities and velocities.  However \citet{Simon07} with a much
larger sample of 212 stars were not able to reproduce this result.

A recently discovered uFd is Leo~V \citep{Belokurov08}, at a distance
of 180~kpc with M$_V = -4.3$.  Fig.~\ref{fig-leov} illustrates how
hard it can be to quantify the structural properties and distinctness
of these small diffuse systems several of which may be embedded in
Galactic scale streams.  Leo~V may be related to Leo~IV, to which it
is very close to both spatially and in velocity.  They may both be the
remnants of the same tidal interaction, but given how metal poor their
stars appear to be they would have to be the outer envelopes of dwarf
galaxies (which are known to have metallicity gradients), like Sgr,
and not disrupted globular clusters.

Thus the nature of some of the uFds still remains a mystery, and there
is likely to be a range of origins for these systems.  The brighter uFds
(M$_V < -5$) are relatively easy to study and do appear to be a low
mass tail to dSphs (e.g., CVn~I) and dI/transition systems (e.g.,
Leo~T). This reduces the sizes of objects in which stars can form 
in the early universe \citep[e.g.,][]{Bovill08, Salvadori09}, and it
follows the trend that these smaller systems 
could barely enriched themselves. This could be 
either due to efficient winds, or inefficient star formation, both of
which could be the result of a low galactic mass.

It remains a matter of conjecture what are the fainter systems such as
Coma and UMa~II (M$_V \sim -4$). From its highly irregular stellar
distribution UMa~II is clearly a disrupted system that sits behind
high velocity cloud complex A \citep{Zucker06a, Belokurov07orph}. Coma
has very similar properties to UMa~II, and it also has an irregular
extended shape \citep{Belokurov07}. UMa~II is one of the few objects
which lie in the gap between globular clusters and dwarf galaxies in
the left hand plot of Fig.~\ref{fig-bing}. This is a region which has
so far not been populated by either galaxies or globular clusters but
if the large dark matter masses of the uFds are
correct then they are more likely to be an extension of the galaxy class
than of the globular cluster class.


\section{Detailed Abundances of Resolved Stars}\label{hrspec}

The detailed chemical abundance patterns in individual stars of a
stellar population provide a fossil record of chemical enrichment over
different timescales.  As generations of stars form and evolve, stars
of various masses contribute different elements to the system, on
timescales directly linked to their mass.  Of course, the information
encoded in these abundance patterns is always integrated over the
lifetime of the system at the time the stars studied were born.  Using
a range of stars as tracers provides snapshots of the chemical
enrichment stage of the gas in the system throughout the SFH of the
galaxy.  This approach also assumes that the chemical composition at
the stellar surface is unaffected by any connection between interior
layers of the star, where material is freshly synthesised, and the
photosphere. This assumption is generally true for main-sequence
stars, but evolved stars (giants or super-giants) will have
experienced mixing episodes that modified the surface composition of
the elements involved in hydrogen burning through the CNO cycle,
i.e. carbon, nitrogen and possibly also oxygen.

These studies require precise measurements of elemental abundances in
individual stars and this can only be done with high-resolution and
reasonably high signal-to-noise spectra.  It is only very recently
that this has become possible beyond our Galaxy. It is efficient
high-resolution spectrographs on 8$-$10m telescopes that have made it
possible to obtain high resolution (R$>$40000) spectra of RGB stars in
nearby dSphs and O, B and A super-giants in more distant dIs.  These
stars typically have magnitudes in the range V=$17-19$.  Before the
VLT and Keck, the chemical composition of extra-galactic
stars could only be measured in super-giants in the nearby Magellanic
Clouds \citep[e.g.,][]{Wolf73,Hill95,Hill97,Venn99}, yielding present
day (at most a few 10$^7$yr ago) measurements of chemical
composition. Looking exclusively at young objects however makes it virtually
impossible to uniquely disentangle how this enrichment
built up over time.

\begin{figure}[ht]
\centerline{\psfig{file=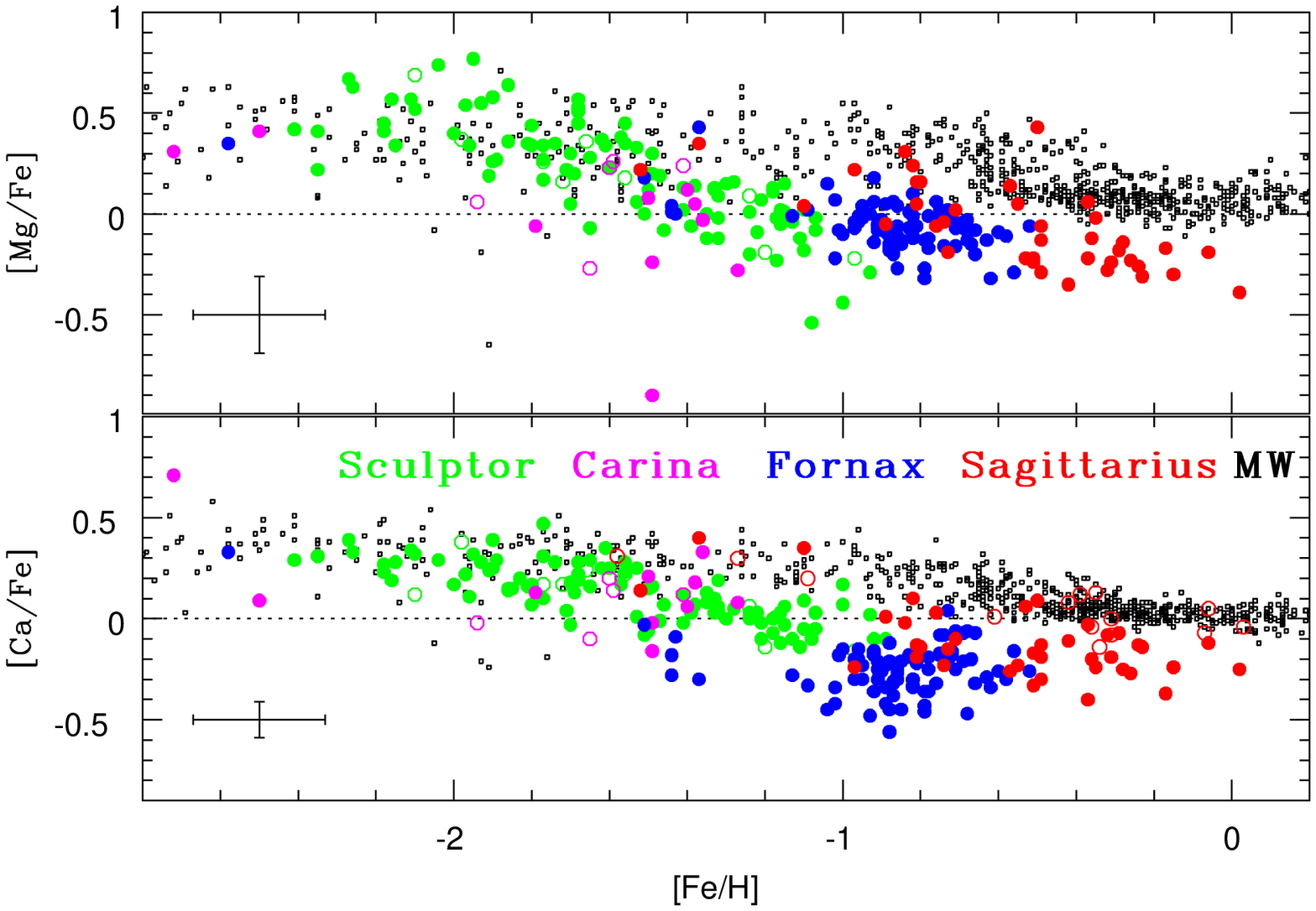,width=12cm}}
\caption{Alpha-elements (Mg and Ca) in four nearby dwarf spheroidal galaxies:
Sgr \cite[red:][]{Sbordone07,Monaco05,McWilliam05conf}, Fnx
\cite[blue:][]{Letarte07phd, Shetrone03}, Scl \cite[green: Hill
et al. in prep;][]{Shetrone03,Geisler05} and Carina
\cite[magenta:][]{Koch08Car,Shetrone03}. Open symbols refer to
single-slit spectroscopy measurements, while filled circles refer to
multi-object spectroscopy. The small black symbols are a compilation
of the MW disk and halo star abundances, from \cite{Venn04}.}
\label{alpha_all}
\end{figure}

\subsection{Dwarf Spheroidal Galaxies}\label{HR_dSph}

The first studies of detailed chemical abundances in dSph galaxies are
those of \citet[][17 stars in Draco, 
Ursa~Min \& Sextans]{Shetrone98,Shetrone01} using 
Keck-HIRES and \citet[][2 stars
in Sgr]{Bonifacio00} using VLT-UVES.  These early works were
shortly followed by similar studies slowly increasing in size
\citep{Shetrone03, Bonifacio04, Sadakane04, Geisler05,
McWilliam05conf}.  The total number of stars probed in individual studies
remained very low (typically only 3 to 6 in any one galaxy except for 
Sgr).  This was because the stars had to be observed
one at a time, and for the most distant dSphs this
required exposure times of up to 5~hours per star.
Nevertheless from these small samples it was already clear that dSph
galaxies follow unique chemical evolution paths, which are distinct
from that of any of the MW components
\citep[e.g.,][]{Shetrone01,Shetrone03,Tolstoy03,Venn04}.

Most recently, high-resolution spectrographs with high multiplex
capabilities have resulted in large samples ($> 80$~stars) of high
resolution spectra of individual stars to determine 
abundances in a relatively short time.  The FLAMES multi-fiber
facility on VLT \citep{Pasquini02} has so far been the most productive
in this domain.
There are a number of FLAMES high resolution spectroscopy studies in
preparation, but some results are already published for Sgr and its
stream \citep[][39 stars]{Monaco05,Sbordone07,Monaco07}, Fnx
\citep[][81 stars]{Letarte07phd}, Carina \citep[][18 stars]{Koch08Car}
and Scl (Hill et al. in preparation, 89 stars).

These new extensive studies not only provide abundances with better
statistics, but they also allow statistical studies over the total
metallicity range in each galaxy. This allows for an almost complete
picture of their chemical evolution over time, with abundance {\em
trends} as a function of metallicity for each system.  Only the most
metal-poor regime in these systems is perhaps still somewhat
under-represented in these samples, although this is in part because
they are rare \citep{Helmi06}, and in part because these large samples
of abundances have been chosen in the inner parts of the
galaxies, where younger and/or more metal-rich populations tend to
dominate \citep{Tolstoy04, Battaglia06}.
New studies to fill in this lack of measured abundances in low
  metallicity stars  are in preparation \citep[e.g.,][]{Aoki09}. 
In the following, we will consider groups of
elements that give particular insights into dwarf galaxy evolution.

\subsubsection{alpha elements}\label{alphas}

The $\alpha$-elements abundances that can easily be measured in RGB
spectra includes O, Mg, Si, Ca and Ti.  Although the $\alpha$-elements
have often been considered as an homogeneous group, and their
abundances are sometimes averaged to produce a single [$\alpha$/Fe]
ratio, their individual nucleosynthetic origin is not always exactly
the same. For example, O and Mg are produced during the hydrostatic He burning
in massive stars, and their yields are not expected to be affected by the
SNII explosion conditions. On the other hand Si, Ca and Ti are mostly
produced during the SNII explosion.  This distinction is also seen in
the observations \citep[e.g.,][]{Fulbright07}, where Si, Ca and Ti
usually track one another, but O and Mg often show different trends
with [Fe/H]. It is therefore generally advisable to treat the three
$\alpha$-elements which are well probed in dwarf galaxies separately.  
Fig.~\ref{alpha_all} shows a compilation of Mg and Ca
 abundances of individual stars in those dSphs with more than 15
measurements.

The apparent paucity of $\alpha$-elements (relative to iron) in dSph
galaxies compared to the MW disk or halo was first noted by
\citet{Shetrone98,Shetrone01,Shetrone03,Tolstoy03,Venn04} from small
samples. Fig.~\ref{alpha_all} shows this convincingly over most of
the metallicity range in each system. However, it also appears that
each of these dSphs starts, at low [Fe/H], with [$\alpha$/Fe] ratios
similar to those in the MW halo at low metallicities. 
These ratios
in the dSphs then evolve down to lower values 
than is seen in the MW at the same metallicities.

The ratio of $\alpha$-elements to iron, [$\alpha$/Fe], is commonly
used to trace the star-formation timescale in a system, because it is
sensitive to the ratio of SNII (massive stars) to SNIa (intermediate 
mass binary systems with mass transfer) that have occurred in the
past. SNIa have a longer time scale than SNII and as soon as they
start to contribute they dominate the iron enrichment
and [$\alpha$/Fe] inevitably decreases. After that, no SFH can ever again
result in enhanced [$\alpha$/Fe], unless coupled with galactic winds removing
only the SNIa ejecta and not that of SNII.
This is seen as
a ``knee'' in a plot of [Fe/H] vs. [$\alpha$/Fe], see
Fig.~\ref{alpha_all}.  The knee position indicates the
metal-enrichment achieved by a system at the time SNIa start to
contribute to the chemical evolution \citep[e.g.,][]{Matteucci90,
Matteucci03}.  This is between $10^8$ and $10^9$yrs after the first star
formation episode.  A galaxy that efficiently produces {\it and
retains} metals over this time frame will reach a higher metallicity
by the time SNIa start to contribute than a galaxy which either loses
significant metals in a galactic wind, or simply does not have a very
high SFR.  The position of this knee is expected to be different for
different dSphs because of the wide variety of SFHs.  In the data
there are already strong hints that not all dSphs have a knee at the
same position.

At present the available data only cover the knee with sufficient
statistics to quantify the position in the Scl dSph, a system
which stopped forming stars 10~Gyr ago, and the knee occurs at
[Fe/H]$\approx -1.8$.  This is the same break-point as the two
kinematically distinct populations in this galaxy \citep{Tolstoy04,
Battaglia07phd}, see Fig.~\ref{fig-batt07}.  This means that the
metal-poor population has formed before any SNIa enrichment took
place, which means on a timescale shorter than 1~Gyr.

In other dSphs the knee is not well defined due to a lack of data, but
limits can be established.  The Sgr dSph has enhanced
[$\alpha$/Fe] up to [Fe/H] $\approx -1.0$, which is significantly more
metal-rich than the position of the knee in the Scl dSph. This is
consistent with what we know of the SFH of Sgr, which has
steadily formed stars over a period of 8$-$10~Gyrs, and only stopped
forming stars about 2$-$3~Gyr ago \citep[e.g.,][]{Dolphin02}.  The
Carina dSph has had an unusually complex SFH, with at least three separate
bursts of star formation \citep{Hurley98}, see Fig.~\ref{carina}. The
abundance measurements in Carina are presently too scarce to have any
hope to confidently detect these episodes in the chemical enrichment
pattern \citep[e.g.,][]{Tolstoy03}.  It appears to possesses
[$\alpha$/Fe] poor stars between [Fe/H] $= -1.7$ and $-2.0$, which
suggests that the knee occurs at lower [Fe/H] than in Scl.  It
seems that Carina has had the least amount of chemical evolution
before the onset of SNIa of all galaxies in Fig.~\ref{alpha_all}.  In
the Fnx dSph, another galaxy with a complex SFH, the sample does
not include a sufficient number of metal-poor stars to determine even
an approximate position of the knee. There are abundances for only five
stars below $\rm [Fe/H]=-1.2$, and only one below [Fe/H $ = -1.5$. The
knee is constrained to be below [Fe/H] $< -1.5$.  From this (small)
sample of dSph galaxies, it appears that the position of the knee
correlates with the total luminosity of the galaxy, and the mean
metallicity of the galaxy. Which suggests that the presently most
luminous galaxies are those that must have formed more stars at the
earliest times and/or retained metals more efficiently than the less
luminous systems.

The abundance ratios observed in all dSphs for stars on the metal-poor
side of the knee, tend to be indistinguishable from those in the
MW halo.  From this small sample it seems that the first
billion years of chemical enrichment gave rise to similar enrichment
patterns in small dwarf galaxies and in the MW halo.  Because
the [$\alpha$/Fe] at early times is sensitive to the IMF of the
massive stars, if [$\alpha$/Fe] in metal-poor stars in dSphs and in
the MW halo (or even the bulge) are similar, then there is no
need to resort to IMF variations between these systems. Poor IMF
sampling has been invoked as a possible cause of lowering
[$\alpha$/Fe] in dwarfs \citep{Tolstoy03, Carigi08}, but these new
large samples suggest that this explanation may no longer 
be necessary, at least in systems as luminous as Sculptor,
 Fornax or Sagittarius.
On the other hand, there is now a hint that the slightly less
  luminous Sextans (M$_V= -9.5$) could display a scatter in the
  $\alpha$/Fe ratios at the lowest metallicities, including 
  [$\alpha$/Fe] close to solar \citep{Aoki09}. Such a scatter is so far
  observed only in this purely old system, and suggests a 
very inhomogeneous metal-enrichment in this system that presumably
never retained much of the metal it produced. The true extent of this scatter in
Sextans remains to be investigated, and extension to other similar
systems is needed before general conclusions can be reached on the
mechanisms leading to the chemical homogeneity -or not- of dwarf galaxies.

At later times, in those stars which formed $\sim$1~Gyr after the
first stars, on the metal-rich side of the knee, the decrease of
[$\alpha$/Fe] with increasing metallicity is very well marked. In
fact, the end points of the evolution in each of the dSphs investigated
has a surprisingly low [$\alpha$/Fe], see Fig.~\ref{alpha_all}.
A natural explanation of these low ratios could involve a sudden decrease
 of star formation, that would make enrichment by massive stars
inefficient and leave SNIa to drive the chemical evolution. This
sudden drop in star formation could be the natural result of galactic
winds which can have a scientific impact on dwarf galaxies with
relatively shallow potential wells (see Section \ref{chem_models}) or
perhaps tidal stripping.
  In this case, one would expect the metal-rich and low [$\alpha$/Fe]
  populations to be predominantly young, corresponding to the residual
  star formation after the sudden decrease. However, the current
  age-determinations for individual giants in these systems are not
  accurate enough to probe this hypothesis \citep[e.g.,][]{Battaglia06}. 

\begin{figure}[ht]
\centerline{\psfig{file=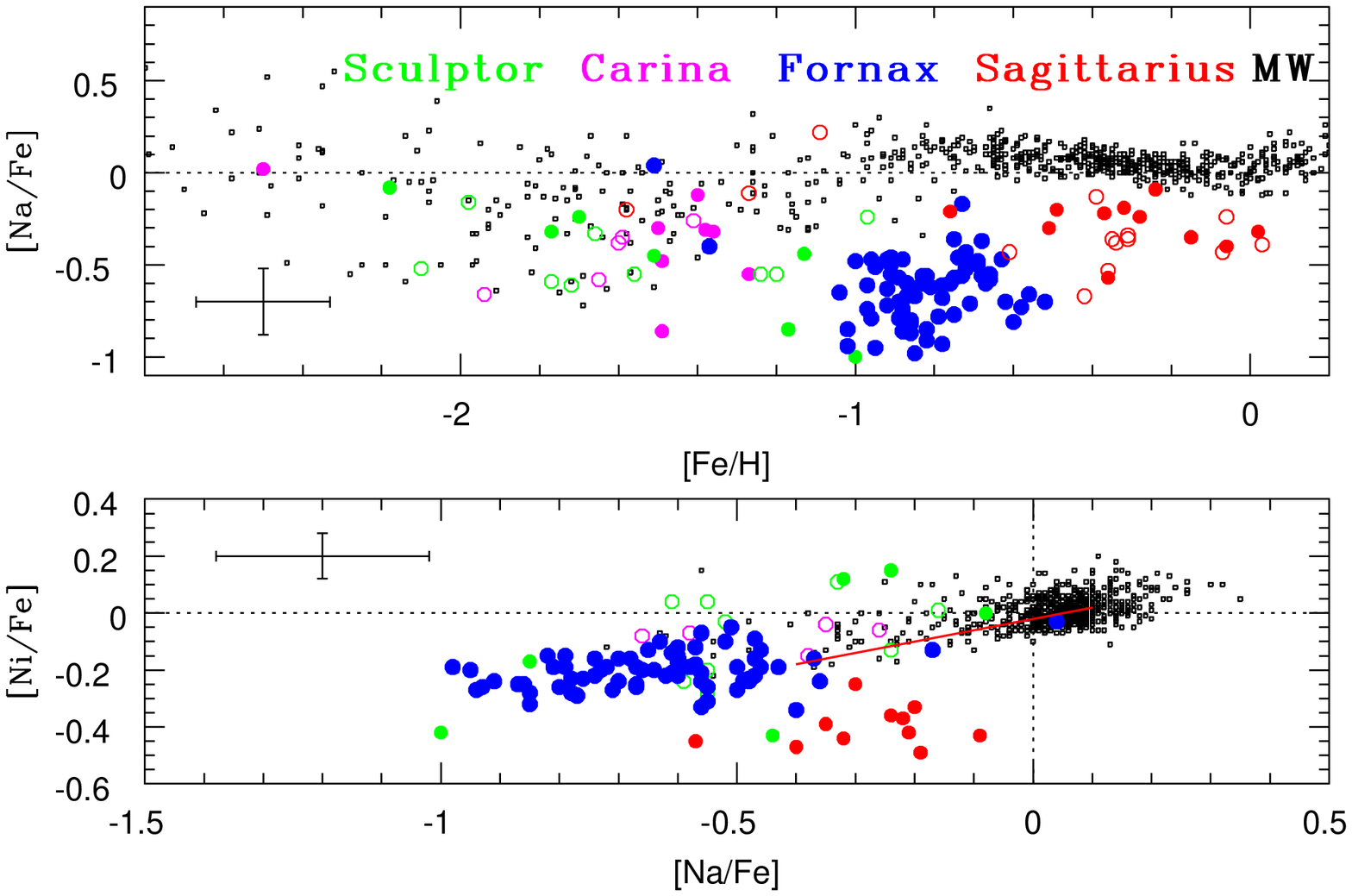,width=12cm}}
\caption{Sodium (above) and nickel (below) 
in  the same four dSphs as in Fig.~\ref{alpha_all}, 
compared to the
MW. Sgr \cite[red:][]{Sbordone07,Monaco05,McWilliam05conf}, Fnx
\cite[blue:][]{Letarte07phd, Shetrone03}, Scl \cite[green: Hill
et al. in prep;][]{Shetrone03,Geisler05} and Carina
\cite[magenta:][]{Koch08Car,Shetrone03}. Open symbols refer to
single-slit spectroscopy measurements, while filled circles refer to
multi-object spectroscopy. The small black symbols are a compilation
of the MW disk and halo star abundances, from \cite{Venn04}.
}
\label{odd_all}
\end{figure}

\subsubsection{Sodium and Nickel}

Another example of the low impact of massive-stars on the chemical
enrichment of dSphs is given by sodium. Fig.~\ref{odd_all} shows the
compilation of dSphs stars compared to the evolution of Na in the
MW. According to stellar current models, Na is mostly produced
in massive stars (during hydrostatic burning) with a
metallicity-dependent yield.  The abundance of Na in metal-poor dSph
stars is apparently not different to the MW halo stars at the
same [Fe/H], but its abundance at later stages in the evolution is
distinct from the MW, dSph producing (or keeping) too little Na to keep on the 
MW trend above [Fe/H] $> -1$.

Sodium and nickel under-abundances have also been remarked upon by
\citet{Nissen97,Nissen09} in a fraction of halo stars which also
display low $\alpha$ abundances, thereby producing a [Na/Fe]$-$[Ni/Fe]
correlation. This correlation is tentatively explained as the common
sensitivity of both elements to neutron-excesses in supernovae.
Fnx is the most striking example that seems to follow the same
slope as the Na-Ni relationship in the MW, but extending the
trend to much lower [Na/Fe] and [Ni/Fe] values \citep{Letarte07phd},
see Fig.~\ref{odd_all}.
Nickel, unlike sodium, is also largely produced in SNIa \citep{Tsujimoto95}, 
so the Ni-Na relation can in theory be modified by SNIa
nucleosynthesis, especially in the metal-rich populations of dwarfs
where the low [$\alpha$/Fe] ratios point towards a strong SNIa
contribution.

\begin{figure}[ht]
\centerline{\psfig{file=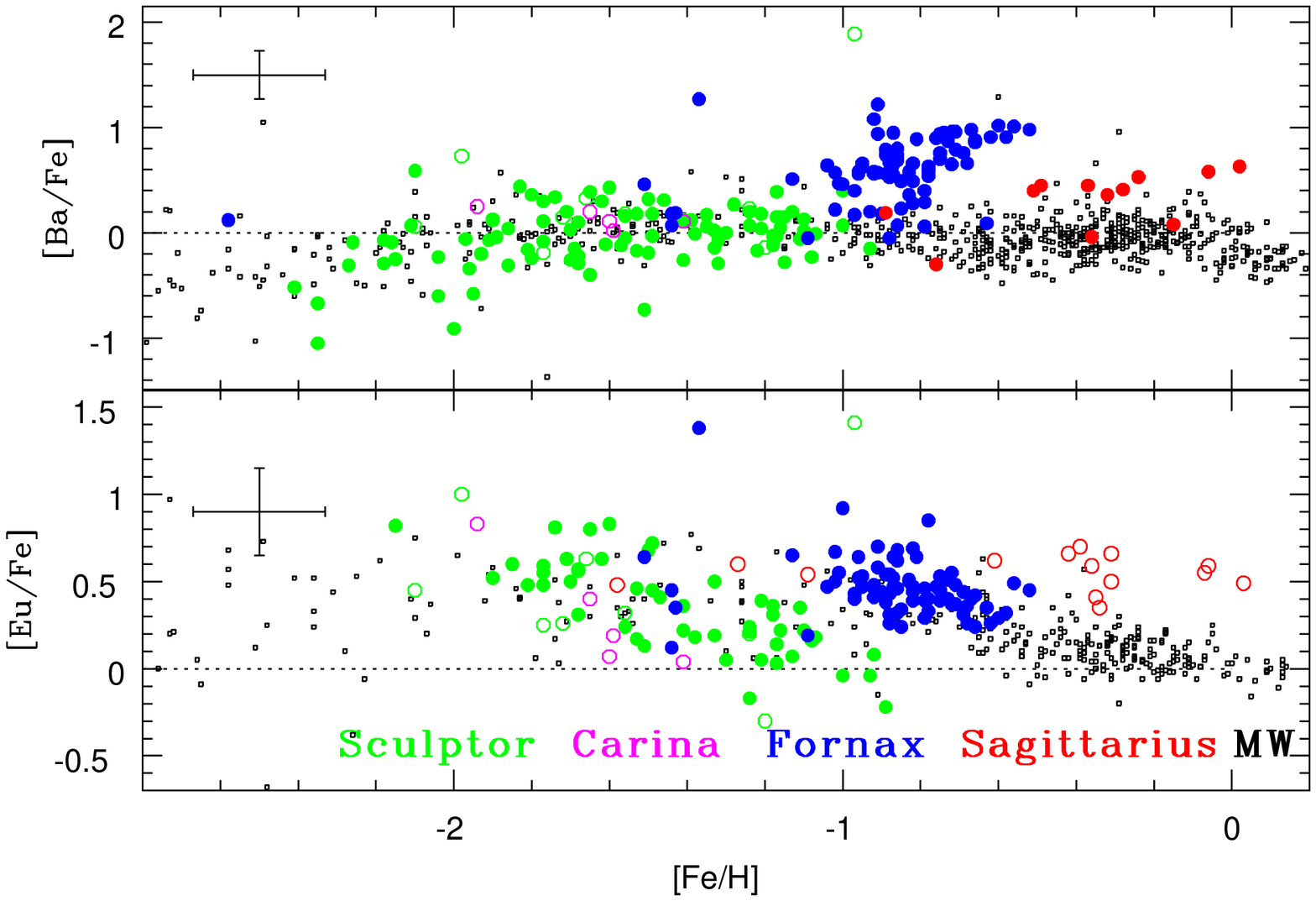,width=12cm}}
\caption{Neutron-capture elements Y, Ba \& Eu in the same four 
dSphs as in Fig.~\ref{alpha_all}, 
compared to the MW. Sgr \cite[red:][]{Sbordone07,Monaco05,McWilliam05conf}, Fnx
\cite[blue:][]{Letarte07phd, Shetrone03}, Scl \cite[green: Hill
et al. in prep;][]{Shetrone03,Geisler05} and Carina
\cite[magenta:][]{Koch08Car,Shetrone03}. Open symbols refer to
single-slit spectroscopy measurements, while filled circles refer to
multi-object spectroscopy. The small black symbols are a compilation
of the MW disk and halo star abundances, from \cite{Venn04}.
}
\label{heavies_sfe}
\end{figure}

\subsubsection{neutron-capture elements}

Despite their complicated nucleosynthetic origin, heavy neutron
capture elements can provide useful insight into the chemical
evolution of galaxies. Nuclei heavier than Z$\sim$30 are produced by
adding neutrons to iron (and other iron-peak) nuclei. Depending on the
rate (relative to $\beta$ decay) at which these captures occur, and
therefore on the neutron densities in the medium, the processes are
called either slow or rapid ({\it s-} or {\it r-}) process. 
The {\it s-}process is well
constrained to occur in low to intermediate-mass ($1-4 \rm M_{\odot}$)
thermally pulsating AGB stars \citep[see][and references
therein]{Travaglio04}, and therefore provide a contribution to
chemical enrichment that is delayed by $\sim 100-300$~Myrs from the
time that the stars were born. Thus {\it s-}process elements can in
principle be used to probe star formation on similar timescales to
[$\alpha$/Fe]. The {\it r-}process production site is clearly associated
with massive star nucleosynthesis. The most plausible candidate being
SNII, although the exact mechanism to provide the very large neutron
densities needed is still under debate, \citep[e.g.,][and references
therein]{Sneden08}.  This means that {\it r-}process elements should
contribute to the chemical enrichment of a galaxy with very little, if
any, delay.  Obviously they need pre-existing Fe-peak seeds and are
therefore not primary elements such as $\alpha$ elements.  One
complication arises from the fact that most neutron-capture elements
(through their multiple isotopes) can be produced by either the {\it s-} or
the {\it r-} process, such as yttrium (Y), barium (Ba) or lanthanum (La).
Among the few exceptions is europium (Eu), which is almost exclusively 
an {\it r-}process product.

\begin{figure}[ht]
\centerline{\psfig{file=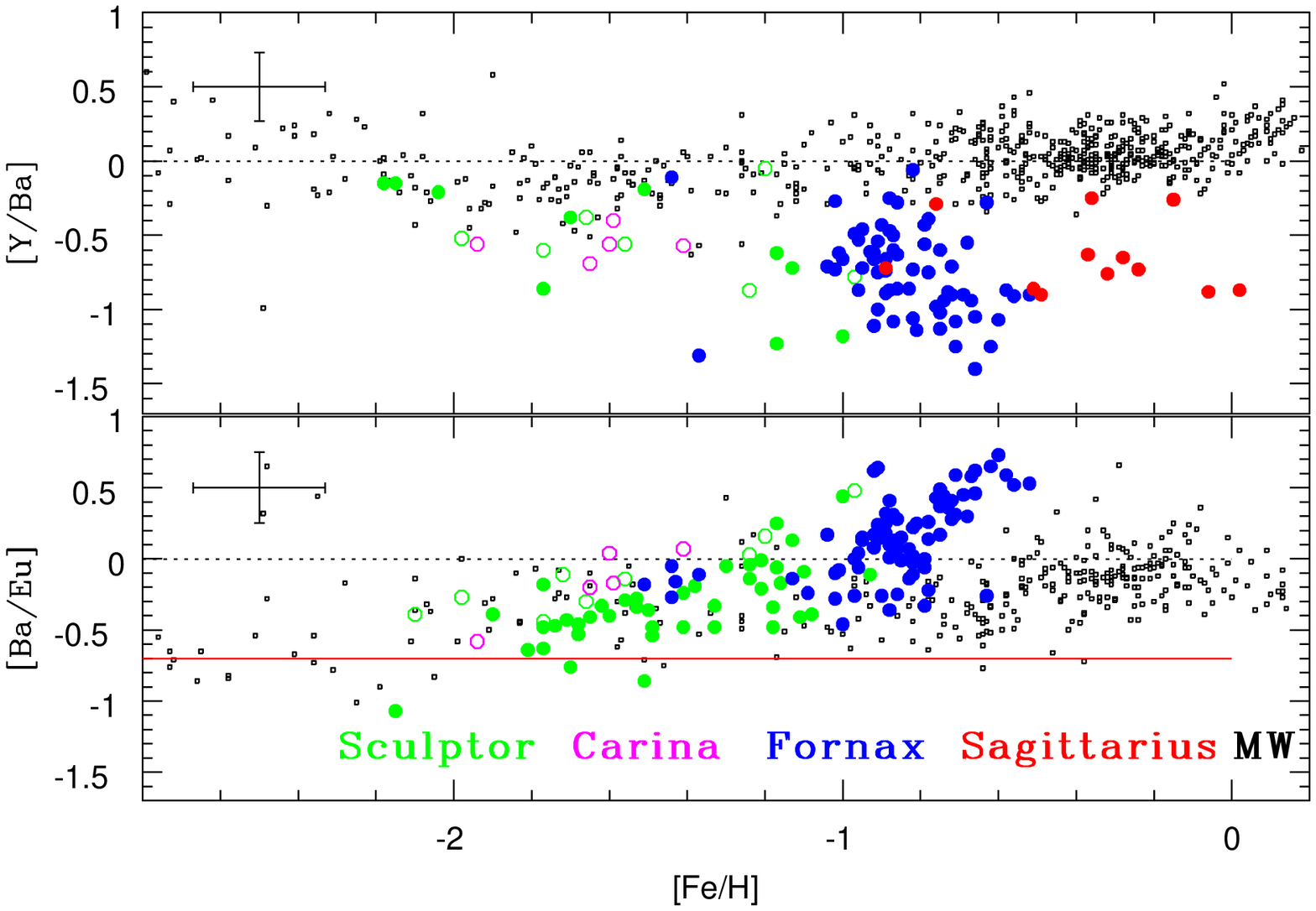,width=12cm}}
\caption{Ratios of {\it r-} to {\it s-} process element production in 
the same four dSphs as in Fig.~\ref{alpha_all}, 
compared to the MW. Sgr \cite[red:][]{Sbordone07,Monaco05,McWilliam05conf}, Fnx
\cite[blue:][]{Letarte07phd, Shetrone03}, Scl \cite[green: Hill
et al. in prep;][]{Shetrone03,Geisler05} and Carina
\cite[magenta:][]{Koch08Car,Shetrone03}. Open symbols refer to
single-slit spectroscopy measurements, while filled circles refer to
multi-object spectroscopy. The small black symbols are a compilation
of the MW disk and halo star abundances, from \cite{Venn04}.
}
\label{heavies_ratios}
\end{figure}

Fig.~\ref{heavies_sfe} compares  Ba and Eu abundances in four
dSph galaxies and in the MW. At first glance, the Eu evolution
in dSph galaxies resembles that of their respective $\alpha$-elements
(see Fig.~\ref{alpha_all}), as expected for an {\it r-}process originating
in massive stars.  In the MW, the Ba and Y are dominated by the
{\it r-}process for [Fe/H] $\lsim -2.0$
\citep[e.g.,][]{Simmerer04,Johnson02}, while the {\it s-}process dominates at
higher metallicities (e.g., more than 80\% of the solar Ba is of
{\it s-}process origin).

At early times (at [Fe/H] $< -1$) there seems to be little difference
between the various dSphs, and the MW halo in
Fig~\ref{heavies_sfe}.  However, there is a hint that at the lowest metallicities
([Fe/H] $< -1.8$), [Ba/Fe] increases in scatter and starts to turn
down. This hint is confirmed in the plot in Section~\ref{HR_uf} which
includes other dSphs, although from much smaller samples
\citep{Shetrone01, Fulbright04, Aoki09}. In fact, this scatter and downturn of
[Ba/Fe] is a well known feature in the MW halo \citep[][and references 
therein]{Francois07, Barklem05}, where it occurs at much lower
metallicities ([Fe/H] $< -3.0$). So far we have extremely low number
statistics for dSphs and these results need to be confirmed in larger
samples of low-metallicity stars.  These low {\it r-}process values at
higher [Fe/H] than in the Galactic halo would either mean that the
dwarf galaxies enriched faster than the halo at the earliest times or
that the site for the {\it r-}process is less common (or less efficient) in
dSphs. The {\it r-}process elements are clearly useful tracers of early time
scales, because unlike the $\alpha$-elements (in the halo and in dSphs)
they show significant scatter in the lowest metallicity stars. The
{\it r-}process is thus produced in much rarer events than the
$\alpha$-elements and so it can be a much finer tracer of time scales
and enrichment (and mixing) processes.

The
ratio of [Ba/Eu], shown in Fig.~\ref{heavies_ratios}, indicates the
fraction of Ba produced by the {\it s-}process to that produced by the
{\it r-}process.  In dSphs, as in the MW, the early evolution of all
neutron-capture elements is dominated by the {\it r-}process \citep[this was
already noted by][]{Shetrone01,Shetrone03}. In each system, however,
the low and intermediate mass AGB stars contribute {\it s-}process elements,
that soon start to dominate the Ba (and other neutron capture
elements) production. The metallicity of this switch from {\it r-}
to {\it s-}process ([Fe/H]$\sim -1.8$, the same as the [$\alpha$/Fe] knee) 
is only somewhat constrained in the Scl dSph.  
This turnover needs to be better constrained in
Scl and even more so in other galaxies to provide timing constraints
on the chemical enrichment rate. It could reveal the metallicity
reached by the system at the time when the {\it s-}process produced in AGBs
starts to contribute.

For the more metal-rich stars ([Fe/H] $> -1$) there is also a
distinctive behaviour of [Ba/Fe] in dSphs (Fig.~\ref{heavies_sfe}).  
In the Scl dSph the [Ba/Fe]
values never leave the MW trend, but this galaxy also has almost no
stars more metal-rich than [Fe/H]$<-1$.  Fnx, on the other hand,
and to a lesser extent Sgr, display large excesses of barium
for [Fe/H]$>-1$.  This is now barium produced by the {\it s-}process, and it
shows the clear dominance of the {\it s-}process at late times in dSphs. 

\begin{figure}[ht]
\centerline{\psfig{file=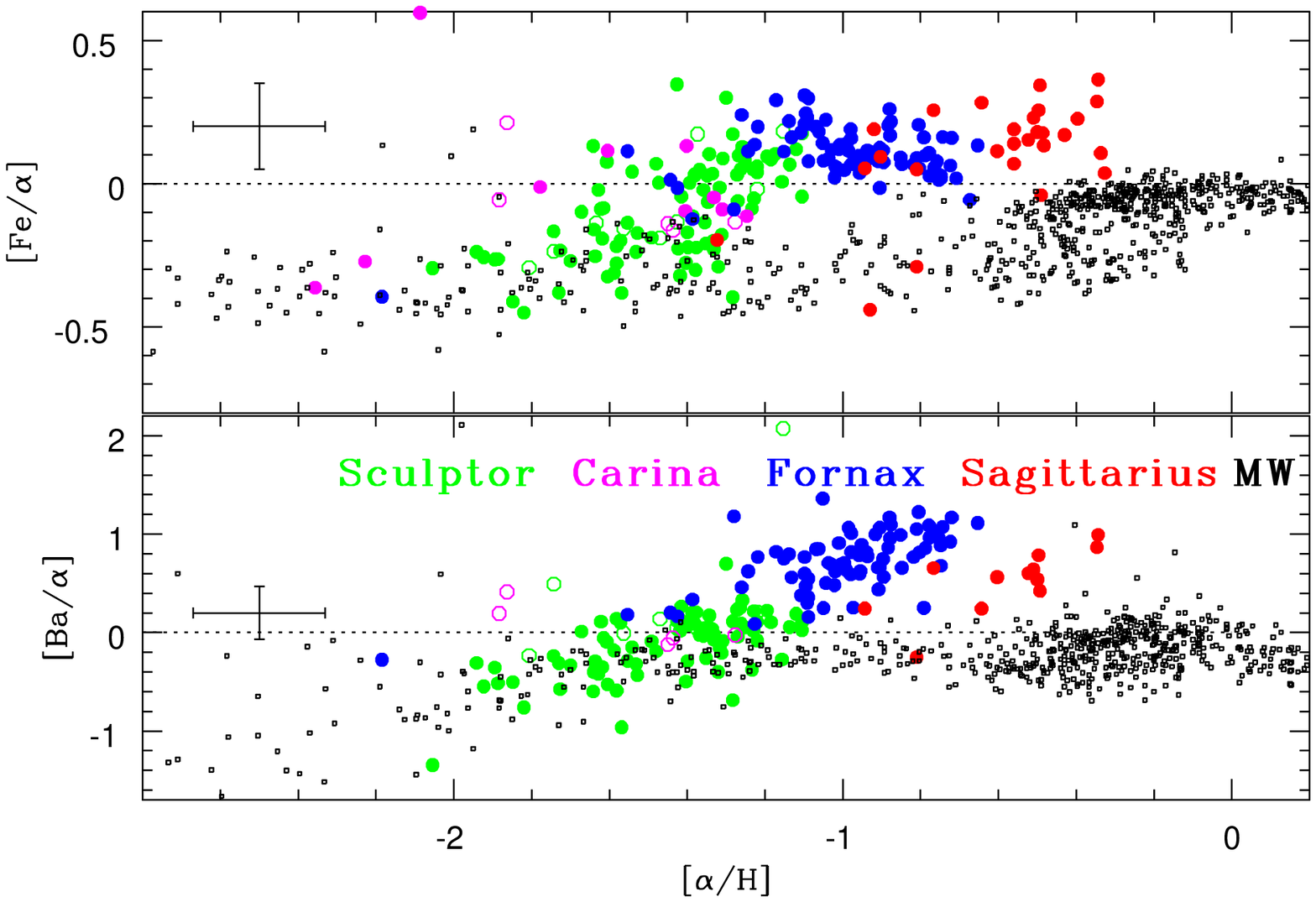,width=12cm}}
\caption{Trends of iron and neutron-capture elements as a function of
  $\alpha$ elements in the same four dSphs as 
  in Fig.~\ref{alpha_all},
 compared to the MW. Sgr \cite[red:][]{Sbordone07,Monaco05,McWilliam05conf}, Fnx
\cite[blue:][]{Letarte07phd, Shetrone03}, Scl \cite[green: Hill
et al. in prep;][]{Shetrone03,Geisler05} and Carina
\cite[magenta:][]{Koch08Car,Shetrone03}. Open symbols refer to
single-slit spectroscopy measurements, while filled circles refer to
multi-object spectroscopy. The small black symbols are a compilation
of the MW disk and halo star abundances, from \cite{Venn04}.
}
\label{heavies_salfa}
\end{figure}

Fig.~\ref{heavies_salfa} shows the trends of [Fe/$\alpha$],
and [Ba/$\alpha$] against [$\alpha$/H].  The fact that 
[$\alpha$/H] keeps increasing significantly after the knee in the 
Scl dSph
demonstrates that even in this system which has no significant
intermediate-age population, there was still ongoing star formation
contributing $\alpha$ enrichment from massive stars well after SNIa
started contributing. 
This is also confirmed by the presence of stars, which have
[Fe/H] $> -1.8$ (the knee), and were therefore formed after SNIa
started exploding.  In Fnx or Sgr, the very flat, extended and high
[Fe/$\alpha$] plateau also shows that massive stars have kept feeding
the chemical enrichment all along the evolution, even though they do
not dominate the Fe enrichment. As for the {\it s-}process, the widely
different behaviour of Scl, Fnx and Sgr and the MW are even
more striking viewed in this representation than they were in 
Fig.~\ref{heavies_sfe}, illustrating the total disconnect of massive stars
nucleosynthesis to Ba, and the strong influence of AGB stars at a time
when massive stars do not drive the metallicity evolution anymore.

\begin{figure}[ht]
\centerline{\psfig{file=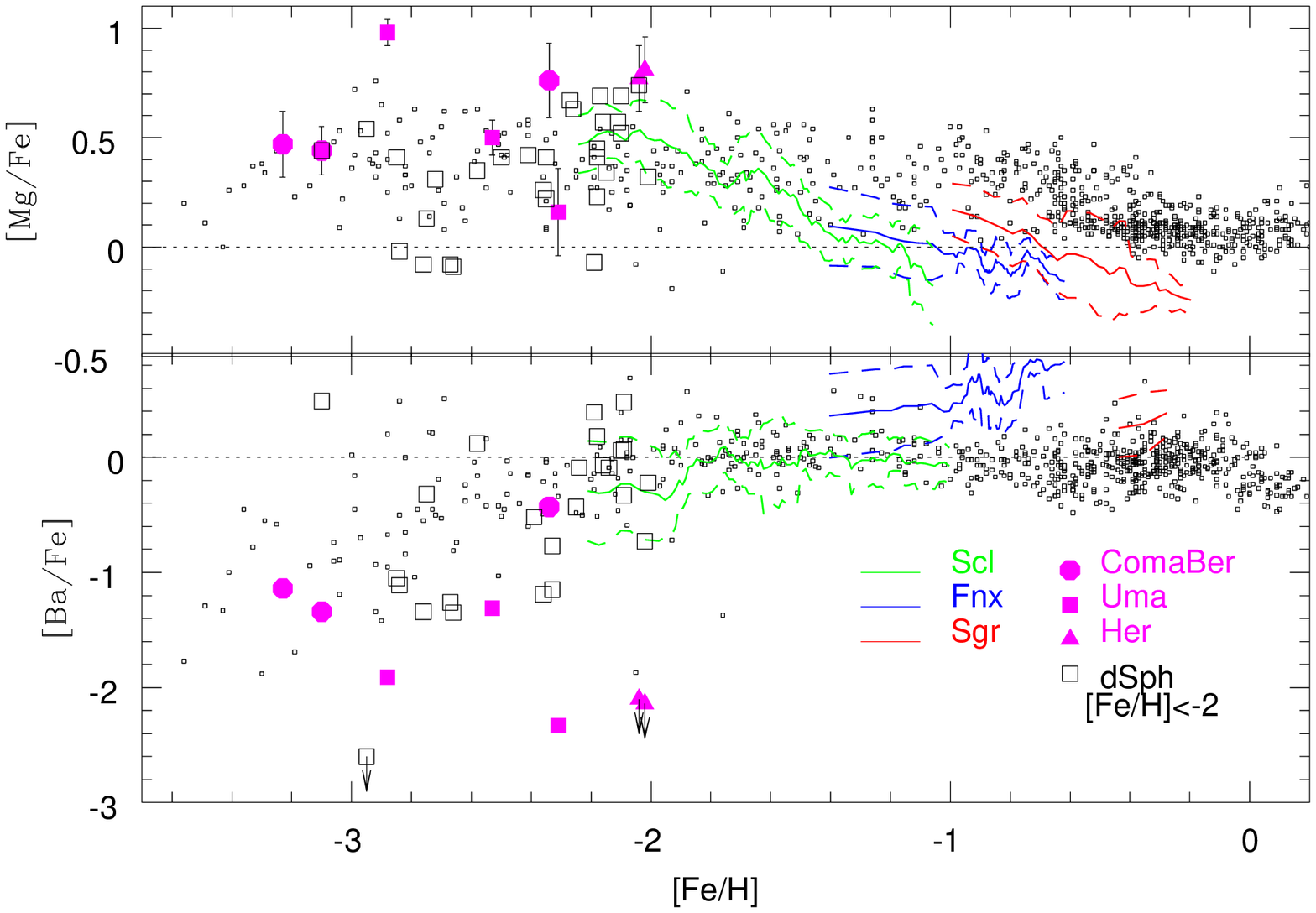,width=12cm}}
\caption{
Mg ($\alpha$ element) and Ba ({\it s-}process element) abundances in dSphs,
uFds and the Galactic halo. The magenta symbols are abundances of stars
in uFds as measured by \citet{Frebel09}, for three
stars in UMa~II (squares) and Coma (circles) and by
\citet[][triangles]{Koch08herc} in Herc. These are compared to the
trends derived from Fig.~\ref{alpha_all} and Fig.~\ref{heavies_sfe} for
Scl, Fnx and Sgr as well as individual stellar abundances for all very
metal-poor stars ([Fe/H]$<-2$) in dSphs \citep[][, black open
squares]{Fulbright04,Sadakane04,Venn04,Koch08Car,Aoki09}, and the MW
from the compilation by \citet{Venn04} and complemented by
\citet{Cayrel04, Francois07}. The dSph trends were derived by a simple
10 points running average on the data for each dSph galaxy with a
sufficient statistics (more than 20 measurements).
}
\label{abund-uf}
\end{figure}

\subsection{Ultra-Faint dwarf galaxies}\label{HR_uf}

Individual stars in the uFds that have recently been discovered around
the MW have so far been little observed at high spectral
resolution. This is probably due to the difficulty in confirming
membership for the brighter stars in these systems.  However, several
groups are currently following up confirmed members (typically
selected from lower resolution Ca~II triplet observations) to derive
abundances. So far, \citet{Koch08herc} have observed two RGB
stars in Herc (M$_V \sim -6.6$) and \citet{Frebel09} 
are following up RGB stars in the even fainter uFds UMa~II and
Coma (both with, M$_V \sim -4.$).  The latter study confirms 
that uFds do contain very metal-poor stars, [Fe/H]$<
-3$, \citep[as found by][]{Kirby08}, 
unlike the more luminous ``classical'' dSphs 
\citep{Helmi06}.  It also appears that these uFds extend the
metallicity-luminosity relation down to the lower luminosities
\citep{Simon07}, see Section~\ref{LR-ufaint}.

The two stars in Herc seem to have particularly peculiar abundance
patterns, with high Mg and O abundances (hydrostatic burning in
massive stars), normal Ca, Ti abundances (explosive nucleosynthesis in
massive stars), and exceedingly unenriched in 
 Ba \citep{Koch08herc}. On the
other hand, elemental ratios in the extremely metal-poor stars in the
two fainter dwarfs UMa~II and Coma \citep{Frebel09} are 
remarkably similar to the MW halo extremely
metal-poor stars. Fig.~\ref{abund-uf} compares Mg and Ba
measurements in faint dSphs, with all more luminous dSph stars that
have [Fe/H]$\lsim -2$ \citep[including a new sample of 6 very metal-poor
stars in Sextans by][]{Aoki09} and the MW. In fact, only Sextans seems
to have scattered and low [Mg/Fe] ratios, while other dSphs and uFds all show
similar [Mg/Fe] enhancements. The similarity between stars
with metallicities below [Fe/H]$\lsim -2.5$ in the MW and faint
dwarfs is seen also in other light elements, such as Na, Sc, Cr, Mn,
Ni or Zn. This may also be true of more luminous dSphs, see
Section~\ref{HR_dSph}.

The overall similarity between all the most metal-poor stars for
element ratios up to the iron-peak can be taken as an indication that
star formation and metal-enrichment, even at the earliest times, and
even in the smallest systems, has proceeded in a similar manner. This
may lead to the net yield of the very first stars. The very low
dispersion found in abundance ratios of these elements in Galactic
extremely metal-poor stars (EMPS), down to metallicities of [Fe/H]
$\sim -4$ came as a surprise \citep{Cayrel04}: since it was thought
that one or a few SN~II were sufficient to enrich the gas to those
metallicities, the expectation was that among EMPS the variety of
metal-production sites (SN~II of different masses) would appear
as dispersed abundance ratios. We are now adding to this puzzle the
fact that these well defined abundance ratios are also achieved by
considerably smaller halos.

The only discrepancy among the most metal-poor stars concerns the
{\it r-}process element Ba, that stands out below the MW halo
distribution both for faint and somewhat more luminous dSph
galaxies. The most extreme low Ba abundances are found so far in
Herc \citep{Koch08herc} and Draco \citep{Fulbright04}, where only
upper limits were detected.

\begin{figure}[ht]
\centerline{\psfig{file=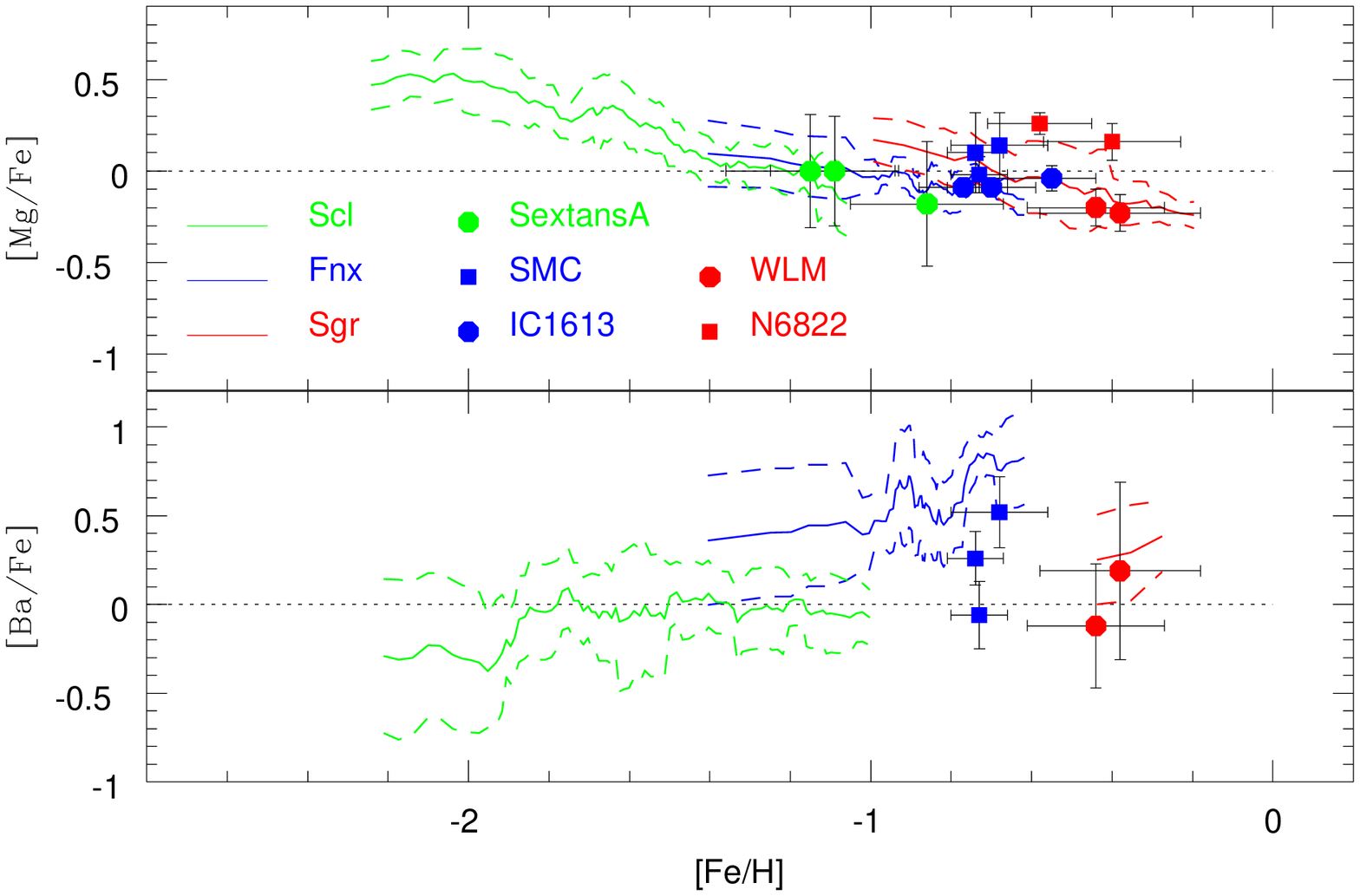,width=12cm}}
\caption{
Mg ($\alpha$ element) and Ba ({\it s-}process element) abundances in
individual super-giants in the dI galaxies Sex~A \citep{Kaufer04}, SMC
\citep{Venn99,Hill97,Luck98}, NGC~6822 \citep{Venn01} and WLM
\citep{Venn03}, compared to the trends derived from 
Fig.~\ref{alpha_all} and Fig.~\ref{heavies_sfe} for Scl, Fnx and Sgr (see
Fig.~\ref{abund-uf} for details on the trends). For the SMC, the
points are in fact the mean and dispersion of the samples studied in
the references given ($\sim 6-10$ A, K and Cepheids super-giants respectively). 
}
\label{abund-di}
\end{figure}

\subsection{Dwarf irregulars}\label{HR_dIrr}

The dIs are all (except the SMC) located at rather large distances
from the MW and so far, the only probes that could be used to
derive chemical abundances in these objects were HII regions
and a few super-giant stars. Both types of probes allow a look-back
time of at most a few 10~Myr, and this is the end-point of a
Hubble-time's worth of chemical evolution for any galaxy.  This
limitation makes it difficult to gather relevant information to
constrain the chemical enrichment over time in these systems.
However, abundances in HII regions and super-giants (see references in
Table~\ref{sfh_late}) 
are useful to understand how dIs fit in the general picture
of dwarf galaxies, and how they compare to larger late-type galaxies.
First, they give the present day metallicity of these systems, and all
are more metal-poor than the MW disk young population, in
agreement with the metallicity-luminosity relation \citep[see for
example][ for a relation based on dIs within 5~Mpc]{vanZee06b}, and
range between $\rm 12+\log(O/H)\sim8.1$ (e.g., NGC~6822, IC~1613) to $\rm 12 +
\log(O/H)\sim7.30$ (Leo~A), or [O/H]$\sim -0.6$ to $-1.4$.  Both HII
regions and super-giants typically agree on the oxygen abundances of
the systems, within the respective measurement uncertainties
\citep{Venn03,Kaufer04}, with little metallicity dispersion within a
galaxy, and no spatial gradient
\citep[e.g.,][]{Kobulnicky97,vanZee06,vanZee06b}. This holds even in
the most metal-poor galaxies, and suggests a very efficient mix of
metals across the galaxy despite the clumpiness of the ISM and ongoing
star-formation. 
The shear within these systems is expected to be very low, and this
has been taken as an indication that mixing occurs in the gaseous hot
phase, before the gas cools down to form new stars
\citep[e.g.,][]{vanZee06}.

A to M type super-giants have a further interest as they provide the
present-day [$\alpha$/Fe] ratios in dIs, which are not accessible from
HII regions 
where typically only light elements (e.g., He, N, O, Ne, S, Ar) can be
measured, and and no iron (nor any
other element that would trace SNIa).

The first dI where abundances of stars were measured was of
course the SMC in our backyard.
The largest samples to date with abundances in SMC are of 
super-giants which can be found in \citet[][K-type stars]{Hill97},
\citet[][F-type stars]{Luck98} or \citet[][A-type
stars]{Venn99}. Similar studies in more distant dIs needed efficient
spectrographs on 8$-$10m telescopes, and at the expense of observing for
many hours a few stars detailed abundances have been observed in
A-type super-giants out to distances of 1.3~Mpc. This work has been
pioneered by K. Venn and collaborators
using A-type stars \citep[][]{Venn01, Venn03,
Kaufer04} in NGC~6822, Sextans~A and WLM.  There has also been a more
recent study using M-type stars in IC~1613 by \citet{Tautvaisiene07}.

Fig.~\ref{abund-di} illustrates the observed low [$\alpha$/Fe] in
these systems, and compares them to the observed trends of older
populations (RGB stars) in dSph galaxies, as defined in
Figs.~\ref{abund-uf} \& \ref{alpha_all}. These low [$\alpha$/Fe] are
expected in galaxies that have formed stars over a long period of
time, however; they clearly occur at much lower metallicities than in
larger systems such as the MW or the LMC, pointing towards an
inefficient metal-enrichment of the galaxy (low star formation and/or
metal-losses through winds). It is interesting to see in
Fig.~\ref{abund-di} how dIs actually prolong the trends of dSph
galaxies, not only for $\alpha$ elements but also for neutron-capture
elements. From these diagnostics, dSphs are entirely consistent with
being dIs that lost their gas at a late stage of their evolution. The
Fnx dSph and the SMC, which are both dominated by intermediate-age
populations, are also quite similar in their chemical enrichment, 
except that Fnx ran out of 
gas (or lost its gas) and stopped star formation about $10^8$yr ago.


\section{Chemical Evolution Models}\label{chem_models}

The detailed evolutionary histories of dwarf galaxies have intrigued
astronomers for decades. One of the main reasons of interest is that 
they are often extremely low metallicity systems,
and thus assumed to be highly unevolved. Their low abundances
of metals and helium, derived from HII region spectra, allow
the determination of the primordial helium abundance with a minimum of
extrapolation \citep[e.g.,][]{Peimbert74, Izotov98, Olive97, Izotov07}
and thus provide insights into Big Bang Nucleosynthesis.  

Most recently dwarf galaxies have been of interest due to their
cosmological importance as potential building blocks of larger
systems.  Nearby dwarf galaxies are the closest we can get to the
detailed study of a primordial system.  They typically have relatively
simple structures and often very low metallicities. We assume that
because small systems are believed to be the first to collapse in the
early universe it was galaxies like these that were the first to
form and are thus potential hosts of the first stars.  Their
widespread distribution throughout the early Universe also makes them
suitable candidates to be able to re-ionise the Universe uniformly
and rapidly \citep[e.g.,][]{Choud08, Stark07}.  Recently there have
been large samples of stellar abundances of individual stars
obtained in nearby dwarf galaxies [e.g.][Hill et al. 2009, in
prep.]\citep{Letarte07phd, Monaco07}.  These will provide a wealth of
information on chemical evolution through time, and determine
the accurate evolutionary path of these small systems and their
contribution to Universal processes such as the build up of metals in
the Universe.

Beatrice Tinsley, beginning in the mid 1960s pioneered the field of
galactic chemical evolution modelling. The cornerstones were laid by
E. Salpeter in 1955 with his paper on the IMF and in 1959 with his
first determination of the effects of stellar evolution on the
metallicity evolution of stellar populations.  This work was extended
by M. Schmidt in 1959 and 1963 to determine universal predictions for
the SFR in a galaxy. Tinsley however provided the first full
description of the theoretical modelling of galactic chemical
evolution and of its relevance to many astrophysical topics.  In 1968
she was already studying the evolutionary properties of galaxies of
different morphological types \citep{Tinsley68}, and with subsequent
seminal papers \citep{Audouze76,Tinsley80} she set the stage for all
future studies of galactic chemical evolution.  Since the late
seventies \citep[e.g.,][]{Lequeux79} a wealth of chemical evolution
models have been computed for dwarf galaxies in general and for
late-type dwarfs in particular \citep[see e.g.,][and references
therein]{Matteucci83,Pilyugin93,Marconi94,Carigi95,Tosi98}.

The predictions of early models of the chemical evolution of dwarf
galaxies were far from unique \citep[as reviewed e.g., by][]{Tosi98},
because few observational constraints were available: the ISM chemical
abundances (helium, nitrogen and oxygen as derived from the emission
lines of HII regions), and the gas and total mass (mainly from 21 cm
radio observations). These data define present-day galaxy properties,
and do not constrain the early epochs.  This allows for little
discrimination between different models which may have very different
paths to the same end point. Star formation laws, IMF and gas flows
could be treated as free parameters and with their uncertainties it
was inconceivable to model the evolution of individual galaxies,
unless unusually rich in observational data. In practice, until
recently, only the Magellanic Clouds were modelled individually
\citep[e.g.,][]{Gilmore91,Pagel98}, even though many constraints were
still missing (e.g., accurate field star metallicity distribution,
detailed abundances in older populations, etc.).  It is fair to ask
how good the predictions of some of these models were because 30 years
later we are still arguing whether or not dIs and BCDs differ only in
the recent SFH, and if BCDs are actually ancient systems with a recent
burst, as predicted by \citet{Lequeux79}, or something entirely
different.

Detailed chemical evolution models of individual dwarf galaxies have
recently become possible as more accurate SFHs become available
combined with large samples of stellar abundances for
individual stars over a range of ages. This provides an accurate
age-metallicity relation which is a key constraint for chemical
evolution models.

\subsection{Explaining Low Metallicity}

One of the major challenges for chemical evolution models of dwarf
galaxies has always been to reconcile their low observed metallicity
with the fairly high SFR of the most metal-poor
systems, many of which are actively star-forming BCDs. Historically,
three mechanisms have been envisaged to accomplish this task
\citep{Matteucci83}:

\begin{itemize}
\item {\it variations in the IMF}; steeper IMF slopes and/or mass range
cut-offs have been proposed to reduce the chemical 
enrichment from massive stars; 

\item {\it accretion of metal free, or
very metal-poor gas}, to dilute the enrichment of the
galaxy; 

\item {\it metal-rich gas outflows}, such as galactic winds,
triggered by supernova explosions in systems with shallow potential wells,
or gas stripping due to interactions with other galaxies or with the
intergalactic medium to efficiently 
remove the metal enriched gas from the system.
\end{itemize}

Whether one of these mechanisms is preferable or a combination of any or
all of them is required is still matter of debate.  
When detailed numerical
models were computed it was immediately recognized that metal enriched
winds are the most straightforward mechanism to recreate the observed
properties of dwarf galaxies \citep{Matteucci85, Pilyugin93,
Marconi94, Carigi95}.  The infall of metal-poor gas can in principle
explain the evolution of gas-rich dwarfs, but gas accretion is also
most likely to trigger more star formation and chemical
enrichment. This can quickly lead to more rapid enrichment of small
dwarf galaxies.  Bottom-heavy IMFs imply abundance ratios for 
elements produced by stars of different mass which are at odds with
the observed values, see section~\ref{alphas}.

\subsection{Galactic Winds}

It was first proposed by \citet{Larson74} that gas could be blown out
by internal energetic events related to star formation, such as
stellar winds and supernovae explosions. These processes can
accelerate metal-rich stellar and supernova ejecta beyond the escape
velocity of small dwarf galaxies \citep[e.g.,][]{Heiles90,
TenorioTagle96, Rieschick03, Fujita04}.  The theory is periodically
further developed \citep[e.g.,][]{Dekel86, Dercole99, MacLow99,
Ferrara00, Legrand01, Tassis03, Marcolini06, Salvadori08}, and
naturally explains the well established correlation between luminosity
and metallicity \citep[e.g.,][]{Skillman89a,Gallazzi05}, as
smaller galaxies are less able to retain their heavy elements. It can
also explain the structural similarities observed by
\citet{Kormendy85b}, and it has even been suggested that many dwarf
galaxies have lost most of the gas mass they originally possessed and
hence follow the structural relations regardless of the current gas
mass fraction \citep{Dekel86, Skillman95}.  However this theory cannot
explain {\it why} some galaxies loose all their gas very early and
some relatively recently.  There has been no global parameter, such as
mass, found to explain this. Hence the influence of tidal effects is
considered to play an important, but hard to verify,
role \citep[e.g.,][]{Lin83}. It may
also be related to the varying initial conditions under which
different galaxies may have formed, or perhaps also the density of the
DM halo in which they reside.

Galactic winds have been predicted by hydrodynamical simulations
\citep{Dercole99, MacLow99} to be able to remove a large fraction of
the elements synthesized by SNII as well as a fraction of the galaxy
ISM.  Thus galactic winds can be quite effective and lead to a
significant reduction of the ISM enrichment.  The strength of this
effect depends both upon the galaxy mass, or the depth of the
potential well, and on the intensity of the star formation, and thus
the number of SN explosions that can be expected.  This is precisely
what is needed by chemical evolution models to reproduce the observed
properties of dwarfs. Moreover, there is increasing observational
evidence for starburst driven metal-enriched outflows
\citep[e.g.,][]{Meurer92, Heckman01, Martin02, Veilleux05,
Westmoquette08}.
Whether early-type dwarfs are connected to late-type dwarfs, as the
extreme consequence of tremendous winds from originally gas-rich
dwarfs or the consequence of gas stripping or ram pressure in harsh
environments is difficult to say. The evidence that gas-poor dwarfs
are preferentially located in denser environments than gas-rich ones
\cite[]{Binggeli90} seems however to favour the stripping scenario.

\subsection{Modelling Individual systems}

Two kinds of models for individual galaxies are commonly used:  
{\it standard} chemical evolution models  and {\it chemo-dynamical} models. 
The standard models follow the evolution of individual elements, taking into
account global parameters such as mass of the system, gas flows and IMF, and
stellar parameters such as their chemical yields and lifetimes, but make very
simplistic assumptions (if any) on stellar and gas dynamics 
\citep[see][for a comprehensive and still 
relevant review]{Tinsley80}. The chemo-dynamical models 
deal with the dynamical processes in great detail. 
Standard models are quite successful in predicting large-scale,
long-term phenomena, but their simplistic treatment of
stellar and supernovae feedback and of gas motions, is an obvious
 drawback.  
Chemo-dynamical models are more able to account for
small-scale, short-term phenomena, but the timescales required to run
hydrodynamic codes and the errors that start to creep in have made them 
less successful to follow galactic scale evolution over more than a
Gyr. The challenge in the next few years is to improve both types of
approaches and get a more realistic insight into how stars and gas
evolve, chemically and dynamically, in their host galaxies.

\subsubsection{Standard Models}

A number of standard models have been computed for nearby dSphs
adopting the individual SFHs derived from deep HST photometry and
comparing the model predictions with the stellar chemical abundances
inferred from new generation spectroscopy. \citet{Carigi02} analysed
Carina, Ursa~Min, Leo~I, and Leo~II and suggested a relation between
the duration of the star formation activity and the size of the dark
matter halo.  Lanfranchi \& Matteucci \cite[e.g.,][and references
therein]{Lanfranchi08} devoted a series of papers to the chemical
evolution of Carina, Draco, Sgr, Sextans, Scl and Ursa~Min 
reaching the conclusion that, to reproduce the observed stellar
abundance ratios and age-metallicity relations, they need low star
formation and high wind efficiencies. They suggest that a connection
between dSphs and BCDs is unlikely.

Due to the lack of stellar spectroscopy available for more distant dI
galaxies
chemical evolution models with the detailed approach applied to nearby
dSphs have been computed only for NGC~6822 \citep{Carigi06}, the
closest dI in the Local Group beyond the Clouds.  Models assuming the SFH
derived from HST CMDs have been computed also for the starburst dwarfs
NGC~1569 and NGC~1705 \citep{Romano06}, a few Mpc outside the Local Group.
Projects are in progress to model the chemical evolution of
the Magellanic Clouds with the level of detail and reliability
achieved so far only for the solar neighbourhood, as soon as their
SFHs and age-metallicity relations are derived
\citep[e.g.,][]{Tosi08}. The situation is expected to improve
significantly with the advent of new generation instruments on HST,
VLT and eventually Extremely Large Telescopes, which will allow to
measure reliable stellar metallicities at larger distances.

\subsubsection{Chemo-dynamical models}

To date chemo-dynamical models have mainly been used to study the
effects of feedback from supernovae explosions in a variety of
conditions. They can analyse in detail the heating and cooling
processes and put important constraints on the onset and fate of
galactic winds, stripping and ram pressure. However, they are not yet
able to follow the evolution of a galaxy over the entire Hubble time
assuming empirically derived SFHs. They have been applied to resolved
starburst dwarfs with SFH derived from HST photometry
\citep[e.g.,][for I~Zw~18 and NGC~1569]{Recchi04,Recchi06} and to a
few nearby dSphs \citep[see][for Draco, and Scl and Fnx]{Fenner06,
Marcolini06,Marcolini08}.

\subsubsection{Model predictions and observed abundances}

The different time scales for the chemical enrichment of elements
produced by different stellar processes are particularly useful to
constrain chemical evolution models.  This is especially true for {\it
r-} and {\it s-} process elements, and for Ba in particular.  Only a
few models \citep[and references therein]{Fenner06, Lanfranchi08} are
recent enough to have their predictions compared with the 
abundance patterns measured in dwarfs (as shown in
Section~\ref{hrspec}).  They thus deserve a few more words of comment.
Both types of models reproduce fairly well the observed properties of
the galaxies they are applied to, although sometimes they need to
assume chemical yields different from those available in the
literature \citep[and references therein]{Lanfranchi08}.

[Ba/Fe] is one of the few elements known to have a very large spread
at low metallicities \citep[e.g.,][]{Francois07}, see
Fig.~\ref{abund-uf}.  At early times Ba is produced by the {\it
r-}process which must be a rare occurrence and thus sensitive probe of
enrichment timescales.  The strong rise of [Ba/Fe] seen in Fnx or Sgr
\citep[and the LMC][]{Pompeia08} is clearly attributable to the {\it
s-}process, that is AGB stellar wind pollution.  This is currently not
well predicted by chemical evolution models of dSphs such as those of
\citet{Lanfranchi08}, and \cite{Fenner06}, probably because of the lack of
adequate stellar yields. In these models, the galactic wind removes
the gas and makes the SFR drop suddenly, preventing
high-mass stars from contributing to the enrichment and thereby
lowering the {\it r-}process contribution to the neutron capture
elements.  But the wind does not prevent low and intermediate mass
AGBs from contributing significantly to the ISM. The decreasing {\it
r-}process contribution is indeed observed in
Fig.~\ref{heavies_ratios} with the continuous rise of [Ba/Eu] in Fnx
in the range $\rm [Fe/H]>-0.8$, but this decrease of the r process
also acts to prevent [Ba/Fe] from rising: in fact, in these models,
[Ba/Fe] (or [La/Fe]) decrease slightly at high metallicities.  The
models of \citet{Fenner06} aim to reproduce the abundances of the Scl
dSph, and indeed there is a turn up of [Ba/Fe] that is linked to the
rise of the {\it s-}process, but [Ba/Fe] at low metallicities in Scl
is strongly underestimated in the models compared to observations.
This is probably due to the strong winds in these models, efficiently
removing metals produced by massive stars, including the {\it
r-}process that makes Ba at low metallicities.  These models also
predict low $\alpha$ elements down to the lowest metallicities in the
systems, which is also not observed.  In these models, the
low-metallicity AGB stars that produce the {\it s-}process responsible
for the [Ba/Fe] upturn, also produce the right [Ba/Y].

Finally, Fig.~\ref{heavies_ratios} shows that, in the domain where the
neutron-capture enrichment is dominated by the {\it s-}process, [Y/Ba]
in dSphs are exceedingly low. The most straight-forward interpretation
assumes that low-metallicity AGB stars dominate the {\it
s-}process. Suggesting that in these stars nucleosynthesis favours
high-mass nuclei over lower mass ones, as the result of less numerous
seed nuclei (iron mostly) being bombarded by similar neutron fluxes to
those at higher metallicities. However, this simple minded explanation
has so far lacked any quantitative prediction to be tested against
observations, owing largely to the uncertainties plaguing the detailed
{\it s-}process computations (thermal pulses in AGBs are a challenge
to model).  Another interpretation put forward by \citet{Lanfranchi08}
is that low [Y/Ba] is reached by
simply decreasing the {\it r-}/{\it s-} fractions at late times (as
above, due to galactic winds loosing preferentially {\it r-}process
elements).  Again, the models are not able to reproduce the steep rise
in heavier {\it s-}process elements such as Ba.  This shows us that
yields inferred from the solar neighbourhood are not
adequate. Although they can reproduce the abundance patterns in the
halo and disk of the MW, they cannot easily also reproduce the abundances
of dSphs, such as Fnx and Sgr, nor the LMC.

\section{Concluding Remarks}

In this review we have provided an overview of the current
understanding of the detailed properties of dwarf galaxies from
studies of their resolved stellar populations. This includes CMD
analysis, to determine accurate SFHs, as well as low and high
resolution spectroscopy, to determine kinematic and chemical
properties of stars over a range of ages.

Most dwarf galaxies in
the Local Group have structural properties similar to each other and
to larger late-type and spheroidal systems (section~\ref{intro}).
Early-type dwarfs tend to extend to fainter magnitudes, with
transition types being found at the faint end of the dI distribution. The
fact that there exists a transition type, intermediate in properties
between a dSph and a dI, supports the idea that there is an
evolutionary pathway.  The transition between early and late
types may indicate the average mass at which
galaxies will always loose their gas, especially if they spend time
in the vicinity of a large galaxy. But of
course this mass will be dependent on the environment that a galaxy
has passed through, which could explain why this is not a sharp
cut-off.  Despite numerous caveats and regardless of size, luminosity
and SFH all dSph and dI galaxies in the LG (and beyond) appear to
overlap along a straight line in the M$_V - \mu_V$, plane (see top
panel, Fig.~\ref{fig-bing}), a relation which is unchanged over a
range of $\approx$15~magnitudes in M$_V$.

The continuity of structural properties from dwarf galaxies to larger
spheroidal and late-type systems is most likely dominated by physical
processes that scale with mass. For example, the efficiency with which
gas and/or metals can be lost to a system during its evolution through
supernova winds and/or interactions.  Thus, early-type dwarf galaxies
in the Local Group must have suffered the largest effect due to
interactions with large galaxies, as has already been suggested from
the morphology-density relation.  Accurate SFHs have been determined
for a range of dwarf galaxy types (see section~\ref{sfhs}), The
different classes of dwarf galaxies have different rates of present
day star formation activity and possibly also different degrees of
past disruption. However the past SFHs of early and late-type systems
bear strong similarities to each other (see
Fig.~\ref{sfh_lcid}). There is evidence for interruptions and
enhancements in the SFHs of dwarf galaxies. This is especially true of
early-types, a few of which have experienced star formation activity
only at the earliest epochs, but most have had extended or recurrent
star formation activity.  No genuinely ``young'' galaxy (of any type)
has ever been found; stars are always found at the oldest lookback
times observed.

The SFHs of BCDs (i.e., comparing Fig.~\ref{sfh_lcid} with
Fig.~\ref{sfh_bcd}), are also broadly similar to dIs.  However the
recent SFRs in BCDs are usually much higher than in dIs.  The SFRs in
BCDs are more similar to those found in active star forming zones with
HII regions in the SMC (Fig.~\ref{sfh_in}) or in the MW, with the
difference that the BCDs are forming stars globally, dominating the
entire galaxy.  All the BCDs which have been studied in detail (see
section~\ref{bcds}) have apparently had their strongest star formation
episode recently, unlike dIs. This is most likely a selection effect
due to the difficulty in finding distant low luminosity dwarf
galaxies, unless they happen to be currently actively forming stars.

Spectroscopy of individual stars has helped to define the detailed
chemical and kinematic properties of stellar populations of different
ages in nearby dSph systems and in comparison to larger systems such
as the MW and the LMC (sections~\ref{lrspec} \& \ref{hrspec}).  The
chemical enrichment of dwarf galaxies seems to be dominated by
effects that are most likely dominated by gas and metal loss.
The least massive systems actually seem to loose such
a large fraction of their metals during star formation episodes that
star formation has a slow effect on the global chemical evolution (see
section~\ref{chem_models}). This means that galaxies with the same
mass but quite different SFHs end up with the same final metallicity,
consistent with the well known mass/luminosity-metallicity relation for
dwarf galaxies.

One clear mismatch in the physical properties between early and late
type dwarfs is that the HI gas in the brighter late-types, such as
SMC, NGC~6822 and IC~1613, is rotating with $\sim$20$-$60kms$^{-1}$.
Such high rotation values are never seen in the stars of early-type
dwarfs.  However, we really do not know if old and young stars in dwarf
galaxies can have different kinematic properties, such as
is seen in the MW.  
The kinematics of late-type galaxies
has always been measured using HI gas, out of which their young
populations are currently forming.  Early-type galaxies, on the other
hand, are of necessity probed using only their old or even ancient
stellar populations.  A careful comparison of the kinematics and
metallicity distributions of {\it equivalent tracers} in early and
late type galaxies has still to be made.

The abundance patterns of RGB stars for large samples of
individual stars, typically in dSph galaxies (section~\ref{hrspec}),
show that there are distinct differences in the chemical evolution paths 
between galaxies. The rate
at which $\alpha-$enrichment occurs varies between 
systems. There are not yet large enough samples in a diversity of
environments to really say how this may or may not relate to the mass
of a system, its SFH or the rate of mass and/or metal 
loss. But stars do retain a
clear abundance signature of the galactic 
environment in which they were
born. These patterns also extend to younger stars in late-type
systems, see Fig.~\ref{abund-di}. 

The hierarchical theory of galaxy
formation contains at its heart the concept of smaller systems
continuously merging to form larger ones. 
This leads to the general
expectation that the properties of the smaller systems will be
reflected in the larger. 
Thus 
the relationship between the properties of individual stars in small
dwarf galaxies around the MW, and stars in the MW is a recurrent theme
(see section~\ref{hrspec}).  From
recent abundance studies of low metallicity stars in dSphs, see
Fig.~\ref{abund-uf}, it seems likely that there exist only narrow windows
of opportunity when the merging of dwarf galaxies to form larger
systems would not lead to inconsistencies.  To properly understand the
constraints that these kinds of data can put on the merger history of
the MW requires more extensive abundance studies of metal poor stars
in dwarf galaxies, as well as a better theoretical understanding of
supernovae yields (including the {\it r-}process) and the mixing of
ejecta into interstellar gas.  It also remains an open question how
the uFds may relate to the merger history of the MW, and if they
can fully account for the deficiencies of the larger types of
dwarf galaxies as building block of the MW.

The Dark Matter content of dwarf galaxies is also of importance for
the verification of cosmological theories, as it indicates how galaxies
we see today which may have lost a significant fraction of their
initial baryons 
relate to structures in cosmological simulations. 
For an accurate determination of the dynamical properties it must be
realised that dwarf galaxies are not the simple systems they were once
thought to be. To make assessments of the total mass
of small galaxies is complicated, not least because the dark matter
halos are likely to extend beyond the baryonic tracers, but also
because there are multiple components in the baryonic matter.

This review shows the inherent complexities that are involved in
understanding even the smallest galaxies. These  
low metallicity systems show a wealth of variety in their properties,
such as luminosity, surface brightness, star formation history (both
past and present), kinematics and abundances. However, 
there is strong evidence that they are all part of a continuous
distribution of galaxies from small to large. 

\vskip 2cm

{\it ... E quindi uscimmo a riveder le stelle.} 

\noindent\small{(Dante Alighieri, La Divina Commedia, Inferno XXXIV, 139}

\clearpage 

\begin{landscape}
\begin{table}
\tiny
\begin{center}
\caption{Dwarf Galaxies with synthetic CMD SFH analyses: 
late and transition types in the Local Group}
\label{sfh_late}
 \begin{minipage}{20cm}
\begin{tabular}{llcclccccccc}
Galaxy & D (kpc) & M$_V$\footnote[1]{From Mateo 1998, updated using new distances, except where otherwise indicated} & r$_{h}$ ($'$)\footnote[2]{Holmberg limit, from, Mateo 1998, except where otherwise indicated}& instrument (fov) & look-back\footnote[3]
{faintest main features visible in the CMD: red giant branch (RGB), Horizontal Branch (HB), Main Sequence Turnoffs $>$~2~Gyr old (MSTO), oldest Main Sequence Turnoffs (oMSTO)} 
& $\le 10$~Myr & $1-8$~Gyr & $\ge 10$~Gyr & \multicolumn{3}{c}{Spectroscopy} \\
       &   &       &          &      &          &              &           &              & LR\footnote[4]{individual RGB stars} & HR$^d$ & HII\\
       &  &     &   &      &          &              &           &              &     &    &    \\
(1)    & (2)   & (3) & (4)      & (5)   & (6)      & (7)          & (8)      & (9)\footnote[5] {If the CMD is ambiguous as to the presence of an HB, but the presence of ancient stars has been confirmed by the measurement of RR~Lyr variable stars, this is noted. A question mark signifies that there is not enough information to determine whether or not an ancient population is present}  & (10) & (11) & (12) \\ 
\\
WLM         & 978$\pm$20~~[1]   & -14.6 & 5.5  & WFPC2~~(160$''$)~~[2] & HB     & $\surd$ & $\surd$ & $\surd$ & [3] & [4] & [5]  \\
Sextans~B   & 1370$\pm$180~~[6] & -14.2 & 3.9  & ESO/2.2m~~(2$'$)~~[7] & RGB     & $\surd$ & $\surd$ & ? & ... & ... & [8]  \\
NGC~3109    & 1300$\pm$200~~[9] & -15.8 & 13.3 & ESO/2.2m~~(2$'$)~~[10] & RGB   & $\surd$ & $\surd$ & ? & ... & [11] & [12]  \\
NGC~6822    & 460$\pm$5~~[13]   & -15.1 & 40   & INT/WFC~~(23$' \times 11'$)~~[14]    & RGB     & $\surd$ & $\surd$ & ? & [15] & [16] & [17]  \\
            &                   &       &      & WFPC2~~(160$''$~~[18]  & HB      & $\surd$ & $\surd$ & RRL[19] & & &  \\
Leo~A       & 800$\pm$40~~[20]  & -11.7 & 3.9  & ACS~~(195$''$)~~[21]  & oMSTO      & $\surd$ & $\surd$ & RRL[20] & [22] & ... & [23]  \\
Sextans~A   & 1320$\pm$40~~[24] & -14.5 & 4.0  & WFPC2~~(160$''$)~~[25]  & HB      & $\surd$ & $\surd$ & ? & ... & [26] & [8,27]  \\
IC~1613     & 721$\pm$5~~[28]   & -14.6 & 11$\pm$3& WFPC2~~(160$''$)~~[29]& MSTO   & $\surd$ & $\surd$ & $\surd$ & ... & [30] & [31]  \\
            &                   &       &      & ACS~~(195$''$)~~[32]    & oMSTO   & $\surd$ & $\surd$ & $\surd$ & & &  \\
SagDIG	    & 1050$\pm$50~~[33] & -12.2 & 1.7  & ACS~~(195$''$)~~[33]    & HB      & $\surd$ & $\surd$ & $\surd$ & ... & ... & [34]  \\
Pegasus     & 919$\pm$30~~[35]  & -12.8 & 3.9  & WFPC2~~(160$''$)~~[36]       & RGB     & $\surd$ & $\surd$ & ? & ... & ... & [37]  \\
DDO~210\footnote[6]{transition types, gas but no star formation}
            & 1071$\pm$39~~[35] & -10.6 & 1.6  & Subaru~~(30$'$)/VLT~~(7$'$)~~[38]  & HB     & x & $\surd$ & $\surd$   & ... & ... & x  \\
LGS~3$^f$   & 620$\pm$20~~[39]  & -9.9  & 14.5$\pm$4.5& WFPC2~~(160$''$)~~[39]& HB  & x & $\surd$ & $\surd$ & [40] & ... & x  \\
            &                   &       &      & ACS~~(195$''$)~~[41]    & oMSTO   & x & $\surd$ & $\surd$ & & &  \\
Phoenix$^f$ & 406$\pm$13~~[42]  & -10.1 & $>$8.6 & WFPC2~~(160$''$)~~[43]& HB      & x & $\surd$ & $\surd$ & [44] & ... & x  \\
Leo T$^f$   & 400$\pm$40~~[45]  & -8.0~~[45]   & 1.4~~[45]& LBT~~(23$'$)~~[46]& HB     & x & $\surd$ & ? & [47] & ... & ... \\
{\it SMC}   & 59.7$\pm$2.2~~[48]& -16.1   & 320 & WFPC2~~(160$''$)~~[49]& oMSTO & $\surd$ & $\surd$ & $\surd$ & [50] & [51] & [52]  \\
            &                   &       &      & LCO1m~~drift scan~~[53]  & MSTO    & $\surd$ & $\surd$ & $\surd$ & & &   \\
            &                   &       &      & WFI~~(30$'$)~~[54]    & oMSTO   & $\surd$ & $\surd$ & $\surd$ & & &  \\
            &                   &       &      & ACS~~(195$''$)~~[55]    & oMSTO   & $\surd$ & $\surd$ & $\surd$ & & & \\
GR~8        & 2200$\pm$400~~[56,57] & -12.3~~[58]     & 1.0~~[58] & WFPC2~~(160$''$)~~[57]       & RGB     & $\surd$ & $\surd$ & ?       & ... & ...  & [59]  \\

\\ 
\end{tabular}
\end{minipage}
\end{center}
\end{table}

\clearpage

[1] \citet{Gieren08}; [2] \citet{Dolphin00a}; [3] \citet{Leaman09}; [4] \citet{Urbaneja08}; \citet{Venn03}; [5] \citet{Lee05}; 
[6] \citet{Sakai97}; [7] \citet{Tosi91}; [8] \citet{Magrini05}; 
[9] \citet{Soszynski06}; [10] \citet{Greggio93}; [11] \citet{Evans06}; [12] \citet{Pena07}; 
[13] \citet{Gieren06}; [14] Gallart et al. 1996a,b; [15] \citet{Tolstoy01}; [16] \citet{Venn01}; [17] \citet{Lee06}; [18] \citet{Wyder01, Wyder03}; [19] \citet{Clementini03}; 
[20] \citet{Dolphin02leoa}; [21] \citet{Cole07}; [22] \citet{Brown07}; [23] \citet{vanZee06}; 
[24] \citet{Dolphin03dist}; [25] \citet{Dolphin03}; [26] \citet{Kaufer04}; 
[27] \citet{Kniazev05}; 
[28] \citet{Pietrzynski06}; [29] \citet{Skillman03}; [30] \citet{Bresolin07}; \citet{Tautvaisiene07}; [31] \citet{Peimbert88}; [32] Skillman et al. 2009 in prep. (LCID);
[33] \citet{Momany05}; [34] \citet{Skillman89b}; 
[35] \citet{McConnachie05}; [36] \citet{Gallagher98}; [37] \citet{Skillman97}; 
[38] \citet{McConnachie06}; 
[39] \citet{Miller01}; [40] \citet{Cook99}; [41] Hildago et al. 2009, in prep. (LCID); 
[42] \citet{Held99}; [43] \citet{Holtzman00, Young07}; [44] \citet{Gallart01, Irwin02}; 
[45] \citet{Irwin07}; [46] \citet{DeJong08}; [47] \citet{Simon07};
[48] \citet{Hilditch05}; [49] \citet{Dolphin01b}; [50] \citet{Carrera08, Harris06}; [51] \citet{Hill97, Venn99, Evans05}; [52] \citet{Vermeij02}; 
[53] \citet{Harris04}; [54] \citet{Noel07}; [55] \citet{Cignoni09}; 
[56] \citet{Tolstoy95};  [57] \citet{Dohm98}; [58] Mateo 1998; [59] \citet{vanZee06};
\end{landscape}

\clearpage

\begin{landscape}
\begin{table}
\tiny
\begin{center}
\caption{Dwarf Galaxies with synthetic CMD SFH analyses: 
early-types in the Local Group}
\label{sfh_early}
 \begin{minipage}{20cm}
\begin{tabular}{llcclccccccc}
Galaxy & D (kpc) & M$_V$\footnote[1]{From Mateo 1998, updated using new distances, except where otherwise indicated} & r$_{h}$ ($'$)\footnote[2]{Holmberg limit, from, Mateo 1998, except where otherwise indicated}& instrument (fov) & look-back\footnote[3]{faintest main features visible in the CMD: tip of the red giant branch (TRGB), Horizontal Branch (HB), oldest Main Sequence Turnoffs (oMSTO)} 
 & $\le 10$~Myr & $1-8$~Gyr & $\ge 10$~Gyr & \multicolumn{3}{c}{Spectroscopy} \\
       &   &       &          &      &          &              &           &              & LR\footnote[4]{individual RGB stars} & HR$^d$ & HII\\
       &  &     &   &      &          &              &           &              &     &    &    \\
(1)    & (2)   & (3) & (4)      & (5)   & (6)      & (7) & (8)      & (9)\footnote[5]{If the CMD is ambiguous as to the presence of an HB, but the presence of ancient stars has been confirmed by the measurement of RR~Lyr variable stars, this is noted} & (10) & (11) & (12) \\ 
\\
Carina   & 101$\pm$5~~[1]  & -9.3  & 28.8$\pm$3.6 & CTIO4m~~(15$'$)~~[2]     & oMSTO & x & $\surd$ & $\surd$ & [3] & [4,5] & x  \\
         &               &       &              & WFPC2~~(160$''$)~~[6,7]\footnote[6]{same data set, different analyses}  & oMSTO & x & $\surd$ & RRL[8] & &  &   \\
Leo~I    & 254$\pm$17~~[9] & -11.9 & 12.6$\pm$1.5 & WFPC2~~(160$''$)~~[6,7,10]$^f$  & oMSTO & x & $\surd$ & RRL[11] & [12] & [4] & x  \\
Leo~II   & 233$\pm$15~~[13]& -9.8  & 8.7$\pm$0.9  & WFPC2~~(160$''$)~~[6,7]$^f$ & oMSTO & x & $\surd$ & $\surd$ & [14] & [15] & x  \\
         &               &       &              & WFCAM~~(14$'$)~~[16]     & HB    & x & $\surd$ & $\surd$ & & & \\
Ursa~Min & 70$\pm$9~~[17]  & -9.0  & 50.6$\pm$3.6 & WFPC2~~(160$''$)~~[6,7]$^f$ & oMSTO & x & $\surd$ & $\surd$ & [18] & [19] & x  \\
         &               &       &              & INT/WFC~~(23$' \times 11'$)~~[20]       & oMSTO & x & $\surd$ & $\surd$ & & & \\
         &               &       &              & KPNO0.9m~~(23$'$)~~[21]  & oMSTO & x & $\surd$ & $\surd$ & & & \\
Draco    & 76$\pm$6~~[22]  & -8.6  & 28.3$\pm$2.4 & INT/WFC~~(23$' \times 11'$)~~[23]       & oMSTO & x & $\surd$ & $\surd$ & [18] & [19, 24] & x  \\
         &               &       &              & WFPC2~~(160$''$)~~[7]      & oMSTO & x & $\surd$ & $\surd$ & & & \\
Sculptor & 85.9$\pm$5~~[25]& -11.2 & 76.5$\pm$5   & WFPC2~~(160$''$)~~[7]      & oMSTO & x & $\surd$ & $\surd$ & [26] & [4, 27] & x  \\
Fornax   & 138$\pm$5~~[28] & -13.2 & 71$\pm$4     & ESO/WFI~~(34$'$)~~[29]       & oMSTO & x & $\surd$ & $\surd$ & [30] & [4, 31] & x  \\
         &               &       &              & FORS~~(7$'$)~~[32]      & oMSTO & x & $\surd$ & $\surd$ &      &      &   \\
Cetus    & 775$\pm$50~~[33]& -10.1~~[33]& 4.8[33]   & ACS~~(195$''$)~~[34]       & oMSTO & x & $\surd$ & $\surd$ & [35]  & ...  & x  \\
Tucana   & 880$\pm$40~~[36]& -9.6  & 3.7$\pm$1.2  & ACS~~(195$''$)~~[37]       & oMSTO & x & $\surd$ & $\surd$ & [38]  & ...  & x  \\

NGC~185 & 616$\pm$26~~[39] & -15.5 & 16$\pm$2     & NOT~~(3.8$'$)~~[40]       & TRGB  & x & $\surd$ & RRL[41] & ... & ... & x   \\
NGC~205 & 824$\pm$27~~[39] & -16.6 & 6.2$\pm$0.2  & NOT~~(3.8$'$)~~[40]       & TRGB  & x & $\surd$ & RRL[42] & [43] & ... & x   \\
BooI    & 62$\pm$3~~[44, 45]   & -5.8~~[45] & 12.8$\pm$0.7[45]& SDSS~~($>$8000$^o$)~~[46]& HB    & x & $\surd$ & $\surd$ & [47, 48]  & [50] & x  \\
CVnI    & 220$\pm$20~~[45] & -7.9~~[45] & 8.5$\pm$0.5[45] & SDSS~~($>$8000$^o$)~~[46]& HB    & x & $\surd$ & $\surd$ & [48, 49] & ... & x  \\
UMaII   & 32$\pm$5~~[45]   & -3.8~~[45] & $\sim$12[45]    & SDSS~~$>$8000$^o$()~~[46]& HB    & x & $\surd$ & $\surd$ & [48] & [51] & x  \\

\\ 
\end{tabular}
 \end{minipage}
\end{center}
\end{table}

\clearpage

[1] \citet{Mateo98a}; [2] \citet{Hurley98}; [3] \citet{Koch06}; [4] \citet{Shetrone03}; [5] \citet{Koch08Car}; [6] \citet{Hernandez00}; [7] \citet{Dolphin02}; [8] \citet{Saha86}; 
[9] \citet{Bellazzini04}; [10] \citet{Gallart99}; [11] \citet{Held01}; [12] \citet{Bosler07, Koch07leo1, Mateo08}; 
[13] \citet{Bellazzini05}; [14] \citet{Koch07leo2}; [15] \citet{Shetrone09}; [16] \citet{Gullieuszik08}
[17] \citet{Nemec88}; [18] \citet{Wilkinson04, Munoz05}; [19] \citet{Shetrone01, Sadakane04}; [20] \citet{Carrera02}; [21] \citet{Ikuta02};
[22] \citet{Bonanos04}; [23] \citet{Aparicio01}; [24] \citet{Smith06, Fulbright04}; 
[25] \citet{Pietrzynski08}; [26] \citet{Tolstoy04, Coleman05, Westfall06, Battaglia08mass}; [27] Hill et al. 2009, in prep.; 
[28] \citet{Rizzi07}; [29] \citet{Coleman08}; [30] \citet{Battaglia06, Walker06fnx}; [31] Letarte et al. 2009, in prep.; [32] \citet{Gallart05conf}
[33] \citet{Whiting99}; [34] Monelli et al. (2009) in prep. (LCID); [35] \citet{Lewis07};
[36] \citet{Castellani96, Saviane96}; [37] Gallart et al. (2009) in prep. (LCID); [38] \citet{Fraternali09}; 
[39] \citet{McConnachie05}; [40] \citet{Martinez99}; [41] \citet{Saha90};
[42] \citet{Saha92}; [43] \citet{Geha06};
[44] \citet{Siegel06}; [45] \citet{Martin08}; [46] \citet{deJong08sdss}; [47] \citet{Munoz06Boo};
[48] \citet{Simon07}; [49] \citet{Ibata06}; [50] \citet{Norris08}; [51] \citet{Frebel09}
\end{landscape}

\noindent{\bf \large Acknowledgements}

The authors are grateful to M. Cignoni, A. Frebel, C. Gallart,
S. Hildago, M. Mateo \& M. Monelli for providing us with data and
analyses ahead of publication or as private communication.

We thank M. Cignoni, D. Romano, A. Cole, G. Battaglia, V. Belokurov 
\& A. Helmi for useful comments and help in the preparation of figures
 and tables.  We
thank F. Matteucci, M. Irwin, R. Sancisi, F. Fraternali 
and P. Jablonka for useful
conversations, and E. Skillman S. Salvadori, M. Breddels,
E. Starkenburg, T. de Boer, P. van der Kruit \& G. Clementini 
for careful comments on the text.
Very detailed and useful comments from the Editor,
J. Kormendy were also highly appreciated.

ET thanks Bologna Observatory for hospitality 
\& Paris Observatory for hospitality and
financial support. ET grateful acknowledges support from 
an NWO-VICI grant.  VH acknowledges the
financial support of Programme National Galaxies (PNG) of CNRS/INSU,
France.

This research has made use of the NASA/IPAC Extragalactic Database
(NED) which is operated by the Jet Propulsion Laboratory, California
Institute of Technology, under contract with the National Aeronautics
and Space Administration.

\bibliographystyle{Astronomy} 
\bibliography{etvhmt} 

\end{document}

%% file: tolstoyhilltosi_resub.bbl
\begin{thebibliography}{}
\expandafter\ifx\csname natexlab\endcsname\relax\def\natexlab#1{#1}\fi

\bibitem[{{Aaronson}(1983)}]{Aaronson83}
{Aaronson} M. 1983.
\newblock \textit{\apjl} 266:L11--L15

\bibitem[{{Adelman-McCarthy} et~al.(2007){Adelman-McCarthy}, {Ag{\"u}eros},
  {Allam}, {Anderson}, {Anderson} et~al.}]{Adelman07}
{Adelman-McCarthy} JK, {Ag{\"u}eros} MA, {Allam} SS, {Anderson} KSJ, {Anderson}
  SF, et~al. 2007.
\newblock \textit{\apjs} 172:634--644

\bibitem[{{Aloisi} et~al.(2007){Aloisi}, {Clementini}, {Tosi}, {Annibali},
  {Contreras} et~al.}]{Aloisi07}
{Aloisi} A, {Clementini} G, {Tosi} M, {Annibali} F, {Contreras} R, et~al. 2007.
\newblock \textit{\apjl} 667:L151--L154

\bibitem[{{Aloisi} et~al.(2003){Aloisi}, {Savaglio}, {Heckman}, {Hoopes},
  {Leitherer} \& {Sembach}}]{Aloisi03}
{Aloisi} A, {Savaglio} S, {Heckman} TM, {Hoopes} CG, {Leitherer} C, {Sembach}
  KR. 2003.
\newblock \textit{\apj} 595:760--778

\bibitem[{{Aloisi}, {Tosi} \& {Greggio}(1999)}]{Aloisi99}
{Aloisi} A, {Tosi} M, {Greggio} L. 1999.
\newblock \textit{\aj} 118:302--322

\bibitem[{{Aloisi} et~al.(2005){Aloisi}, {van der Marel}, {Mack}, {Leitherer},
  {Sirianni} \& {Tosi}}]{Aloisi05}
{Aloisi} A, {van der Marel} RP, {Mack} J, {Leitherer} C, {Sirianni} M, {Tosi}
  M. 2005.
\newblock \textit{\apjl} 631:L45--L48

\bibitem[{{Angeretti} et~al.(2005){Angeretti}, {Tosi}, {Greggio}, {Sabbi},
  {Aloisi} \& {Leitherer}}]{Angeretti05}
{Angeretti} L, {Tosi} M, {Greggio} L, {Sabbi} E, {Aloisi} A, {Leitherer} C.
  2005.
\newblock \textit{\aj} 129:2203--2216

\bibitem[{{Annibali} et~al.(2003){Annibali}, {Greggio}, {Tosi}, {Aloisi} \&
  {Leitherer}}]{Annibali03}
{Annibali} F, {Greggio} L, {Tosi} M, {Aloisi} A, {Leitherer} C. 2003.
\newblock \textit{\aj} 126:2752--2773

\bibitem[{{Aoki} et~al.(2009){Aoki}, {Arimoto}, {Sadakane}, {Tolstoy},
  {Battaglia} et~al.}]{Aoki09}
{Aoki} W, {Arimoto} N, {Sadakane} K, {Tolstoy} E, {Battaglia} G, et~al. 2009.
\newblock \textit{A\&A, in press, arXiv:0904.4307}

\bibitem[{{Aparicio}, {Carrera} \&
  {Mart{\'{\i}}nez-Delgado}(2001)}]{Aparicio01}
{Aparicio} A, {Carrera} R, {Mart{\'{\i}}nez-Delgado} D. 2001.
\newblock \textit{\aj} 122:2524--2537

\bibitem[{{Aparicio} \& {Gallart}(2004)}]{Aparicio04}
{Aparicio} A, {Gallart} C. 2004.
\newblock \textit{\aj} 128:1465--1477

\bibitem[{{Aparicio} et~al.(1996){Aparicio}, {Gallart}, {Chiosi} \&
  {Bertelli}}]{Aparicio96}
{Aparicio} A, {Gallart} C, {Chiosi} C, {Bertelli} G. 1996.
\newblock \textit{\apjl} 469:L97+

\bibitem[{{Audouze} \& {Tinsley}(1976)}]{Audouze76}
{Audouze} J, {Tinsley} BM. 1976.
\newblock \textit{\araa} 14:43--79

\bibitem[{{Baade}(1944{\natexlab{a}})}]{Baade44a}
{Baade} W. 1944{\natexlab{a}}.
\newblock \textit{\apj} 100:147--+

\bibitem[{{Baade}(1944{\natexlab{b}})}]{Baade44b}
{Baade} W. 1944{\natexlab{b}}.
\newblock \textit{\apj} 100:137--+

\bibitem[{{Barklem} et~al.(2005){Barklem}, {Christlieb}, {Beers}, {Hill},
  {Bessell} et~al.}]{Barklem05}
{Barklem} PS, {Christlieb} N, {Beers} TC, {Hill} V, {Bessell} MS, et~al. 2005.
\newblock \textit{\aap} 439:129--151

\bibitem[{Battaglia(2007)}]{Battaglia07phd}
Battaglia G. 2007.
\newblock \textit{Chemistry and Kinematics of stars in Local Group galaxies}.
\newblock Ph.D. thesis, University of Groningen

\bibitem[{{Battaglia} et~al.(2008{\natexlab{a}}){Battaglia}, {Helmi},
  {Tolstoy}, {Irwin}, {Hill} \& {Jablonka}}]{Battaglia08mass}
{Battaglia} G, {Helmi} A, {Tolstoy} E, {Irwin} M, {Hill} V, {Jablonka} P.
  2008{\natexlab{a}}.
\newblock \textit{\apjl} 681:L13--L16

\bibitem[{{Battaglia} et~al.(2008{\natexlab{b}}){Battaglia}, {Irwin},
  {Tolstoy}, {Hill}, {Helmi} et~al.}]{Battaglia08cat}
{Battaglia} G, {Irwin} M, {Tolstoy} E, {Hill} V, {Helmi} A, et~al.
  2008{\natexlab{b}}.
\newblock \textit{\mnras} 383:183--199

\bibitem[{{Battaglia} et~al.(2006){Battaglia}, {Tolstoy}, {Helmi}, {Irwin},
  {Letarte} et~al.}]{Battaglia06}
{Battaglia} G, {Tolstoy} E, {Helmi} A, {Irwin} MJ, {Letarte} B, et~al. 2006.
\newblock \textit{\aap} 459:423--440

\bibitem[{{Beers} \& {Christlieb}(2005)}]{Beers05}
{Beers} TC, {Christlieb} N. 2005.
\newblock \textit{\araa} 43:531--580

\bibitem[{{Bekki} \& {Freeman}(2003)}]{Bekki03}
{Bekki} K, {Freeman} KC. 2003.
\newblock \textit{\mnras} 346:L11--L15

\bibitem[{{Bellazzini}, {Gennari} \& {Ferraro}(2005)}]{Bellazzini05}
{Bellazzini} M, {Gennari} N, {Ferraro} FR. 2005.
\newblock \textit{\mnras} 360:185--193

\bibitem[{{Bellazzini} et~al.(2004){Bellazzini}, {Gennari}, {Ferraro} \&
  {Sollima}}]{Bellazzini04}
{Bellazzini} M, {Gennari} N, {Ferraro} FR, {Sollima} A. 2004.
\newblock \textit{\mnras} 354:708--712

\bibitem[{{Belokurov} et~al.(2007{\natexlab{a}}){Belokurov}, {Evans}, {Irwin},
  {Lynden-Bell}, {Yanny} et~al.}]{Belokurov07orph}
{Belokurov} V, {Evans} NW, {Irwin} MJ, {Lynden-Bell} D, {Yanny} B, et~al.
  2007{\natexlab{a}}.
\newblock \textit{\apj} 658:337--344

\bibitem[{{Belokurov} et~al.(2008){Belokurov}, {Walker}, {Evans}, {Faria},
  {Gilmore} et~al.}]{Belokurov08}
{Belokurov} V, {Walker} MG, {Evans} NW, {Faria} DC, {Gilmore} G, et~al. 2008.
\newblock \textit{\apjl} 686:L83--L86

\bibitem[{{Belokurov} et~al.(2009){Belokurov}, {Walker}, {Evans}, {Gilmore},
  {Irwin} et~al.}]{Belokurov09}
{Belokurov} V, {Walker} MG, {Evans} NW, {Gilmore} G, {Irwin} M, et~al. 2009.
\newblock \textit{submitted to MNRAS, arXiv:0903.0818}

\bibitem[{{Belokurov} et~al.(2007{\natexlab{b}}){Belokurov}, {Zucker}, {Evans},
  {Kleyna}, {Koposov} et~al.}]{Belokurov07}
{Belokurov} V, {Zucker} DB, {Evans} NW, {Kleyna} JT, {Koposov} S, et~al.
  2007{\natexlab{b}}.
\newblock \textit{\apj} 654:897--906

\bibitem[{{Binggeli}(1994)}]{Binggeli94}
{Binggeli} B. 1994.
\newblock In \textit{Dwarf Galaxies}, eds. G~{Meylan}, P~{Prugniel}

\bibitem[{{Binggeli}, {Tarenghi} \& {Sandage}(1990)}]{Binggeli90}
{Binggeli} B, {Tarenghi} M, {Sandage} A. 1990.
\newblock \textit{\aap} 228:42--60

\bibitem[{{Bonanos} et~al.(2004){Bonanos}, {Stanek}, {Szentgyorgyi}, {Sasselov}
  \& {Bakos}}]{Bonanos04}
{Bonanos} AZ, {Stanek} KZ, {Szentgyorgyi} AH, {Sasselov} DD, {Bakos} G{\'A}.
  2004.
\newblock \textit{\aj} 127:861--867

\bibitem[{{Bonifacio} et~al.(2000){Bonifacio}, {Hill}, {Molaro}, {Pasquini},
  {Di Marcantonio} \& {Santin}}]{Bonifacio00}
{Bonifacio} P, {Hill} V, {Molaro} P, {Pasquini} L, {Di Marcantonio} P, {Santin}
  P. 2000.
\newblock \textit{\aap} 359:663--668

\bibitem[{{Bonifacio} et~al.(2004){Bonifacio}, {Sbordone}, {Marconi},
  {Pasquini} \& {Hill}}]{Bonifacio04}
{Bonifacio} P, {Sbordone} L, {Marconi} G, {Pasquini} L, {Hill} V. 2004.
\newblock \textit{\aap} 414:503--514

\bibitem[{{Bosler}, {Smecker-Hane} \& {Stetson}(2007)}]{Bosler07}
{Bosler} TL, {Smecker-Hane} TA, {Stetson} PB. 2007.
\newblock \textit{\mnras} 378:318--338

\bibitem[{{Bovill} \& {Ricotti}(2009)}]{Bovill08}
{Bovill} MS, {Ricotti} M. 2009.
\newblock \textit{\apj} 693:1859--1870

\bibitem[{{Bresolin} et~al.(2007){Bresolin}, {Urbaneja}, {Gieren},
  {Pietrzy{\'n}ski} \& {Kudritzki}}]{Bresolin07}
{Bresolin} F, {Urbaneja} MA, {Gieren} W, {Pietrzy{\'n}ski} G, {Kudritzki} RP.
  2007.
\newblock \textit{\apj} 671:2028--2039

\bibitem[{{Brown} et~al.(2007){Brown}, {Geller}, {Kenyon} \& {Kurtz}}]{Brown07}
{Brown} WR, {Geller} MJ, {Kenyon} SJ, {Kurtz} MJ. 2007.
\newblock \textit{\apj} 666:231--235

\bibitem[{{Carigi}, {Col{\'{\i}}n} \& {Peimbert}(2006)}]{Carigi06}
{Carigi} L, {Col{\'{\i}}n} P, {Peimbert} M. 2006.
\newblock \textit{\apj} 644:924--939

\bibitem[{{Carigi} et~al.(1995){Carigi}, {Colin}, {Peimbert} \&
  {Sarmiento}}]{Carigi95}
{Carigi} L, {Colin} P, {Peimbert} M, {Sarmiento} A. 1995.
\newblock \textit{\apj} 445:98--107

\bibitem[{{Carigi} \& {Hernandez}(2008)}]{Carigi08}
{Carigi} L, {Hernandez} X. 2008.
\newblock \textit{\mnras} 390:582--594

\bibitem[{{Carigi}, {Hernandez} \& {Gilmore}(2002)}]{Carigi02}
{Carigi} L, {Hernandez} X, {Gilmore} G. 2002.
\newblock \textit{\mnras} 334:117--128

\bibitem[{{Carrera} et~al.(2002){Carrera}, {Aparicio},
  {Mart{\'{\i}}nez-Delgado} \& {Alonso-Garc{\'{\i}}a}}]{Carrera02}
{Carrera} R, {Aparicio} A, {Mart{\'{\i}}nez-Delgado} D, {Alonso-Garc{\'{\i}}a}
  J. 2002.
\newblock \textit{\aj} 123:3199--3209

\bibitem[{{Carrera} et~al.(2008){Carrera}, {Gallart}, {Aparicio}, {Costa},
  {M{\'e}ndez} \& {No{\"e}l}}]{Carrera08}
{Carrera} R, {Gallart} C, {Aparicio} A, {Costa} E, {M{\'e}ndez} RA, {No{\"e}l}
  NED. 2008.
\newblock \textit{\aj} 136:1039--1048

\bibitem[{{Castellani}, {Marconi} \& {Buonanno}(1996)}]{Castellani96}
{Castellani} M, {Marconi} G, {Buonanno} R. 1996.
\newblock \textit{\aap} 310:715--721

\bibitem[{{Cayrel} et~al.(2004){Cayrel}, {Depagne}, {Spite}, {Hill}, {Spite}
  et~al.}]{Cayrel04}
{Cayrel} R, {Depagne} E, {Spite} M, {Hill} V, {Spite} F, et~al. 2004.
\newblock \textit{\aap} 416:1117--1138

\bibitem[{{Chiosi} et~al.(2006){Chiosi}, {Vallenari}, {Held}, {Rizzi} \&
  {Moretti}}]{Chiosi06}
{Chiosi} E, {Vallenari} A, {Held} EV, {Rizzi} L, {Moretti} A. 2006.
\newblock \textit{\aap} 452:179--193

\bibitem[{{Choudhury}, {Ferrara} \& {Gallerani}(2008)}]{Choud08}
{Choudhury} TR, {Ferrara} A, {Gallerani} S. 2008.
\newblock \textit{\mnras} 385:L58--L62

\bibitem[{{Cignoni} et~al.(2009){Cignoni}, {Sabbi}, {Nota}, {Tosi},
  {Degl'Innocenti} et~al.}]{Cignoni09}
{Cignoni} M, {Sabbi} E, {Nota} A, {Tosi} M, {Degl'Innocenti} S, et~al. 2009.
\newblock \textit{\aj} 137:3668--3684

\bibitem[{{Cioni} et~al.(2008){Cioni}, {Bekki}, {Clementini}, {de Blok},
  {Emerson} et~al.}]{Cioni08}
{Cioni} MRL, {Bekki} K, {Clementini} G, {de Blok} WJG, {Emerson} JP, et~al.
  2008.
\newblock \textit{Publications of the Astronomical Society of Australia}
  25:121--128

\bibitem[{{Clementini} et~al.(2003){Clementini}, {Held}, {Baldacci} \&
  {Rizzi}}]{Clementini03}
{Clementini} G, {Held} EV, {Baldacci} L, {Rizzi} L. 2003.
\newblock \textit{\apjl} 588:L85--L88

\bibitem[{{Cole} et~al.(2007){Cole}, {Skillman}, {Tolstoy}, {Gallagher},
  {Aparicio} et~al.}]{Cole07}
{Cole} AA, {Skillman} ED, {Tolstoy} E, {Gallagher} III JS, {Aparicio} A, et~al.
  2007.
\newblock \textit{\apjl} 659:L17--L20

\bibitem[{{Coleman}, {Da Costa} \& {Bland-Hawthorn}(2005)}]{Coleman05}
{Coleman} MG, {Da Costa} GS, {Bland-Hawthorn} J. 2005.
\newblock \textit{\aj} 130:1065--1082

\bibitem[{{Coleman} \& {de Jong}(2008)}]{Coleman08}
{Coleman} MG, {de Jong} JTA. 2008.
\newblock \textit{\apj} 685:933--946

\bibitem[{{Cook} et~al.(1999){Cook}, {Mateo}, {Olszewski}, {Vogt}, {Stubbs} \&
  {Diercks}}]{Cook99}
{Cook} KH, {Mateo} M, {Olszewski} EW, {Vogt} SS, {Stubbs} C, {Diercks} A. 1999.
\newblock \textit{\pasp} 111:306--312

\bibitem[{{Dall'Ora} et~al.(2006){Dall'Ora}, {Clementini}, {Kinemuchi},
  {Ripepi}, {Marconi} et~al.}]{DallOra06}
{Dall'Ora} M, {Clementini} G, {Kinemuchi} K, {Ripepi} V, {Marconi} M, et~al.
  2006.
\newblock \textit{\apjl} 653:L109--L112

\bibitem[{{de Jong} et~al.(2008{\natexlab{a}}){de Jong}, {Harris}, {Coleman},
  {Martin}, {Bell} et~al.}]{DeJong08}
{de Jong} JTA, {Harris} J, {Coleman} MG, {Martin} NF, {Bell} EF, et~al.
  2008{\natexlab{a}}.
\newblock \textit{\apj} 680:1112--1119

\bibitem[{{de Jong} et~al.(2008{\natexlab{b}}){de Jong}, {Rix}, {Martin},
  {Zucker}, {Dolphin} et~al.}]{deJong08sdss}
{de Jong} JTA, {Rix} HW, {Martin} NF, {Zucker} DB, {Dolphin} AE, et~al.
  2008{\natexlab{b}}.
\newblock \textit{\aj} 135:1361--1383

\bibitem[{{Dekel} \& {Silk}(1986)}]{Dekel86}
{Dekel} A, {Silk} J. 1986.
\newblock \textit{\apj} 303:39--55

\bibitem[{{D'Ercole} \& {Brighenti}(1999)}]{Dercole99}
{D'Ercole} A, {Brighenti} F. 1999.
\newblock \textit{\mnras} 309:941--954

\bibitem[{{Dohm-Palmer} et~al.(1998){Dohm-Palmer}, {Skillman}, {Gallagher},
  {Tolstoy}, {Mateo} et~al.}]{Dohm98}
{Dohm-Palmer} RC, {Skillman} ED, {Gallagher} J, {Tolstoy} E, {Mateo} M, et~al.
  1998.
\newblock \textit{\aj} 116:1227--1243

\bibitem[{{Dohm-Palmer} et~al.(2002){Dohm-Palmer}, {Skillman}, {Mateo}, {Saha},
  {Dolphin} et~al.}]{Dohm02}
{Dohm-Palmer} RC, {Skillman} ED, {Mateo} M, {Saha} A, {Dolphin} A, et~al. 2002.
\newblock \textit{\aj} 123:813--831

\bibitem[{{Dohm-Palmer} et~al.(1997){Dohm-Palmer}, {Skillman}, {Saha},
  {Tolstoy}, {Mateo} et~al.}]{Dohm97}
{Dohm-Palmer} RC, {Skillman} ED, {Saha} A, {Tolstoy} E, {Mateo} M, et~al. 1997.
\newblock \textit{\aj} 114:2527--+

\bibitem[{{Dolphin}(1997)}]{Dolphin97}
{Dolphin} A. 1997.
\newblock \textit{New Astronomy} 2:397--409

\bibitem[{{Dolphin}(2000)}]{Dolphin00a}
{Dolphin} AE. 2000.
\newblock \textit{\apj} 531:804--812

\bibitem[{{Dolphin}(2002)}]{Dolphin02}
{Dolphin} AE. 2002.
\newblock \textit{\mnras} 332:91--108

\bibitem[{{Dolphin} et~al.(2002){Dolphin}, {Saha}, {Claver}, {Skillman}, {Cole}
  et~al.}]{Dolphin02leoa}
{Dolphin} AE, {Saha} A, {Claver} J, {Skillman} ED, {Cole} AA, et~al. 2002.
\newblock \textit{\aj} 123:3154--3198

\bibitem[{{Dolphin} et~al.(2003{\natexlab{a}}){Dolphin}, {Saha}, {Skillman},
  {Dohm-Palmer}, {Tolstoy} et~al.}]{Dolphin03dist}
{Dolphin} AE, {Saha} A, {Skillman} ED, {Dohm-Palmer} RC, {Tolstoy} E, et~al.
  2003{\natexlab{a}}.
\newblock \textit{\aj} 125:1261--1290

\bibitem[{{Dolphin} et~al.(2003{\natexlab{b}}){Dolphin}, {Saha}, {Skillman},
  {Dohm-Palmer}, {Tolstoy} et~al.}]{Dolphin03}
{Dolphin} AE, {Saha} A, {Skillman} ED, {Dohm-Palmer} RC, {Tolstoy} E, et~al.
  2003{\natexlab{b}}.
\newblock \textit{\aj} 126:187--196

\bibitem[{{Dolphin} et~al.(2001){Dolphin}, {Walker}, {Hodge}, {Mateo},
  {Olszewski} et~al.}]{Dolphin01b}
{Dolphin} AE, {Walker} AR, {Hodge} PW, {Mateo} M, {Olszewski} EW, et~al. 2001.
\newblock \textit{\apj} 562:303--313

\bibitem[{{Dolphin} et~al.(2005){Dolphin}, {Weisz}, {Skillman} \&
  {Holtzman}}]{Dolphin05}
{Dolphin} AE, {Weisz} DR, {Skillman} ED, {Holtzman} JA. 2005.
\newblock \textit{Invited review, Resolved Stellar Populations,
  arXiv:astro-ph/0506430}

\bibitem[{{Eggen}, {Lynden-Bell} \& {Sandage}(1962)}]{Eggen62}
{Eggen} OJ, {Lynden-Bell} D, {Sandage} AR. 1962.
\newblock \textit{\apj} 136:748--+

\bibitem[{{Evans} et~al.(2006){Evans}, {Bresolin}, {Urbaneja},
  {Peitrzy{\'n}ski}, {Gieren} \& {Kudritzki}}]{Evans06}
{Evans} C, {Bresolin} F, {Urbaneja} M, {Peitrzy{\'n}ski} G, {Gieren} W,
  {Kudritzki} RP. 2006.
\newblock \textit{The Messenger} 126:5--+

\bibitem[{{Evans} et~al.(2005){Evans}, {Smartt}, {Lennon}, {Dufton}, {Hunter}
  et~al.}]{Evans05}
{Evans} C, {Smartt} S, {Lennon} D, {Dufton} P, {Hunter} I, et~al. 2005.
\newblock \textit{The Messenger} 122:36--38

\bibitem[{{Evstigneeva} et~al.(2008){Evstigneeva}, {Drinkwater}, {Peng},
  {Hilker}, {DePropris} et~al.}]{Evstig08}
{Evstigneeva} EA, {Drinkwater} MJ, {Peng} CY, {Hilker} M, {DePropris} R, et~al.
  2008.
\newblock \textit{\aj} 136:461--478

\bibitem[{{Faber} \& {Lin}(1983)}]{Faber83}
{Faber} SM, {Lin} DNC. 1983.
\newblock \textit{\apjl} 266:L17--L20

\bibitem[{{Faber} et~al.(1997){Faber}, {Tremaine}, {Ajhar}, {Byun}, {Dressler}
  et~al.}]{Faber97}
{Faber} SM, {Tremaine} S, {Ajhar} EA, {Byun} YI, {Dressler} A, et~al. 1997.
\newblock \textit{\aj} 114:1771--+

\bibitem[{{Fagotto} et~al.(1994{\natexlab{a}}){Fagotto}, {Bressan}, {Bertelli}
  \& {Chiosi}}]{Fagotto94a}
{Fagotto} F, {Bressan} A, {Bertelli} G, {Chiosi} C. 1994{\natexlab{a}}.
\newblock \textit{\aaps} 104:365--376

\bibitem[{{Fagotto} et~al.(1994{\natexlab{b}}){Fagotto}, {Bressan}, {Bertelli}
  \& {Chiosi}}]{Fagotto94b}
{Fagotto} F, {Bressan} A, {Bertelli} G, {Chiosi} C. 1994{\natexlab{b}}.
\newblock \textit{\aap} 105:29--38

\bibitem[{{Fenner} et~al.(2006){Fenner}, {Gibson}, {Gallino} \&
  {Lugaro}}]{Fenner06}
{Fenner} Y, {Gibson} BK, {Gallino} R, {Lugaro} M. 2006.
\newblock \textit{\apj} 646:184--191

\bibitem[{{Ferrara} \& {Tolstoy}(2000)}]{Ferrara00}
{Ferrara} A, {Tolstoy} E. 2000.
\newblock \textit{\mnras} 313:291--309

\bibitem[{{Ferraro} et~al.(1989){Ferraro}, {Fusi Pecci}, {Tosi} \&
  {Buonanno}}]{Ferraro89}
{Ferraro} FR, {Fusi Pecci} F, {Tosi} M, {Buonanno} R. 1989.
\newblock \textit{\mnras} 241:433--452

\bibitem[{{Fran{\c c}ois} et~al.(2007){Fran{\c c}ois}, {Depagne}, {Hill},
  {Spite}, {Spite} et~al.}]{Francois07}
{Fran{\c c}ois} P, {Depagne} E, {Hill} V, {Spite} M, {Spite} F, et~al. 2007.
\newblock \textit{\aap} 476:935--950

\bibitem[{{Fraternali} et~al.(2009){Fraternali}, {Tolstoy}, {Irwin} \&
  {Cole}}]{Fraternali09}
{Fraternali} F, {Tolstoy} E, {Irwin} M, {Cole} A. 2009.
\newblock \textit{A\&A, in press arXiv:0903.4635}

\bibitem[{{Frebel} et~al.(2009){Frebel}, {Simon}, {Geha} \&
  {Willman}}]{Frebel09}
{Frebel} A, {Simon} JD, {Geha} M, {Willman} B. 2009.
\newblock \textit{submitted to ApJ arXiv:0902.2395}

\bibitem[{{Freeman} \& {Bland-Hawthorn}(2002)}]{Freeman02}
{Freeman} K, {Bland-Hawthorn} J. 2002.
\newblock \textit{\araa} 40:487--537

\bibitem[{{Freeman}(1970)}]{Freeman70}
{Freeman} KC. 1970.
\newblock \textit{\apj} 160:811--+

\bibitem[{{Fujita} et~al.(2004){Fujita}, {Mac Low}, {Ferrara} \&
  {Meiksin}}]{Fujita04}
{Fujita} A, {Mac Low} MM, {Ferrara} A, {Meiksin} A. 2004.
\newblock \textit{\apj} 613:159--179

\bibitem[{{Fulbright}, {McWilliam} \& {Rich}(2007)}]{Fulbright07}
{Fulbright} JP, {McWilliam} A, {Rich} RM. 2007.
\newblock \textit{\apj} 661:1152--1179

\bibitem[{{Fulbright}, {Rich} \& {Castro}(2004)}]{Fulbright04}
{Fulbright} JP, {Rich} RM, {Castro} S. 2004.
\newblock \textit{\apj} 612:447--453

\bibitem[{{Gallagher} et~al.(1998){Gallagher}, {Tolstoy}, {Dohm-Palmer},
  {Skillman}, {Cole} et~al.}]{Gallagher98}
{Gallagher} JS, {Tolstoy} E, {Dohm-Palmer} RC, {Skillman} ED, {Cole} AA, et~al.
  1998.
\newblock \textit{\aj} 115:1869--1887

\bibitem[{{Gallagher}, {Hunter} \& {Tutukov}(1984)}]{Gallagher84a}
{Gallagher} III JS, {Hunter} DA, {Tutukov} AV. 1984.
\newblock \textit{\apj} 284:544--556

\bibitem[{{Gallart} et~al.(1996){Gallart}, {Aparicio}, {Bertelli} \&
  {Chiosi}}]{Gallart96a}
{Gallart} C, {Aparicio} A, {Bertelli} G, {Chiosi} C. 1996.
\newblock \textit{\aj} 112:1950--+

\bibitem[{{Gallart} et~al.(2005){Gallart}, {Aparicio}, {Zinn}, {Buonanno},
  {Hardy} \& {Marconi}}]{Gallart05conf}
{Gallart} C, {Aparicio} A, {Zinn} R, {Buonanno} R, {Hardy} E, {Marconi} G.
  2005.
\newblock In \textit{IAU Colloq. 198: Near-fields cosmology with dwarf
  elliptical galaxies}, eds. H~{Jerjen}, B~{Binggeli}

\bibitem[{{Gallart} et~al.(1999){Gallart}, {Freedman}, {Aparicio}, {Bertelli}
  \& {Chiosi}}]{Gallart99}
{Gallart} C, {Freedman} WL, {Aparicio} A, {Bertelli} G, {Chiosi} C. 1999.
\newblock \textit{\aj} 118:2245--2261

\bibitem[{{Gallart} et~al.(2001){Gallart}, {Mart{\'{\i}}nez-Delgado},
  {G{\'o}mez-Flechoso} \& {Mateo}}]{Gallart01}
{Gallart} C, {Mart{\'{\i}}nez-Delgado} D, {G{\'o}mez-Flechoso} MA, {Mateo} M.
  2001.
\newblock \textit{\aj} 121:2572--2583

\bibitem[{{Gallart} \& {The Lcid Team}(2007)}]{Gallart07}
{Gallart} C, {The Lcid Team}. 2007.
\newblock In \textit{IAU Symposium}, eds. A~{Vazdekis}, RF~{Peletier}, vol. 241
  of \textit{IAU Symposium}

\bibitem[{{Gallart}, {Zoccali} \& {Aparicio}(2005)}]{Gallart05}
{Gallart} C, {Zoccali} M, {Aparicio} A. 2005.
\newblock \textit{\araa} 43:387--434

\bibitem[{{Gallazzi} et~al.(2005){Gallazzi}, {Charlot}, {Brinchmann}, {White}
  \& {Tremonti}}]{Gallazzi05}
{Gallazzi} A, {Charlot} S, {Brinchmann} J, {White} SDM, {Tremonti} CA. 2005.
\newblock \textit{\mnras} 362:41--58

\bibitem[{{Geha} et~al.(2006){Geha}, {Guhathakurta}, {Rich} \&
  {Cooper}}]{Geha06}
{Geha} M, {Guhathakurta} P, {Rich} RM, {Cooper} MC. 2006.
\newblock \textit{\aj} 131:332--342

\bibitem[{{Geha} et~al.(2009){Geha}, {Willman}, {Simon}, {Strigari}, {Kirby}
  et~al.}]{Geha08}
{Geha} M, {Willman} B, {Simon} JD, {Strigari} LE, {Kirby} EN, et~al. 2009.
\newblock \textit{\apj} 692:1464--1475

\bibitem[{{Geisler} et~al.(2005){Geisler}, {Smith}, {Wallerstein}, {Gonzalez}
  \& {Charbonnel}}]{Geisler05}
{Geisler} D, {Smith} VV, {Wallerstein} G, {Gonzalez} G, {Charbonnel} C. 2005.
\newblock \textit{\aj} 129:1428--1442

\bibitem[{{Gieren} et~al.(2006){Gieren}, {Pietrzy{\'n}ski}, {Nalewajko},
  {Soszy{\'n}ski}, {Bresolin} et~al.}]{Gieren06}
{Gieren} W, {Pietrzy{\'n}ski} G, {Nalewajko} K, {Soszy{\'n}ski} I, {Bresolin}
  F, et~al. 2006.
\newblock \textit{\apj} 647:1056--1064

\bibitem[{{Gieren} et~al.(2008){Gieren}, {Pietrzy{\'n}ski}, {Szewczyk},
  {Soszy{\'n}ski}, {Bresolin} et~al.}]{Gieren08}
{Gieren} W, {Pietrzy{\'n}ski} G, {Szewczyk} O, {Soszy{\'n}ski} I, {Bresolin} F,
  et~al. 2008.
\newblock \textit{\apj} 683:611--619

\bibitem[{{Gilmore} et~al.(2007){Gilmore}, {Wilkinson}, {Wyse}, {Kleyna},
  {Koch} et~al.}]{Gilmore07}
{Gilmore} G, {Wilkinson} MI, {Wyse} RFG, {Kleyna} JT, {Koch} A, et~al. 2007.
\newblock \textit{\apj} 663:948--959

\bibitem[{{Gilmore} \& {Wyse}(1991)}]{Gilmore91}
{Gilmore} G, {Wyse} RFG. 1991.
\newblock \textit{\apjl} 367:L55--L58

\bibitem[{{Glatt} et~al.(2008){Glatt}, {Gallagher}, {Grebel}, {Nota}, {Sabbi}
  et~al.}]{Glatt08}
{Glatt} K, {Gallagher} III JS, {Grebel} EK, {Nota} A, {Sabbi} E, et~al. 2008.
\newblock \textit{\aj} 135:1106--1116

\bibitem[{{Grebel}(1999)}]{Grebel99}
{Grebel} EK. 1999.
\newblock In \textit{The Stellar Content of Local Group Galaxies}, eds.
  P~{Whitelock}, R~{Cannon}, vol. 192 of \textit{IAU Symposium}

\bibitem[{{Grebel}, {Gallagher} \& {Harbeck}(2003)}]{Grebel03}
{Grebel} EK, {Gallagher} III JS, {Harbeck} D. 2003.
\newblock \textit{\aj} 125:1926--1939

\bibitem[{{Greggio} et~al.(1993){Greggio}, {Marconi}, {Tosi} \&
  {Focardi}}]{Greggio93}
{Greggio} L, {Marconi} G, {Tosi} M, {Focardi} P. 1993.
\newblock \textit{\aj} 105:894--932

\bibitem[{{Greggio} et~al.(1998){Greggio}, {Tosi}, {Clampin}, {de Marchi},
  {Leitherer} et~al.}]{Greggio98}
{Greggio} L, {Tosi} M, {Clampin} M, {de Marchi} G, {Leitherer} C, et~al. 1998.
\newblock \textit{\apj} 504:725--+

\bibitem[{{Grocholski} et~al.(2008){Grocholski}, {Aloisi}, {van der Marel},
  {Mack}, {Annibali} et~al.}]{Grocholski08}
{Grocholski} AJ, {Aloisi} A, {van der Marel} RP, {Mack} J, {Annibali} F, et~al.
  2008.
\newblock \textit{\apjl} 686:L79--L82

\bibitem[{{Gullieuszik} et~al.(2008){Gullieuszik}, {Held}, {Rizzi}, {Girardi},
  {Marigo} \& {Momany}}]{Gullieuszik08}
{Gullieuszik} M, {Held} EV, {Rizzi} L, {Girardi} L, {Marigo} P, {Momany} Y.
  2008.
\newblock \textit{\mnras} 388:1185--1197

\bibitem[{{Harris} \& {Zaritsky}(2004)}]{Harris04}
{Harris} J, {Zaritsky} D. 2004.
\newblock \textit{\aj} 127:1531--1544

\bibitem[{{Harris} \& {Zaritsky}(2006)}]{Harris06}
{Harris} J, {Zaritsky} D. 2006.
\newblock \textit{\aj} 131:2514--2524

\bibitem[{{Harris}(1996)}]{Harris96}
{Harris} WE. 1996.
\newblock \textit{\aj} 112:1487--+

\bibitem[{{Heckman} et~al.(2001){Heckman}, {Sembach}, {Meurer}, {Strickland},
  {Martin} et~al.}]{Heckman01}
{Heckman} TM, {Sembach} KR, {Meurer} GR, {Strickland} DK, {Martin} CL, et~al.
  2001.
\newblock \textit{\apj} 554:1021--1034

\bibitem[{{Heiles}(1990)}]{Heiles90}
{Heiles} C. 1990.
\newblock \textit{\apj} 354:483--491

\bibitem[{{Held} et~al.(2001){Held}, {Clementini}, {Rizzi}, {Momany}, {Saviane}
  \& {Di Fabrizio}}]{Held01}
{Held} EV, {Clementini} G, {Rizzi} L, {Momany} Y, {Saviane} I, {Di Fabrizio} L.
  2001.
\newblock \textit{\apjl} 562:L39--L42

\bibitem[{{Held}, {Saviane} \& {Momany}(1999)}]{Held99}
{Held} EV, {Saviane} I, {Momany} Y. 1999.
\newblock \textit{\aap} 345:747--759

\bibitem[{{Helmi} et~al.(2006){Helmi}, {Irwin}, {Tolstoy}, {Battaglia}, {Hill}
  et~al.}]{Helmi06}
{Helmi} A, {Irwin} MJ, {Tolstoy} E, {Battaglia} G, {Hill} V, et~al. 2006.
\newblock \textit{\apjl} 651:L121--L124

\bibitem[{{Hernandez}, {Gilmore} \& {Valls-Gabaud}(2000)}]{Hernandez00}
{Hernandez} X, {Gilmore} G, {Valls-Gabaud} D. 2000.
\newblock \textit{\mnras} 317:831--842

\bibitem[{{Hilditch}, {Howarth} \& {Harries}(2005)}]{Hilditch05}
{Hilditch} RW, {Howarth} ID, {Harries} TJ. 2005.
\newblock \textit{\mnras} 357:304--324

\bibitem[{{Hill}, {Andrievsky} \& {Spite}(1995)}]{Hill95}
{Hill} V, {Andrievsky} S, {Spite} M. 1995.
\newblock \textit{\aap} 293:347--359

\bibitem[{{Hill}, {Barbuy} \& {Spite}(1997)}]{Hill97}
{Hill} V, {Barbuy} B, {Spite} M. 1997.
\newblock \textit{\aap} 323:461--468

\bibitem[{{Hodge} \& {Miller}(1995)}]{Hodge95}
{Hodge} P, {Miller} BW. 1995.
\newblock \textit{\apj} 451:176--+

\bibitem[{{Hodge}(1971)}]{Hodge71}
{Hodge} PW. 1971.
\newblock \textit{\araa} 9:35--+

\bibitem[{{Holtzman}, {Afonso} \& {Dolphin}(2006)}]{Holtzman06}
{Holtzman} JA, {Afonso} C, {Dolphin} A. 2006.
\newblock \textit{\apjs} 166:534--548

\bibitem[{{Holtzman}, {Smith} \& {Grillmair}(2000)}]{Holtzman00}
{Holtzman} JA, {Smith} GH, {Grillmair} C. 2000.
\newblock \textit{\aj} 120:3060--3069

\bibitem[{{Hunter} \& {Elmegreen}(2004)}]{Hunter04}
{Hunter} DA, {Elmegreen} BG. 2004.
\newblock \textit{\aj} 128:2170--2205

\bibitem[{{Hunter} \& {Elmegreen}(2006)}]{Hunter06}
{Hunter} DA, {Elmegreen} BG. 2006.
\newblock \textit{\apjs} 162:49--79

\bibitem[{{Hunter} \& {Gallagher}(1986)}]{Hunter86}
{Hunter} DA, {Gallagher} III JS. 1986.
\newblock \textit{\pasp} 98:5--28

\bibitem[{{Hurley-Keller}, {Mateo} \& {Nemec}(1998)}]{Hurley98}
{Hurley-Keller} D, {Mateo} M, {Nemec} J. 1998.
\newblock \textit{\aj} 115:1840--1855

\bibitem[{{Ibata} et~al.(2006){Ibata}, {Chapman}, {Irwin}, {Lewis} \&
  {Martin}}]{Ibata06}
{Ibata} R, {Chapman} S, {Irwin} M, {Lewis} G, {Martin} N. 2006.
\newblock \textit{\mnras} 373:L70--L74

\bibitem[{{Ikuta} \& {Arimoto}(2002)}]{Ikuta02}
{Ikuta} C, {Arimoto} N. 2002.
\newblock \textit{\aap} 391:55--65

\bibitem[{{Irwin} \& {Hatzidimitriou}(1995)}]{Irwin95}
{Irwin} M, {Hatzidimitriou} D. 1995.
\newblock \textit{\mnras} 277:1354--1378

\bibitem[{{Irwin} \& {Tolstoy}(2002)}]{Irwin02}
{Irwin} M, {Tolstoy} E. 2002.
\newblock \textit{\mnras} 336:643--648

\bibitem[{{Irwin} et~al.(2007){Irwin}, {Belokurov}, {Evans}, {Ryan-Weber}, {de
  Jong} et~al.}]{Irwin07}
{Irwin} MJ, {Belokurov} V, {Evans} NW, {Ryan-Weber} EV, {de Jong} JTA, et~al.
  2007.
\newblock \textit{\apjl} 656:L13--L16

\bibitem[{{Izotov} \& {Thuan}(1998)}]{Izotov98}
{Izotov} YI, {Thuan} TX. 1998.
\newblock \textit{\apj} 500:188--+

\bibitem[{{Izotov} \& {Thuan}(1999)}]{Izotov99}
{Izotov} YI, {Thuan} TX. 1999.
\newblock \textit{\apj} 511:639--659

\bibitem[{{Izotov}, {Thuan} \& {Stasi{\'n}ska}(2007)}]{Izotov07}
{Izotov} YI, {Thuan} TX, {Stasi{\'n}ska} G. 2007.
\newblock \textit{\apj} 662:15--38

\bibitem[{{Johnson} \& {Bolte}(2002)}]{Johnson02}
{Johnson} JA, {Bolte} M. 2002.
\newblock \textit{\apj} 579:616--625

\bibitem[{{Kallivayalil}, {van der Marel} \& {Alcock}(2006)}]{Kallivayalil06}
{Kallivayalil} N, {van der Marel} RP, {Alcock} C. 2006.
\newblock \textit{\apj} 652:1213--1229

\bibitem[{{Kaufer} et~al.(2004){Kaufer}, {Venn}, {Tolstoy}, {Pinte} \&
  {Kudritzki}}]{Kaufer04}
{Kaufer} A, {Venn} KA, {Tolstoy} E, {Pinte} C, {Kudritzki} RP. 2004.
\newblock \textit{\aj} 127:2723--2737

\bibitem[{{Kirby}, {Guhathakurta} \& {Sneden}(2008)}]{Kirby08meth}
{Kirby} EN, {Guhathakurta} P, {Sneden} C. 2008.
\newblock \textit{\apj} 682:1217--1233

\bibitem[{{Kirby} et~al.(2008){Kirby}, {Simon}, {Geha}, {Guhathakurta} \&
  {Frebel}}]{Kirby08}
{Kirby} EN, {Simon} JD, {Geha} M, {Guhathakurta} P, {Frebel} A. 2008.
\newblock \textit{\apjl} 685:L43--L46

\bibitem[{{Kleyna} et~al.(2005){Kleyna}, {Wilkinson}, {Evans} \&
  {Gilmore}}]{Kleyna05}
{Kleyna} JT, {Wilkinson} MI, {Evans} NW, {Gilmore} G. 2005.
\newblock \textit{\apjl} 630:L141--L144

\bibitem[{{Kniazev} et~al.(2005){Kniazev}, {Grebel}, {Pustilnik}, {Pramskij} \&
  {Zucker}}]{Kniazev05}
{Kniazev} AY, {Grebel} EK, {Pustilnik} SA, {Pramskij} AG, {Zucker} DB. 2005.
\newblock \textit{\aj} 130:1558--1573

\bibitem[{{Kobulnicky} \& {Skillman}(1997)}]{Kobulnicky97}
{Kobulnicky} HA, {Skillman} ED. 1997.
\newblock \textit{\apj} 489:636--+

\bibitem[{{Koch} et~al.(2008{\natexlab{a}}){Koch}, {Grebel}, {Gilmore}, {Wyse},
  {Kleyna} et~al.}]{Koch08Car}
{Koch} A, {Grebel} EK, {Gilmore} GF, {Wyse} RFG, {Kleyna} JT, et~al.
  2008{\natexlab{a}}.
\newblock \textit{\aj} 135:1580--1597

\bibitem[{{Koch} et~al.(2007{\natexlab{a}}){Koch}, {Grebel}, {Kleyna},
  {Wilkinson}, {Harbeck} et~al.}]{Koch07leo2}
{Koch} A, {Grebel} EK, {Kleyna} JT, {Wilkinson} MI, {Harbeck} DR, et~al.
  2007{\natexlab{a}}.
\newblock \textit{\aj} 133:270--283

\bibitem[{{Koch} et~al.(2006){Koch}, {Grebel}, {Wyse}, {Kleyna}, {Wilkinson}
  et~al.}]{Koch06}
{Koch} A, {Grebel} EK, {Wyse} RFG, {Kleyna} JT, {Wilkinson} MI, et~al. 2006.
\newblock \textit{\aj} 131:895--911

\bibitem[{{Koch} et~al.(2008{\natexlab{b}}){Koch}, {McWilliam}, {Grebel},
  {Zucker} \& {Belokurov}}]{Koch08herc}
{Koch} A, {McWilliam} A, {Grebel} EK, {Zucker} DB, {Belokurov} V.
  2008{\natexlab{b}}.
\newblock \textit{\apjl} 688:L13--L16

\bibitem[{{Koch} et~al.(2007{\natexlab{b}}){Koch}, {Wilkinson}, {Kleyna},
  {Gilmore}, {Grebel} et~al.}]{Koch07leo1}
{Koch} A, {Wilkinson} MI, {Kleyna} JT, {Gilmore} GF, {Grebel} EK, et~al.
  2007{\natexlab{b}}.
\newblock \textit{\apj} 657:241--261

\bibitem[{{Koposov} et~al.(2008){Koposov}, {Belokurov}, {Evans}, {Hewett},
  {Irwin} et~al.}]{Koposov08}
{Koposov} S, {Belokurov} V, {Evans} NW, {Hewett} PC, {Irwin} MJ, et~al. 2008.
\newblock \textit{\apj} 686:279--291

\bibitem[{{Kormendy}(1985)}]{Kormendy85b}
{Kormendy} J. 1985.
\newblock \textit{\apj} 295:73--79

\bibitem[{{Kormendy} et~al.(2008){Kormendy}, {Fisher}, {Cornell} \&
  {Bender}}]{Kormendy08}
{Kormendy} J, {Fisher} DB, {Cornell} ME, {Bender} R. 2008.
\newblock \textit{ApJS, in press, arXiv:0810.1681}

\bibitem[{{Kuehn} et~al.(2008){Kuehn}, {Kinemuchi}, {Ripepi}, {Clementini},
  {Dall'Ora} et~al.}]{Kuehn08}
{Kuehn} C, {Kinemuchi} K, {Ripepi} V, {Clementini} G, {Dall'Ora} M, et~al.
  2008.
\newblock \textit{\apjl} 674:L81--L84

\bibitem[{{Kunth} \& {{\"O}stlin}(2000)}]{Kunth00}
{Kunth} D, {{\"O}stlin} G. 2000.
\newblock \textit{\aapr} 10:1--79

\bibitem[{{Lanfranchi}, {Matteucci} \& {Cescutti}(2008)}]{Lanfranchi08}
{Lanfranchi} GA, {Matteucci} F, {Cescutti} G. 2008.
\newblock \textit{\aap} 481:635--644

\bibitem[{{Larson}(1974)}]{Larson74}
{Larson} RB. 1974.
\newblock \textit{\mnras} 169:229--246

\bibitem[{{Leaman} et~al.(2009){Leaman}, {Cole}, {Venn}, {Tolstoy}, {Irwin}
  et~al.}]{Leaman09}
{Leaman} R, {Cole} AA, {Venn} KA, {Tolstoy} E, {Irwin} MJ, et~al. 2009.
\newblock \textit{accepted in ApJ arXiv0904.0657}

\bibitem[{{Lebouteiller} et~al.(2004){Lebouteiller}, {Kunth}, {Lequeux},
  {Lecavelier des Etangs}, {D{\'e}sert} et~al.}]{Lebouteiller04}
{Lebouteiller} V, {Kunth} D, {Lequeux} J, {Lecavelier des Etangs} A,
  {D{\'e}sert} JM, et~al. 2004.
\newblock \textit{\aap} 415:55--61

\bibitem[{{Lee}, {Skillman} \& {Venn}(2005)}]{Lee05}
{Lee} H, {Skillman} ED, {Venn} KA. 2005.
\newblock \textit{\apj} 620:223--237

\bibitem[{{Lee}, {Skillman} \& {Venn}(2006)}]{Lee06}
{Lee} H, {Skillman} ED, {Venn} KA. 2006.
\newblock \textit{\apj} 642:813--833

\bibitem[{{Lee} et~al.(1999){Lee}, {Joo}, {Sohn}, {Rey}, {Lee} \&
  {Walker}}]{Lee99}
{Lee} YW, {Joo} JM, {Sohn} YJ, {Rey} SC, {Lee} HC, {Walker} AR. 1999.
\newblock \textit{\nat} 402:55--57

\bibitem[{{Legrand} et~al.(2001){Legrand}, {Tenorio-Tagle}, {Silich}, {Kunth}
  \& {Cervi{\~n}o}}]{Legrand01}
{Legrand} F, {Tenorio-Tagle} G, {Silich} S, {Kunth} D, {Cervi{\~n}o} M. 2001.
\newblock \textit{\apj} 560:630--635

\bibitem[{{Lequeux} et~al.(1979){Lequeux}, {Peimbert}, {Rayo}, {Serrano} \&
  {Torres-Peimbert}}]{Lequeux79}
{Lequeux} J, {Peimbert} M, {Rayo} JF, {Serrano} A, {Torres-Peimbert} S. 1979.
\newblock \textit{\aap} 80:155--166

\bibitem[{Letarte(2007)}]{Letarte07phd}
Letarte B. 2007.
\newblock \textit{Chemical analysis of the Fornax dwarf galaxy}.
\newblock Ph.D. thesis, University of Groningen

\bibitem[{{Lewis} et~al.(2007){Lewis}, {Ibata}, {Chapman}, {McConnachie},
  {Irwin} et~al.}]{Lewis07}
{Lewis} GF, {Ibata} RA, {Chapman} SC, {McConnachie} A, {Irwin} MJ, et~al. 2007.
\newblock \textit{\mnras} 375:1364--1370

\bibitem[{{Lin} \& {Faber}(1983)}]{Lin83}
{Lin} DNC, {Faber} SM. 1983.
\newblock \textit{\apjl} 266:L21--L25

\bibitem[{{Lo}, {Sargent} \& {Young}(1993)}]{Lo93}
{Lo} KY, {Sargent} WLW, {Young} K. 1993.
\newblock \textit{\aj} 106:507--529

\bibitem[{{Luck} et~al.(1998){Luck}, {Moffett}, {Barnes} \& {Gieren}}]{Luck98}
{Luck} RE, {Moffett} TJ, {Barnes} III TG, {Gieren} WP. 1998.
\newblock \textit{\aj} 115:605--+

\bibitem[{{Mac Low} \& {Ferrara}(1999)}]{MacLow99}
{Mac Low} MM, {Ferrara} A. 1999.
\newblock \textit{\apj} 513:142--155

\bibitem[{{Mackey} et~al.(2006){Mackey}, {Huxor}, {Ferguson}, {Tanvir}, {Irwin}
  et~al.}]{Mackey06}
{Mackey} AD, {Huxor} A, {Ferguson} AMN, {Tanvir} NR, {Irwin} M, et~al. 2006.
\newblock \textit{\apjl} 653:L105--L108

\bibitem[{{Magrini} et~al.(2005){Magrini}, {Leisy}, {Corradi}, {Perinotto},
  {Mampaso} \& {V{\'{\i}}lchez}}]{Magrini05}
{Magrini} L, {Leisy} P, {Corradi} RLM, {Perinotto} M, {Mampaso} A,
  {V{\'{\i}}lchez} JM. 2005.
\newblock \textit{\aap} 443:115--132

\bibitem[{{Mapelli} et~al.(2007){Mapelli}, {Ripamonti}, {Tolstoy},
  {Sigurdsson}, {Irwin} \& {Battaglia}}]{Mapelli07}
{Mapelli} M, {Ripamonti} E, {Tolstoy} E, {Sigurdsson} S, {Irwin} MJ,
  {Battaglia} G. 2007.
\newblock \textit{\mnras} 380:1127--1140

\bibitem[{{Marcolini} et~al.(2008){Marcolini}, {D'Ercole}, {Battaglia} \&
  {Gibson}}]{Marcolini08}
{Marcolini} A, {D'Ercole} A, {Battaglia} G, {Gibson} BK. 2008.
\newblock \textit{\mnras} 386:2173--2180

\bibitem[{{Marcolini} et~al.(2006){Marcolini}, {D'Ercole}, {Brighenti} \&
  {Recchi}}]{Marcolini06}
{Marcolini} A, {D'Ercole} A, {Brighenti} F, {Recchi} S. 2006.
\newblock \textit{\mnras} 371:643--658

\bibitem[{{Marconi}, {Matteucci} \& {Tosi}(1994)}]{Marconi94}
{Marconi} G, {Matteucci} F, {Tosi} M. 1994.
\newblock \textit{\mnras} 270:35--+

\bibitem[{{Marconi} et~al.(1995){Marconi}, {Tosi}, {Greggio} \&
  {Focardi}}]{Marconi95}
{Marconi} G, {Tosi} M, {Greggio} L, {Focardi} P. 1995.
\newblock \textit{\aj} 109:173--199

\bibitem[{{Martin}, {Kobulnicky} \& {Heckman}(2002)}]{Martin02}
{Martin} CL, {Kobulnicky} HA, {Heckman} TM. 2002.
\newblock \textit{\apj} 574:663--692

\bibitem[{{Martin} et~al.(2008{\natexlab{a}}){Martin}, {Coleman}, {De Jong},
  {Rix}, {Bell} et~al.}]{Martin08CanVen}
{Martin} NF, {Coleman} MG, {De Jong} JTA, {Rix} HW, {Bell} EF, et~al.
  2008{\natexlab{a}}.
\newblock \textit{\apjl} 672:L13--L16

\bibitem[{{Martin} et~al.(2008{\natexlab{b}}){Martin}, {Coleman}, {De Jong},
  {Rix}, {Bell} et~al.}]{Martin08}
{Martin} NF, {Coleman} MG, {De Jong} JTA, {Rix} HW, {Bell} EF, et~al.
  2008{\natexlab{b}}.
\newblock \textit{\apjl} 672:L13--L16

\bibitem[{{Martin}, {de Jong} \& {Rix}(2008)}]{Martin08struct}
{Martin} NF, {de Jong} JTA, {Rix} HW. 2008.
\newblock \textit{\apj} 684:1075--1092

\bibitem[{{Mart{\'{\i}}nez-Delgado}, {Aparicio} \&
  {Gallart}(1999)}]{Martinez99}
{Mart{\'{\i}}nez-Delgado} D, {Aparicio} A, {Gallart} C. 1999.
\newblock \textit{\aj} 118:2229--2244

\bibitem[{{Mateo}(1994)}]{Mateo94}
{Mateo} M. 1994.
\newblock In \textit{Dwarf Galaxies}, eds. G~{Meylan}, P~{Prugniel}

\bibitem[{{Mateo}(2008)}]{Mateo08rev}
{Mateo} M. 2008.
\newblock \textit{The Messenger} 134:3--8

\bibitem[{{Mateo}, {Hurley-Keller} \& {Nemec}(1998)}]{Mateo98a}
{Mateo} M, {Hurley-Keller} D, {Nemec} J. 1998.
\newblock \textit{\aj} 115:1856--1868

\bibitem[{{Mateo}, {Olszewski} \& {Walker}(2008)}]{Mateo08}
{Mateo} M, {Olszewski} EW, {Walker} MG. 2008.
\newblock \textit{\apj} 675:201--233

\bibitem[{{Mateo}(1998)}]{Mateo98}
{Mateo} ML. 1998.
\newblock \textit{\araa} 36:435--506

\bibitem[{{Matteucci}(2003)}]{Matteucci03}
{Matteucci} F. 2003.
\newblock \textit{\apss} 284:539--548

\bibitem[{{Matteucci} \& {Brocato}(1990)}]{Matteucci90}
{Matteucci} F, {Brocato} E. 1990.
\newblock \textit{\apj} 365:539--543

\bibitem[{{Matteucci} \& {Chiosi}(1983)}]{Matteucci83}
{Matteucci} F, {Chiosi} C. 1983.
\newblock \textit{\aap} 123:121--134

\bibitem[{{Matteucci} \& {Tosi}(1985)}]{Matteucci85}
{Matteucci} F, {Tosi} M. 1985.
\newblock \textit{\mnras} 217:391--405

\bibitem[{{McConnachie} et~al.(2006){McConnachie}, {Arimoto}, {Irwin} \&
  {Tolstoy}}]{McConnachie06}
{McConnachie} AW, {Arimoto} N, {Irwin} M, {Tolstoy} E. 2006.
\newblock \textit{\mnras} 373:715--728

\bibitem[{{McConnachie} \& {Irwin}(2006)}]{McC06}
{McConnachie} AW, {Irwin} MJ. 2006.
\newblock \textit{\mnras} 365:1263--1276

\bibitem[{{McConnachie} et~al.(2005){McConnachie}, {Irwin}, {Ferguson},
  {Ibata}, {Lewis} \& {Tanvir}}]{McConnachie05}
{McConnachie} AW, {Irwin} MJ, {Ferguson} AMN, {Ibata} RA, {Lewis} GF, {Tanvir}
  N. 2005.
\newblock \textit{\mnras} 356:979--997

\bibitem[{{McCumber}, {Garnett} \& {Dufour}(2005)}]{Mccumber05}
{McCumber} MP, {Garnett} DR, {Dufour} RJ. 2005.
\newblock \textit{\aj} 130:1083--1096

\bibitem[{{McWilliam} \& {Smecker-Hane}(2005)}]{McWilliam05conf}
{McWilliam} A, {Smecker-Hane} TA. 2005.
\newblock In \textit{Cosmic Abundances as Records of Stellar Evolution and
  Nucleosynthesis}, eds. TG~{Barnes} III, FN~{Bash}, vol. 336 of
  \textit{Astronomical Society of the Pacific Conference Series}

\bibitem[{{Meurer} et~al.(1992){Meurer}, {Freeman}, {Dopita} \&
  {Cacciari}}]{Meurer92}
{Meurer} GR, {Freeman} KC, {Dopita} MA, {Cacciari} C. 1992.
\newblock \textit{\aj} 103:60--80

\bibitem[{{Miller} et~al.(2001){Miller}, {Dolphin}, {Lee}, {Kim} \&
  {Hodge}}]{Miller01}
{Miller} BW, {Dolphin} AE, {Lee} MG, {Kim} SC, {Hodge} P. 2001.
\newblock \textit{\apj} 562:713--726

\bibitem[{{Momany} et~al.(2005){Momany}, {Held}, {Saviane}, {Bedin},
  {Gullieuszik} et~al.}]{Momany05}
{Momany} Y, {Held} EV, {Saviane} I, {Bedin} LR, {Gullieuszik} M, et~al. 2005.
\newblock \textit{\aap} 439:111--127

\bibitem[{{Momany} et~al.(2007){Momany}, {Held}, {Saviane}, {Zaggia}, {Rizzi}
  \& {Gullieuszik}}]{Momany07}
{Momany} Y, {Held} EV, {Saviane} I, {Zaggia} S, {Rizzi} L, {Gullieuszik} M.
  2007.
\newblock \textit{\aap} 468:973--978

\bibitem[{{Monaco} et~al.(2007){Monaco}, {Bellazzini}, {Bonifacio}, {Buzzoni},
  {Ferraro} et~al.}]{Monaco07}
{Monaco} L, {Bellazzini} M, {Bonifacio} P, {Buzzoni} A, {Ferraro} FR, et~al.
  2007.
\newblock \textit{\aap} 464:201--209

\bibitem[{{Monaco} et~al.(2005){Monaco}, {Bellazzini}, {Bonifacio}, {Ferraro},
  {Marconi} et~al.}]{Monaco05}
{Monaco} L, {Bellazzini} M, {Bonifacio} P, {Ferraro} FR, {Marconi} G, et~al.
  2005.
\newblock \textit{\aap} 441:141--151

\bibitem[{{Mu{\~n}oz} et~al.(2006{\natexlab{a}}){Mu{\~n}oz}, {Carlin},
  {Frinchaboy}, {Nidever}, {Majewski} \& {Patterson}}]{Munoz06Boo}
{Mu{\~n}oz} RR, {Carlin} JL, {Frinchaboy} PM, {Nidever} DL, {Majewski} SR,
  {Patterson} RJ. 2006{\natexlab{a}}.
\newblock \textit{\apjl} 650:L51--L54

\bibitem[{{Mu{\~n}oz} et~al.(2005){Mu{\~n}oz}, {Frinchaboy}, {Majewski},
  {Kuhn}, {Chou} et~al.}]{Munoz05}
{Mu{\~n}oz} RR, {Frinchaboy} PM, {Majewski} SR, {Kuhn} JR, {Chou} MY, et~al.
  2005.
\newblock \textit{\apjl} 631:L137--L141

\bibitem[{{Mu{\~n}oz} et~al.(2006{\natexlab{b}}){Mu{\~n}oz}, {Majewski},
  {Zaggia}, {Kunkel}, {Frinchaboy} et~al.}]{Munoz06}
{Mu{\~n}oz} RR, {Majewski} SR, {Zaggia} S, {Kunkel} WE, {Frinchaboy} PM, et~al.
  2006{\natexlab{b}}.
\newblock \textit{\apj} 649:201--223

\bibitem[{{Nemec}, {Wehlau} \& {Mendes de Oliveira}(1988)}]{Nemec88}
{Nemec} JM, {Wehlau} A, {Mendes de Oliveira} C. 1988.
\newblock \textit{\aj} 96:528--559

\bibitem[{{Nissen} \& {Schuster}(1997)}]{Nissen97}
{Nissen} PE, {Schuster} WJ. 1997.
\newblock \textit{\aap} 326:751--762

\bibitem[{{Nissen} \& {Schuster}(2009)}]{Nissen09}
{Nissen} PE, {Schuster} WJ. 2009.
\newblock In \textit{IAU Symposium}, eds. J~{Andersen}, J~{Bland-Hawthorn},
  B~{Nordstr{\"o}m}, vol. 254 of \textit{IAU Symposium}

\bibitem[{{No{\"e}l} et~al.(2007){No{\"e}l}, {Gallart}, {Costa} \&
  {M{\'e}ndez}}]{Noel07}
{No{\"e}l} NED, {Gallart} C, {Costa} E, {M{\'e}ndez} RA. 2007.
\newblock \textit{\aj} 133:2037--2052

\bibitem[{{Norris} et~al.(2008){Norris}, {Gilmore}, {Wyse}, {Wilkinson},
  {Belokurov} et~al.}]{Norris08}
{Norris} JE, {Gilmore} G, {Wyse} RFG, {Wilkinson} MI, {Belokurov} V, et~al.
  2008.
\newblock \textit{\apjl} 689:L113--L116

\bibitem[{{Olive} et~al.(1995){Olive}, {Rood}, {Schramm}, {Truran} \&
  {Vangioni-Flam}}]{Olive95}
{Olive} KA, {Rood} RT, {Schramm} DN, {Truran} J, {Vangioni-Flam} E. 1995.
\newblock \textit{\apj} 444:680--685

\bibitem[{{Olive}, {Steigman} \& {Skillman}(1997)}]{Olive97}
{Olive} KA, {Steigman} G, {Skillman} ED. 1997.
\newblock \textit{\apj} 483:788--+

\bibitem[{{Olszewski}(1998)}]{EdO98}
{Olszewski} EW. 1998.
\newblock In \textit{Galactic Halos}, ed. D~{Zaritsky}, vol. 136 of
  \textit{Astronomical Society of the Pacific Conference Series}

\bibitem[{{Pagel} \& {Edmunds}(1981)}]{Pagel81}
{Pagel} BEJ, {Edmunds} MG. 1981.
\newblock \textit{\araa} 19:77--113

\bibitem[{{Pagel} \& {Tautvaisiene}(1998)}]{Pagel98}
{Pagel} BEJ, {Tautvaisiene} G. 1998.
\newblock \textit{\mnras} 299:535--544

\bibitem[{{Pancino} et~al.(2000){Pancino}, {Ferraro}, {Bellazzini}, {Piotto} \&
  {Zoccali}}]{Pancino00}
{Pancino} E, {Ferraro} FR, {Bellazzini} M, {Piotto} G, {Zoccali} M. 2000.
\newblock \textit{\apjl} 534:L83--L87

\bibitem[{{Pasquini} et~al.(2002){Pasquini}, {Avila}, {Blecha}, {Cacciari},
  {Cayatte} et~al.}]{Pasquini02}
{Pasquini} L, {Avila} G, {Blecha} A, {Cacciari} C, {Cayatte} V, et~al. 2002.
\newblock \textit{The Messenger} 110:1--9

\bibitem[{{Pe{\~n}a}, {Stasi{\'n}ska} \& {Richer}(2007)}]{Pena07}
{Pe{\~n}a} M, {Stasi{\'n}ska} G, {Richer} MG. 2007.
\newblock \textit{\aap} 476:745--758

\bibitem[{{Peimbert}, {Bohigas} \& {Torres-Peimbert}(1988)}]{Peimbert88}
{Peimbert} M, {Bohigas} J, {Torres-Peimbert} S. 1988.
\newblock \textit{Revista Mexicana de Astronomia y Astrofisica} 16:45--54

\bibitem[{{Peimbert} \& {Torres-Peimbert}(1974)}]{Peimbert74}
{Peimbert} M, {Torres-Peimbert} S. 1974.
\newblock \textit{\apj} 193:327--333

\bibitem[{{Pietrzy{\'n}ski} et~al.(2006){Pietrzy{\'n}ski}, {Gieren},
  {Soszy{\'n}ski}, {Bresolin}, {Kudritzki} et~al.}]{Pietrzynski06}
{Pietrzy{\'n}ski} G, {Gieren} W, {Soszy{\'n}ski} I, {Bresolin} F, {Kudritzki}
  RP, et~al. 2006.
\newblock \textit{\apj} 642:216--224

\bibitem[{{Pietrzy{\'n}ski} et~al.(2008){Pietrzy{\'n}ski}, {Gieren},
  {Szewczyk}, {Walker}, {Rizzi} et~al.}]{Pietrzynski08}
{Pietrzy{\'n}ski} G, {Gieren} W, {Szewczyk} O, {Walker} A, {Rizzi} L, et~al.
  2008.
\newblock \textit{\aj} 135:1993--1997

\bibitem[{{Pilyugin}(1993)}]{Pilyugin93}
{Pilyugin} LS. 1993.
\newblock \textit{\aap} 277:42--+

\bibitem[{{Pompeia} et~al.(2008){Pompeia}, {Hill}, {Spite}, {Cole}, {Primas}
  et~al.}]{Pompeia08}
{Pompeia} L, {Hill} V, {Spite} M, {Cole} A, {Primas} F, et~al. 2008.
\newblock \textit{\aa}

\bibitem[{{Queloz}, {Dubath} \& {Pasquini}(1995)}]{Queloz95}
{Queloz} D, {Dubath} P, {Pasquini} L. 1995.
\newblock \textit{\aap} 300:31--+

\bibitem[{{Recchi} et~al.(2006){Recchi}, {Hensler}, {Angeretti} \&
  {Matteucci}}]{Recchi06}
{Recchi} S, {Hensler} G, {Angeretti} L, {Matteucci} F. 2006.
\newblock \textit{\aap} 445:875--888

\bibitem[{{Recchi} et~al.(2004){Recchi}, {Matteucci}, {D'Ercole} \&
  {Tosi}}]{Recchi04}
{Recchi} S, {Matteucci} F, {D'Ercole} A, {Tosi} M. 2004.
\newblock \textit{\aap} 426:37--51

\bibitem[{{Richstone} \& {Tremaine}(1986)}]{Richstone86}
{Richstone} DO, {Tremaine} S. 1986.
\newblock \textit{\aj} 92:72--74

\bibitem[{{Rieschick} \& {Hensler}(2003)}]{Rieschick03}
{Rieschick} A, {Hensler} G. 2003.
\newblock \textit{\apss} 284:861--864

\bibitem[{{Rizzi} et~al.(2007){Rizzi}, {Held}, {Saviane}, {Tully} \&
  {Gullieuszik}}]{Rizzi07}
{Rizzi} L, {Held} EV, {Saviane} I, {Tully} RB, {Gullieuszik} M. 2007.
\newblock \textit{\mnras} 380:1255--1260

\bibitem[{{Romano}, {Tosi} \& {Matteucci}(2006)}]{Romano06}
{Romano} D, {Tosi} M, {Matteucci} F. 2006.
\newblock \textit{\mnras} 365:759--778

\bibitem[{{Ryan-Weber} et~al.(2008){Ryan-Weber}, {Begum}, {Oosterloo}, {Pal},
  {Irwin} et~al.}]{RyanWeber08}
{Ryan-Weber} EV, {Begum} A, {Oosterloo} T, {Pal} S, {Irwin} MJ, et~al. 2008.
\newblock \textit{\mnras} 384:535--540

\bibitem[{{Sabbi} et~al.(2007){Sabbi}, {Sirianni}, {Nota}, {Tosi}, {Gallagher}
  et~al.}]{Sabbi07}
{Sabbi} E, {Sirianni} M, {Nota} A, {Tosi} M, {Gallagher} J, et~al. 2007.
\newblock \textit{\aj} 133:44--57

\bibitem[{{Sadakane} et~al.(2004){Sadakane}, {Arimoto}, {Ikuta}, {Aoki},
  {Jablonka} \& {Tajitsu}}]{Sadakane04}
{Sadakane} K, {Arimoto} N, {Ikuta} C, {Aoki} W, {Jablonka} P, {Tajitsu} A.
  2004.
\newblock \textit{\pasj} 56:1041--1058

\bibitem[{{Saha} \& {Hoessel}(1990)}]{Saha90}
{Saha} A, {Hoessel} JG. 1990.
\newblock \textit{\aj} 99:97--148

\bibitem[{{Saha}, {Hoessel} \& {Krist}(1992)}]{Saha92}
{Saha} A, {Hoessel} JG, {Krist} J. 1992.
\newblock \textit{\aj} 103:84--+

\bibitem[{{Saha}, {Monet} \& {Seitzer}(1986)}]{Saha86}
{Saha} A, {Monet} DG, {Seitzer} P. 1986.
\newblock \textit{\aj} 92:302--327

\bibitem[{{Sakai}, {Madore} \& {Freedman}(1997)}]{Sakai97}
{Sakai} S, {Madore} BF, {Freedman} WL. 1997.
\newblock \textit{\apj} 480:589--+

\bibitem[{{Salvadori} \& {Ferrara}(2009)}]{Salvadori09}
{Salvadori} S, {Ferrara} A. 2009.
\newblock \textit{\mnras} 395:L6--L10

\bibitem[{{Salvadori}, {Ferrara} \& {Schneider}(2008)}]{Salvadori08}
{Salvadori} S, {Ferrara} A, {Schneider} R. 2008.
\newblock \textit{\mnras} 386:348--358

\bibitem[{{Saviane}, {Held} \& {Piotto}(1996)}]{Saviane96}
{Saviane} I, {Held} EV, {Piotto} G. 1996.
\newblock \textit{\aap} 315:40--51

\bibitem[{{Sbordone} et~al.(2007){Sbordone}, {Bonifacio}, {Buonanno},
  {Marconi}, {Monaco} \& {Zaggia}}]{Sbordone07}
{Sbordone} L, {Bonifacio} P, {Buonanno} R, {Marconi} G, {Monaco} L, {Zaggia} S.
  2007.
\newblock \textit{\aap} 465:815--824

\bibitem[{{Schoerck} et~al.(2008){Schoerck}, {Christlieb}, {Cohen}, {Beers},
  {Shectman} et~al.}]{Schoerck08}
{Schoerck} T, {Christlieb} N, {Cohen} JG, {Beers} TC, {Shectman} S, et~al.
  2008.
\newblock \textit{submitted to A\&A arXiv:0809.1172}

\bibitem[{{Schulte-Ladbeck} et~al.(2000){Schulte-Ladbeck}, {Hopp}, {Greggio} \&
  {Crone}}]{Schulte00}
{Schulte-Ladbeck} RE, {Hopp} U, {Greggio} L, {Crone} MM. 2000.
\newblock \textit{\aj} 120:1713--1730

\bibitem[{{Schulte-Ladbeck} et~al.(2001){Schulte-Ladbeck}, {Hopp}, {Greggio},
  {Crone} \& {Drozdovsky}}]{Schulte01}
{Schulte-Ladbeck} RE, {Hopp} U, {Greggio} L, {Crone} MM, {Drozdovsky} IO. 2001.
\newblock \textit{\aj} 121:3007--3025

\bibitem[{{Searle}, {Sargent} \& {Bagnuolo}(1973)}]{Searle73}
{Searle} L, {Sargent} WLW, {Bagnuolo} WG. 1973.
\newblock \textit{\apj} 179:427--438

\bibitem[{{Seiden}, {Schulman} \& {Gerola}(1979)}]{Seiden79}
{Seiden} PE, {Schulman} LS, {Gerola} H. 1979.
\newblock \textit{\apj} 232:702--706

\bibitem[{{Shetrone} et~al.(2003){Shetrone}, {Venn}, {Tolstoy}, {Primas},
  {Hill} \& {Kaufer}}]{Shetrone03}
{Shetrone} M, {Venn} KA, {Tolstoy} E, {Primas} F, {Hill} V, {Kaufer} A. 2003.
\newblock \textit{\aj} 125:684--706

\bibitem[{{Shetrone}, {Bolte} \& {Stetson}(1998)}]{Shetrone98}
{Shetrone} MD, {Bolte} M, {Stetson} PB. 1998.
\newblock \textit{\aj} 115:1888--1893

\bibitem[{{Shetrone}, {C{\^o}t{\'e}} \& {Sargent}(2001)}]{Shetrone01}
{Shetrone} MD, {C{\^o}t{\'e}} P, {Sargent} WLW. 2001.
\newblock \textit{\apj} 548:592--608

\bibitem[{{Shetrone} et~al.(2009){Shetrone}, {Siegel}, {Cook} \&
  {Bosler}}]{Shetrone09}
{Shetrone} MD, {Siegel} MH, {Cook} DO, {Bosler} T. 2009.
\newblock \textit{\aj} 137:62--71

\bibitem[{{Siegel}(2006)}]{Siegel06}
{Siegel} MH. 2006.
\newblock \textit{\apjl} 649:L83--L86

\bibitem[{{Siegel}, {Shetrone} \& {Irwin}(2008)}]{Siegel08}
{Siegel} MH, {Shetrone} MD, {Irwin} M. 2008.
\newblock \textit{\aj} 135:2084--2094

\bibitem[{{Simmerer} et~al.(2004){Simmerer}, {Sneden}, {Cowan}, {Collier},
  {Woolf} \& {Lawler}}]{Simmerer04}
{Simmerer} J, {Sneden} C, {Cowan} JJ, {Collier} J, {Woolf} VM, {Lawler} JE.
  2004.
\newblock \textit{\apj} 617:1091--1114

\bibitem[{{Simon} \& {Geha}(2007)}]{Simon07}
{Simon} JD, {Geha} M. 2007.
\newblock \textit{\apj} 670:313--331

\bibitem[{{Skillman} \& {Bender}(1995)}]{Skillman95}
{Skillman} ED, {Bender} R. 1995.
\newblock In \textit{Revista Mexicana de Astronomia y Astrofisica Conference
  Series}, eds. M~{Pena}, S~{Kurtz}, vol.~3 of \textit{Revista Mexicana de
  Astronomia y Astrofisica Conference Series}

\bibitem[{{Skillman}, {Bomans} \& {Kobulnicky}(1997)}]{Skillman97}
{Skillman} ED, {Bomans} DJ, {Kobulnicky} HA. 1997.
\newblock \textit{\apj} 474:205--+

\bibitem[{{Skillman} \& {Gallart}(2002)}]{Skillman02}
{Skillman} ED, {Gallart} C. 2002.
\newblock In \textit{Observed HR Diagrams and Stellar Evolution}, eds.
  T~{Lejeune}, J~{Fernandes}, vol. 274 of \textit{Astronomical Society of the
  Pacific Conference Series}

\bibitem[{{Skillman}, {Kennicutt} \& {Hodge}(1989)}]{Skillman89a}
{Skillman} ED, {Kennicutt} RC, {Hodge} PW. 1989.
\newblock \textit{\apj} 347:875--882

\bibitem[{{Skillman}, {Terlevich} \& {Melnick}(1989)}]{Skillman89b}
{Skillman} ED, {Terlevich} R, {Melnick} J. 1989.
\newblock \textit{\mnras} 240:563--572

\bibitem[{{Skillman} et~al.(2003){Skillman}, {Tolstoy}, {Cole}, {Dolphin},
  {Saha} et~al.}]{Skillman03}
{Skillman} ED, {Tolstoy} E, {Cole} AA, {Dolphin} AE, {Saha} A, et~al. 2003.
\newblock \textit{\apj} 596:253--272

\bibitem[{{Smecker-Hane} et~al.(1994){Smecker-Hane}, {Stetson}, {Hesser} \&
  {Lehnert}}]{Smecker94}
{Smecker-Hane} TA, {Stetson} PB, {Hesser} JE, {Lehnert} MD. 1994.
\newblock \textit{\aj} 108:507--513

\bibitem[{{Smecker-Hane} et~al.(1996){Smecker-Hane}, {Stetson}, {Hesser} \&
  {Vandenberg}}]{Smecker96}
{Smecker-Hane} TA, {Stetson} PB, {Hesser} JE, {Vandenberg} DA. 1996.
\newblock In \textit{From Stars to Galaxies: the Impact of Stellar Physics on
  Galaxy Evolution}, eds. C~{Leitherer}, U~{Fritze-von-Alvensleben},
  J~{Huchra}, vol.~98 of \textit{Astronomical Society of the Pacific Conference
  Series}

\bibitem[{{Smith} et~al.(2006){Smith}, {Siegel}, {Shetrone} \&
  {Winnick}}]{Smith06}
{Smith} GH, {Siegel} MH, {Shetrone} MD, {Winnick} R. 2006.
\newblock \textit{\pasp} 118:1361--1372

\bibitem[{{Sneden}, {Cowan} \& {Gallino}(2008)}]{Sneden08}
{Sneden} C, {Cowan} JJ, {Gallino} R. 2008.
\newblock \textit{\araa} 46:241--288

\bibitem[{{Soszy{\'n}ski} et~al.(2006){Soszy{\'n}ski}, {Gieren},
  {Pietrzy{\'n}ski}, {Bresolin}, {Kudritzki} \& {Storm}}]{Soszynski06}
{Soszy{\'n}ski} I, {Gieren} W, {Pietrzy{\'n}ski} G, {Bresolin} F, {Kudritzki}
  RP, {Storm} J. 2006.
\newblock \textit{\apj} 648:375--382

\bibitem[{{Stark} et~al.(2007){Stark}, {Bunker}, {Ellis}, {Eyles} \&
  {Lacy}}]{Stark07}
{Stark} DP, {Bunker} AJ, {Ellis} RS, {Eyles} LP, {Lacy} M. 2007.
\newblock \textit{\apj} 659:84--97

\bibitem[{{Starkenburg} et~al.(2008){Starkenburg}, {Hill}, {Irwin}, {Battaglia}
  \& A.}]{Starkenburg08}
{Starkenburg} E, {Hill} V, {Irwin} M, {Battaglia} G, A. H. 2008.
\newblock \textit{GARCON08, poster,
  http://www.mpa-garching.mpg.de/mpa/conferences/garcon08/Posters/Starkenburg.%
pdf}

\bibitem[{{Suntzeff} et~al.(1993){Suntzeff}, {Mateo}, {Terndrup}, {Olszewski},
  {Geisler} \& {Weller}}]{Suntzeff93}
{Suntzeff} NB, {Mateo} M, {Terndrup} DM, {Olszewski} EW, {Geisler} D, {Weller}
  W. 1993.
\newblock \textit{\apj} 418:208--+

\bibitem[{{Tammann}(1994)}]{Tammann94}
{Tammann} GA. 1994.
\newblock In \textit{European Southern Observatory Astrophysics Symposia}, eds.
  G~{Meylan}, P~{Prugniel}, vol.~49 of \textit{European Southern Observatory
  Astrophysics Symposia}

\bibitem[{{Tassis} et~al.(2003){Tassis}, {Abel}, {Bryan} \&
  {Norman}}]{Tassis03}
{Tassis} K, {Abel} T, {Bryan} GL, {Norman} ML. 2003.
\newblock \textit{\apj} 587:13--24

\bibitem[{{Tautvai{\v s}ien{\.e}} et~al.(2007){Tautvai{\v s}ien{\.e}},
  {Geisler}, {Wallerstein}, {Borissova}, {Bizyaev} et~al.}]{Tautvaisiene07}
{Tautvai{\v s}ien{\.e}} G, {Geisler} D, {Wallerstein} G, {Borissova} J,
  {Bizyaev} D, et~al. 2007.
\newblock \textit{\aj} 134:2318--2327

\bibitem[{{Tenorio-Tagle}(1996)}]{TenorioTagle96}
{Tenorio-Tagle} G. 1996.
\newblock \textit{\aj} 111:1641--+

\bibitem[{{Thuan}, {Lecavelier des Etangs} \& {Izotov}(2002)}]{Thuan02}
{Thuan} TX, {Lecavelier des Etangs} A, {Izotov} YI. 2002.
\newblock \textit{\apj} 565:941--951

\bibitem[{{Tinsley}(1968)}]{Tinsley68}
{Tinsley} BM. 1968.
\newblock \textit{\apj} 151:547--+

\bibitem[{{Tinsley}(1980)}]{Tinsley80}
{Tinsley} BM. 1980.
\newblock \textit{Fundamentals of Cosmic Physics} 5:287--388

\bibitem[{{Tolstoy}(1996)}]{Tolstoy96a}
{Tolstoy} E. 1996.
\newblock \textit{\apj} 462:684--+

\bibitem[{{Tolstoy}(2003)}]{Tolstoy03hst}
{Tolstoy} E. 2003.
\newblock In \textit{A Decade of Hubble Space Telescope Science}, eds.
  M~{Livio}, K~{Noll}, M~{Stiavelli}

\bibitem[{{Tolstoy} et~al.(2006){Tolstoy}, {Hill}, {Irwin}, {Helmi},
  {Battaglia} et~al.}]{Tolstoy06}
{Tolstoy} E, {Hill} V, {Irwin} M, {Helmi} A, {Battaglia} G, et~al. 2006.
\newblock \textit{The Messenger} 123:33--+

\bibitem[{{Tolstoy} \& {Irwin}(2000)}]{Tolstoy00}
{Tolstoy} E, {Irwin} M. 2000.
\newblock \textit{\mnras} 318:1241--1248

\bibitem[{{Tolstoy} et~al.(2001){Tolstoy}, {Irwin}, {Cole}, {Pasquini},
  {Gilmozzi} \& {Gallagher}}]{Tolstoy01}
{Tolstoy} E, {Irwin} MJ, {Cole} AA, {Pasquini} L, {Gilmozzi} R, {Gallagher} JS.
  2001.
\newblock \textit{\mnras} 327:918--938

\bibitem[{{Tolstoy} et~al.(2004){Tolstoy}, {Irwin}, {Helmi}, {Battaglia},
  {Jablonka} et~al.}]{Tolstoy04}
{Tolstoy} E, {Irwin} MJ, {Helmi} A, {Battaglia} G, {Jablonka} P, et~al. 2004.
\newblock \textit{\apjl} 617:L119--L122

\bibitem[{{Tolstoy} \& {Saha}(1996)}]{Tolstoy96b}
{Tolstoy} E, {Saha} A. 1996.
\newblock \textit{\apj} 462:672--+

\bibitem[{{Tolstoy} et~al.(1995){Tolstoy}, {Saha}, {Hoessel} \&
  {Danielson}}]{Tolstoy95}
{Tolstoy} E, {Saha} A, {Hoessel} JG, {Danielson} GE. 1995.
\newblock \textit{\aj} 109:579--587

\bibitem[{{Tolstoy} et~al.(2003){Tolstoy}, {Venn}, {Shetrone}, {Primas}, {Hill}
  et~al.}]{Tolstoy03}
{Tolstoy} E, {Venn} KA, {Shetrone} M, {Primas} F, {Hill} V, et~al. 2003.
\newblock \textit{\aj} 125:707--726

\bibitem[{{Tosi}(1998)}]{Tosi98}
{Tosi} M. 1998.
\newblock In \textit{Dwarf Galaxies and Cosmology}, eds. V~{Thuan}
  T.X.~{Balkowski} C.~{Cayatte}, J~{Thanh Van}, vol.~33 of \textit{Rencontres
  de Moriond}

\bibitem[{{Tosi}(2007{\natexlab{a}})}]{Tosi07hst}
{Tosi} M. 2007{\natexlab{a}}.
\newblock \textit{41st ESLAB Symposium "The impact of HST on European
  Astronomy", arXiv:0707.3057}

\bibitem[{{Tosi}(2007{\natexlab{b}})}]{Tosi07}
{Tosi} M. 2007{\natexlab{b}}.
\newblock In \textit{From Stars to Galaxies: Building the Pieces to Build Up
  the Universe}, eds. A~{Vallenari}, R~{Tantalo}, L~{Portinari}, A~{Moretti},
  vol. 374 of \textit{Astronomical Society of the Pacific Conference Series}

\bibitem[{{Tosi} et~al.(2008){Tosi}, {Gallagher}, {Sabbi}, {Glatt}, {Grebel}
  et~al.}]{Tosi08}
{Tosi} M, {Gallagher} J, {Sabbi} E, {Glatt} K, {Grebel} EK, et~al. 2008.
\newblock \textit{IAU Symposium 255: Low-Metallicity Star Formation: From the
  First Stars to Dwarf Galaxies, arXiv:0808.1182} 808

\bibitem[{{Tosi} et~al.(1991){Tosi}, {Greggio}, {Marconi} \&
  {Focardi}}]{Tosi91}
{Tosi} M, {Greggio} L, {Marconi} G, {Focardi} P. 1991.
\newblock \textit{\aj} 102:951--974

\bibitem[{{Travaglio} et~al.(2004){Travaglio}, {Gallino}, {Arnone}, {Cowan},
  {Jordan} \& {Sneden}}]{Travaglio04}
{Travaglio} C, {Gallino} R, {Arnone} E, {Cowan} J, {Jordan} F, {Sneden} C.
  2004.
\newblock \textit{\apj} 601:864--884

\bibitem[{{Tsujimoto} et~al.(1995){Tsujimoto}, {Nomoto}, {Yoshii}, {Hashimoto},
  {Yanagida} \& {Thielemann}}]{Tsujimoto95}
{Tsujimoto} T, {Nomoto} K, {Yoshii} Y, {Hashimoto} M, {Yanagida} S,
  {Thielemann} FK. 1995.
\newblock \textit{\mnras} 277:945--958

\bibitem[{{Urbaneja} et~al.(2008){Urbaneja}, {Kudritzki}, {Bresolin},
  {Przybilla}, {Gieren} \& {Pietrzy{\'n}ski}}]{Urbaneja08}
{Urbaneja} MA, {Kudritzki} RP, {Bresolin} F, {Przybilla} N, {Gieren} W,
  {Pietrzy{\'n}ski} G. 2008.
\newblock \textit{\apj} 684:118--135

\bibitem[{{Vallenari}, {Schmidtobreick} \& {Bomans}(2005)}]{Vallenari05}
{Vallenari} A, {Schmidtobreick} L, {Bomans} DJ. 2005.
\newblock \textit{\aap} 435:821--829

\bibitem[{{van Zee} \& {Haynes}(2006)}]{vanZee06b}
{van Zee} L, {Haynes} MP. 2006.
\newblock \textit{\apj} 636:214--239

\bibitem[{{van Zee}, {Skillman} \& {Haynes}(2006)}]{vanZee06}
{van Zee} L, {Skillman} ED, {Haynes} MP. 2006.
\newblock \textit{\apj} 637:269--282

\bibitem[{{Veilleux}, {Cecil} \& {Bland-Hawthorn}(2005)}]{Veilleux05}
{Veilleux} S, {Cecil} G, {Bland-Hawthorn} J. 2005.
\newblock \textit{\araa} 43:769--826

\bibitem[{{Venn}(1999)}]{Venn99}
{Venn} KA. 1999.
\newblock \textit{\apj} 518:405--421

\bibitem[{{Venn} et~al.(2004{\natexlab{a}}){Venn}, {Irwin}, {Shetrone}, {Tout},
  {Hill} \& {Tolstoy}}]{Venn04}
{Venn} KA, {Irwin} M, {Shetrone} MD, {Tout} CA, {Hill} V, {Tolstoy} E.
  2004{\natexlab{a}}.
\newblock \textit{\aj} 128:1177--1195

\bibitem[{{Venn} et~al.(2001){Venn}, {Lennon}, {Kaufer}, {McCarthy},
  {Przybilla} et~al.}]{Venn01}
{Venn} KA, {Lennon} DJ, {Kaufer} A, {McCarthy} JK, {Przybilla} N, et~al. 2001.
\newblock \textit{\apj} 547:765--776

\bibitem[{{Venn} et~al.(2004{\natexlab{b}}){Venn}, {Tolstoy}, {Kaufer} \&
  {Kudritzki}}]{Venn04conf}
{Venn} KA, {Tolstoy} E, {Kaufer} A, {Kudritzki} RP. 2004{\natexlab{b}}.
\newblock In \textit{Origin and Evolution of the Elements}, eds. A~{McWilliam},
  M~{Rauch}

\bibitem[{{Venn} et~al.(2003){Venn}, {Tolstoy}, {Kaufer}, {Skillman},
  {Clarkson} et~al.}]{Venn03}
{Venn} KA, {Tolstoy} E, {Kaufer} A, {Skillman} ED, {Clarkson} SM, et~al. 2003.
\newblock \textit{\aj} 126:1326--1345

\bibitem[{{Vermeij} \& {van der Hulst}(2002)}]{Vermeij02}
{Vermeij} R, {van der Hulst} JM. 2002.
\newblock \textit{\aap} 391:1081--1095

\bibitem[{{Walker}, {Mateo} \& {Olszewski}(2009)}]{Walker09}
{Walker} MG, {Mateo} M, {Olszewski} EW. 2009.
\newblock \textit{\aj} 137:3100--3108

\bibitem[{{Walker} et~al.(2006{\natexlab{a}}){Walker}, {Mateo}, {Olszewski},
  {Bernstein}, {Wang} \& {Woodroofe}}]{Walker06fnx}
{Walker} MG, {Mateo} M, {Olszewski} EW, {Bernstein} R, {Wang} X, {Woodroofe} M.
  2006{\natexlab{a}}.
\newblock \textit{\aj} 131:2114--2139

\bibitem[{{Walker} et~al.(2006{\natexlab{b}}){Walker}, {Mateo}, {Olszewski},
  {Pal}, {Sen} \& {Woodroofe}}]{Walker06sext}
{Walker} MG, {Mateo} M, {Olszewski} EW, {Pal} JK, {Sen} B, {Woodroofe} M.
  2006{\natexlab{b}}.
\newblock \textit{\apjl} 642:L41--L44

\bibitem[{{Westfall} et~al.(2006){Westfall}, {Majewski}, {Ostheimer},
  {Frinchaboy}, {Kunkel} et~al.}]{Westfall06}
{Westfall} KB, {Majewski} SR, {Ostheimer} JC, {Frinchaboy} PM, {Kunkel} WE,
  et~al. 2006.
\newblock \textit{\aj} 131:375--406

\bibitem[{{Westmoquette}, {Smith} \& {Gallagher}(2008)}]{Westmoquette08}
{Westmoquette} MS, {Smith} LJ, {Gallagher} JS. 2008.
\newblock \textit{\mnras} 383:864--878

\bibitem[{{Whiting}, {Hau} \& {Irwin}(1999)}]{Whiting99}
{Whiting} AB, {Hau} GKT, {Irwin} M. 1999.
\newblock \textit{\aj} 118:2767--2774

\bibitem[{{Wilkinson} et~al.(2004){Wilkinson}, {Kleyna}, {Evans}, {Gilmore},
  {Irwin} \& {Grebel}}]{Wilkinson04}
{Wilkinson} MI, {Kleyna} JT, {Evans} NW, {Gilmore} GF, {Irwin} MJ, {Grebel} EK.
  2004.
\newblock \textit{\apjl} 611:L21--L24

\bibitem[{{Willman} et~al.(2005){Willman}, {Dalcanton}, {Martinez-Delgado},
  {West}, {Blanton} et~al.}]{Willman05a}
{Willman} B, {Dalcanton} JJ, {Martinez-Delgado} D, {West} AA, {Blanton} MR,
  et~al. 2005.
\newblock \textit{\apjl} 626:L85--L88

\bibitem[{{Wirth} \& {Gallagher}(1984)}]{Wirth84}
{Wirth} A, {Gallagher} III JS. 1984.
\newblock \textit{\apj} 282:85--94

\bibitem[{{Wolf}(1973)}]{Wolf73}
{Wolf} B. 1973.
\newblock \textit{\aap} 28:335--348

\bibitem[{{Wyder}(2001)}]{Wyder01}
{Wyder} TK. 2001.
\newblock \textit{\aj} 122:2490--2523

\bibitem[{{Wyder}(2003)}]{Wyder03}
{Wyder} TK. 2003.
\newblock \textit{\aj} 125:3097--3110

\bibitem[{{Young} \& {Lo}(1997)}]{Young97}
{Young} LM, {Lo} KY. 1997.
\newblock \textit{\apj} 490:710--+

\bibitem[{{Young} et~al.(2007){Young}, {Skillman}, {Weisz} \&
  {Dolphin}}]{Young07}
{Young} LM, {Skillman} ED, {Weisz} DR, {Dolphin} AE. 2007.
\newblock \textit{\apj} 659:331--338

\bibitem[{{Young} et~al.(2003){Young}, {van Zee}, {Lo}, {Dohm-Palmer} \&
  {Beierle}}]{Young03}
{Young} LM, {van Zee} L, {Lo} KY, {Dohm-Palmer} RC, {Beierle} ME. 2003.
\newblock \textit{\apj} 592:111--128

\bibitem[{{Zucker} et~al.(2006{\natexlab{a}}){Zucker}, {Belokurov}, {Evans},
  {Kleyna}, {Irwin} et~al.}]{Zucker06a}
{Zucker} DB, {Belokurov} V, {Evans} NW, {Kleyna} JT, {Irwin} MJ, et~al.
  2006{\natexlab{a}}.
\newblock \textit{\apjl} 650:L41--L44

\bibitem[{{Zucker} et~al.(2006{\natexlab{b}}){Zucker}, {Belokurov}, {Evans},
  {Wilkinson}, {Irwin} et~al.}]{Zucker06b}
{Zucker} DB, {Belokurov} V, {Evans} NW, {Wilkinson} MI, {Irwin} MJ, et~al.
  2006{\natexlab{b}}.
\newblock \textit{\apjl} 643:L103--L106

\end{thebibliography}
